\documentclass[aps,twocolumn,prl,floatfix,showpacs,citeautoscript,superscriptaddress,longbibliography]{revtex4-1}

\makeatletter
\def\@hangfrom@section#1#2#3{\@hangfrom{#1#2}#3}
\def\@hangfroms@section#1#2{#1#2}
\makeatother

\usepackage[none]{hyphenat}
\usepackage{amssymb}
\usepackage{amsmath}
\usepackage{graphicx}
\usepackage{bm}
\usepackage{times}
\usepackage{color}
\usepackage[dvipsnames,svgnames,table]{xcolor}
\usepackage{hyperref}
\usepackage{fancyhdr}
\usepackage{enumitem}
\usepackage[english]{babel}
\usepackage[caption=false]{subfig}
\usepackage{braket}
\usepackage{scalerel}
\usepackage{graphicx}
\usepackage{mathrsfs}
\usepackage{MnSymbol}
\usepackage{lipsum}
\usepackage{soul}
\hypersetup{
pdfstartview={FitH},
colorlinks=true,    
linkcolor=NavyBlue, 
citecolor=Blue,   
filecolor=NavyBlue, 
urlcolor=NavyBlue   
}

\graphicspath{{/Users/nicolodefenu/Dropbox/generalized_higgs/Figs}}

\begin{document}

\title{Equilibrium Parametric Amplification in Raman-Cavity Hybrids}

\author{H. P. {Ojeda Collado}}
\email{hojedaco@physnet.uni-hamburg.de }
\affiliation{Center for Optical Quantum Technologies and Institute for Quantum Physics, University of Hamburg, 22761 Hamburg, Germany}
\affiliation{The Hamburg Center for Ultrafast Imaging, Luruper Chaussee 149, 22761 Hamburg, Germany}
\author{Marios H. Michael}
\email{marios.michael@mpsd.mpg.de}
\affiliation{Max Planck Institute for the Structure and Dynamics of Matter, Luruper Chausse 149, 22761 Hamburg, Germany}
\author{Jim {Skulte}}
\affiliation{Center for Optical Quantum Technologies and Institute for Quantum Physics, University of Hamburg, 22761 Hamburg, Germany}
\affiliation{The Hamburg Center for Ultrafast Imaging, Luruper Chaussee 149, 22761 Hamburg, Germany}
\author{Angel Rubio}
\affiliation{Max Planck Institute for the Structure and Dynamics of Matter, Luruper Chausse 149, 22761 Hamburg, Germany}
\affiliation{Center for Computational Quantum Physics, The Flatiron Institute, 162 Fifth Avenue, New York, New York 10010, USA}
\author{Ludwig Mathey}
\affiliation{Center for Optical Quantum Technologies and Institute for Quantum Physics, University of Hamburg, 22761 Hamburg, Germany}
\affiliation{The Hamburg Center for Ultrafast Imaging, Luruper Chaussee 149, 22761 Hamburg, Germany}

\begin{abstract}
Parametric resonances and amplification have led to extraordinary photoinduced phenomena in pump-probe experiments. While these phenomena manifest themselves in out-of-equilibrium settings, here, we present the striking result of parametric amplification in equilibrium. We demonstrate that quantum and thermal fluctuations of a Raman-active mode amplifies light inside a cavity, at equilibrium, when the Raman mode frequency is twice the cavity mode frequency. This noise-driven amplification leads to the creation of an unusual parametric Raman polariton, intertwining the Raman mode with cavity squeezing fluctuations, with smoking gun signatures in Raman spectroscopy. In the resonant regime, we show the emergence of not only quantum light amplification but also localization and static shift of the Raman mode. Apart from the fundamental interest of equilibrium parametric amplification our study suggests a resonant mechanism for controlling Raman modes and thus matter properties by cavity fluctuations. We conclude by outlining how to compute the Raman-cavity coupling, and suggest possible experimental realizations.
\end{abstract}

\date{\today}

\maketitle

{\em Introduction.-} 
Driving condensed matter with light provides a methodology of controlling its properties in a dynamic fashion, in contrast to the established, static methods, as reflected in recent scientific studies~\cite{de_la_torre_colloquium_2021,basov_towards_2017,Bloch_22,Kennes2023}. In this effort, driving matter with laser light has proved to be a remarkably versatile tool in engineering properties of quantum materials such as controlling ferro-electricity~\cite{Nova19}, magnetism~\cite{disa_photo-induced_2023,siegrist_light-wave_2019,Samuel_21, Emil_20}, superconductivity~\cite{mitrano_possible_2016,budden_evidence_2021, rowe_giant_2023,chattopadhyay_mechanisms_2023,von_hoegen_amplification_2022, michael_parametric_2020}, topological features~\cite{mciver_light-induced_2020} and charge ordering~\cite{kogar_light-induced_2020,dolgirev_self-similar_2020}. Even more interestingly, driving with light has provided the possibility to create novel nonequilibrium states. A notable example includes photonic time crystals~\cite{Lyubarov_2022,michael_generalized_2022,michael2023theory,haque2023terahertz, dolgirev_periodic_2022}, materials that can function as parametric amplifiers for light. Another example is time crystals, denoting a robust, collective dynamical many-body state, in which the response of observables oscillates subharmonically~\cite{Else2020,Zhang2017,Choi2017,Kessler2021,Kongkhambut2021,Taheri2022,Zaletel2023,HP_2021,HP_2023}.

The conceptual approach of dynamical control with light can be extended to the equilibrium domain through cavity-matter hybrids, see e.g.~\cite{schlawin_cavity_2022,curtis2023local,Ruggi_18}. This advancement of control via light involves replacing laser driving by quantum light fluctuations which are strongly coupled to matter through cavities~\cite{jarc_tunable_2022}, plasmonic resonators~\cite{appugliese_breakdown_2022}, surfaces hosting surface phonon polaritons~\cite{Lenk_2022, eckhardt_theory_2023} and photonic crystals~\cite{Baydin_23}. The feasibility of this approach has been demonstrated experimentally, with examples including manipulation of transport~\cite{orgiu_conductivity_2015}, control of superconducting properties~\cite{thomas_exploring_2019}, magnetism~\cite{thomas_large_2021}, topological features~\cite{appugliese_breakdown_2022}, cavity control of chemical reactivity~\cite{thomas_tilting_2019,nagarajan_chemistry_2021,Sclaefer_19} and cavity-enhanced spectroscopy~\cite{Fainstein95,Sulzer22,Melanie16,Gagliardi13}.

In this Letter, we demonstrate that quantum and thermal \emph{noise} of a Raman-active mode, can amplify cavity fluctuations in equilibrium.
We emphasize that parametric amplification generally occurs in driven systems while here we present it in the context of an equilibrium amplification process. This amplification can in turn be used to resonantly control properties of matter and  constitutes a novel method of light control, for Raman-cavity systems.

Our starting point is the nonlinear coupling between Raman active collective modes and light. Here Raman-active modes could be Raman phonons~\cite{forst_nonlinear_2011,Schmidt,Sanker}, molecular vibrations~\cite{chattopadhyay_mechanisms_2023}, Higgs modes in superconductors~\cite{Matsunaga2014,Buzzi21,HP_2018} and amplitude modes in charge density waves~\cite{dolgirev_self-similar_2020} that are even under inversion symmetry, and the electric field of a local cavity mode~\cite{juraschek_cavity_2021}. Therefore, at leading order, the Raman-light Hamiltonian reads $H_{Raman-light} = \lambda Q E_{cav}^2$
where $E_{cav}$ is the electric field in the cavity, $Q$ is the coordinate of the Raman mode and $\lambda$ is the light-matter coupling. This quadratic coupling includes parametrically resonant processes of the type $\hat{a}^\dag \hat{a}^\dag \hat{b} + \hat{a} \hat{a} \hat{b}^\dag$
where $\hat{a}$ and $\hat{b}$ are the photon and Raman annihilation operators respectively. In the presence of coherent Raman oscillations, $\langle b(t) \rangle = A_0 e^{i \omega_R t}$, at the Raman frequency $\omega_R$, the above coupling leads to exponential growth of the light field when the cavity frequency $\omega_c$ satisfies the parametric resonant condition, $2 \omega_c = \omega_R $. This observation naturally leads to the question: can a randomly fluctuating field coming from Raman quantum fluctuations also amplify light? We find that the answer is yes which we demonstrate below.

\begin{figure}[!tb]
\hspace*{0.cm} 
\includegraphics[width=0.5\textwidth]{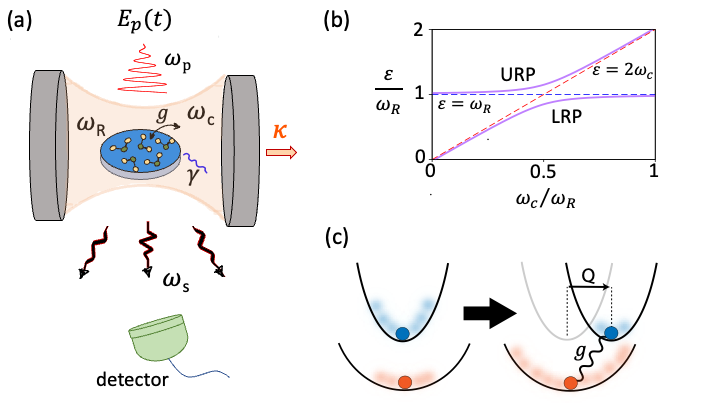}
\caption{Parametric Raman polaritons. (a) Sketch of a Raman medium (blue) coupled to a single cavity mode of frequency $\omega_c$ (light-red shading) with a constant coupling $g$. Light can leak out of the cavity at a total decay rate $\kappa$ while $\gamma$ represents the damping associated with the Raman mode $\omega_R$. The wavy lines at the top and bottom represent a Raman spectroscopy scheme in which a probe field $E_p(t)$ of frequency $\omega_p$ is sent into the sample, and a detector (in green) collects the scattered photons at different frequencies $\omega_s$ giving information about the hybrid Raman-cavity system. (b) Representative Raman spectrum (in purple) in which an avoided crossing appears at the resonant condition $\omega_R=2\omega_c$ indicating the existence of two branches, the upper (URP) and lower (LRP) Raman polaritons. (c) Sketch of how Raman (blue) and cavity (red) properties are modified due to the coupling $g$. Full circles represent the equilibrium position and shaded regions indicate fluctuations. In the resonant regime the cavity fluctuations are amplified while the Raman mode is statically shifted and localized, i.e. its fluctuations decrease.
} 
\label{fig:cartoon}
\end{figure}

To study this phenomenon, we use an open truncated-Wigner approximation method (open TWA)~\cite{polkovnikov_phase_2010,cosme_time_2019,Skulte2023} to simulate the semiclassical dynamics of the Raman-cavity hybrids in the quantum fluctuation regime. We also determine the signatures of equilibrium parametric amplification using Raman spectroscopy (Fig.~\ref{fig:cartoon} (a)). We find that a prominent feature of parametric resonance and equilibrium light amplification is the appearance of two Raman polariton branches as shown in Fig.~\ref{fig:cartoon}(b). We call this polariton parametric Raman polariton and its formation is attributed to the \emph{nonlinear} process of mixing \emph{squeezed photon fluctuations} with the Raman coordinate. This is substantially different to the existent polariton panorama where polaritons typically arise from a \emph{linear} coupling between matter degrees of freedom and light~\cite{Polaritonpanorama}. To quantify the coupling between the Raman mode and the cavity, we compute the resonance splitting between the upper and lower parametric Raman polariton. This is a nonlinear extension of the usual Rabi splitting in the case of infrared active phonon polaritons~\cite{jarc_tunable_2022}. We use a Gaussian theory to provide an analytical expression for this splitting as a function of the coupling strength which agrees well with the simulations.

The key features of the parametric Raman polariton are as follows: (i) The vacuum fluctuations of the cavity mode are amplified. (ii) The fluctuations of the Raman are reduced, in response to the amplification of the cavity fluctuations. This corresponds to a localization of the Raman mode. (iii) The average position of the Raman mode is statically shifted due to the cavity fluctuations as shown schematically in Fig.~\ref{fig:cartoon}(c). These observations suggest that this mechanism can be used to resonantly modify and control both the Raman mode and the cavity in equilibrium. Furthermore we conclude by proposing realistic experimental set-ups where this phenomenon can be observed.

{\em Raman-Cavity Model and Raman spectroscopy.-} We consider a model, in which cavity field fluctuations are locally coupled to a Raman coordinate. The Hamiltonian for this system is given by:
\begin{equation}
    \frac{H}{\hbar}=\omega_c \hat{a}^{\dagger}\hat{a}+\omega_R \hat{b}^{\dagger}\hat{b}+g\left (\hat{b}^{\dagger}+\hat{b}\right )\left (\hat{a}^{\dagger}+\hat{a}\right)^{2}+\frac{g_4}{4}\left (\hat{a}^{\dagger}+\hat{a}\right )^{4}.
    \label{eq:ham}
\end{equation}
The cavity creation (annihilation) operator is $\hat{a}^{\dagger}$ ($\hat{a}$) and $\omega_c$ is the cavity frequency. $\hat{b}^{\dagger}$ and $\hat{b}$ are the creation and annihilation operators for the Raman mode of frequency $\omega_R$ and $g$ is the coupling strength between the cavity and Raman mode. To connect these operators  to the electric field in the cavity we require that $\hat{E}_{cav}^2 = E^2_0 \left(\hat{a} + \hat{a}^\dag \right)^2$, where $E^2_0$ is the nonzero quantum noise of the electric field in the cavity that can be measured experimentally~\cite{appugliese_breakdown_2022}, while the cavity itself is in equilibrium, $\left<\hat{E}_{cav}\right> = 0$ and the Raman coordinate is given by $\hat{Q} = \frac{\hat{b}^\dag + \hat{b}}{\sqrt{2 \omega_R}}$. The last term of strength $g_4$ is an $\hat{E}_{cav}^4$ type of nonlinearity necessary to make the system stable for finite coupling $g < \sqrt{g_4 \omega_R}/2$, a condition that is found analytically in Ref.~\cite{SupInfo}. 

We propose Raman spectroscopy as a natural probe for Raman polaritons. The spectroscopic protocol consists of an incoming probe laser of frequency $\omega_p$ that can be scattered to free space as outgoing photons with frequency $\omega_s$ after interacting with the Raman medium through a stimulated Raman scattering (SRS) [see Fig.~\ref{fig:cartoon} (a)].

The probe is assumed to be a coherent light-source with an associated electric field $E_p(t)=E^{0}_p sin(\omega_p t)$ whereas the scattered photons are described by the Hamiltonian $H_s/\hbar=\omega_s\hat{a}_s^{\dagger}\hat{a}_s$. The SRS Hamiltonian can be written as 
\begin{equation}
    \frac{H_p}{\hbar}=g_s E_p(t)\left (\hat{b}^{\dagger}+\hat{b}\right )\left (\hat{a}_s^{\dagger}+\hat{a}_s\right)
\end{equation}
where $\hat{a}_s^{\dagger}$ ($\hat{a}_s$) is the creation (annihilation) operator for scattered photons and $g_s$ the coupling between the Raman mode and photons being scattered. The requirement for weak probing is that $g_s E^{0}_p$ is much smaller than $g$, as we will choose in the following.

Considering the total Hamiltonian $H_t=H+H_p+H_s$ we derive the corresponding Heisenberg-Langevin equations of motion and use the open TWA method to solve the dynamics. This semiclassical phase-space method captures the lowest order quantum effects beyond mean-field treatment as demonstrated in different contexts~\cite{polkovnikov_phase_2010,cosme_time_2019,kongkhambut_observation_2022,Skulte2023} and consists of sampling the initial states from the corresponding Wigner distribution to take into account the quantum uncertainty. 
The equations of motion for the complex fields $a$, $b$ and $a_s$ associated with cavity, Raman motion and scattered photons operators are given by
\begin{align}
\begin{split}i\partial_{t}{a} =& \omega_c a + 2 g (a+a^{*})(b+b^{*})+ g_4 (a+a^{*})^{3}\\
&- i\kappa a + i\xi_{a} , 
\end{split}\label{eq:motion1} \\
\begin{split}i\partial_{t}{b} =& \omega_R b + g (a+a^{*})^2 + g_s E_p(t) (a_{s}+a_{s}^{*})\\
& - i\gamma (b-b^{*})/2 - \xi_{b} ,
\end{split} \label{eq:motion2}\\
i\partial_{t}{a_{s}} =& \omega_s a_{s}+ g_s E_p(t) (b+b^{*}) -i\kappa_s a_{s}+ i \xi_{s}, 
 \label{eq:motion3}
\end{align}
Here $\kappa$, $\gamma$ and $\kappa_s$ are the decay rates associated with the cavity, Raman and scattered photon field while $\xi_{a}$,  $\xi_{b}$ and $\xi_{s}$ are sources of Gaussian noise obeying the autocorrelation relations $\left<\xi_a^{*}(t_1)\xi_a(t_2)\right>=\kappa\delta(t_1-t_2)$,  $\left<\xi_b(t_1)\xi_b(t_2)\right>=\gamma \delta(t_1-t_2)$ and $\left<\xi_s^{*}(t_1)\xi_s(t_2)\right>=\kappa_s\delta(t_1-t_2)$.  
$\xi_a$ and $\xi_s$ are complex valued whereas $\xi_b$ is real-valued and, along with the damping term, enters only in the equation of motion for the imaginary part of the Raman field which is associated with the momentum of the Raman mode. The choice for the Raman mode is motivated by the Brownian motion in which the frictional force is proportional to the velocity.

We simulate the quantum Langevin Eqs.~(\ref{eq:motion1})-(\ref{eq:motion3}) using a stochastic ordinary differential equation solver and compute relevant observables in the steady state. To initiate the dynamics we ramp out $g$ from zero to a finite value and wait for the steady state before turning on the probe field $E_p (t)$. In particular, we define the Raman spectrum as the number of scattered photons $n_s=|a_s|^2$ as a function of their frequencies which is computed after a certain time of exposure to the probe field~\cite{SupInfo}.
\begin{figure}[!tb]
\hspace*{-0.3cm} 
\includegraphics[width=0.5\textwidth]{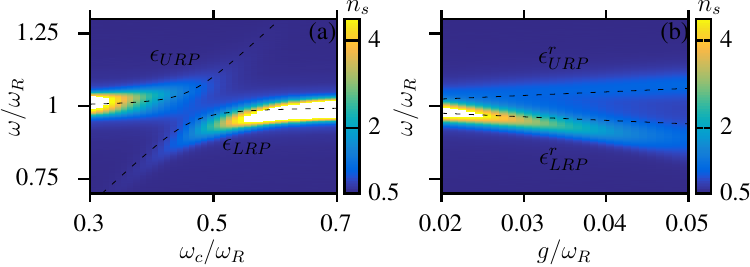}
\caption{(a) Raman spectra $n_s\left(\omega\right)$ for different cavity frequencies $\omega_c$. We consider a probe laser of strength $g_s E_p^0=0.04\omega_R$ and frequency $\omega_p=5\omega_R$ with $g=0.04\omega_R$, $g_4=\kappa=\gamma=\kappa_s=0.01\omega_R$.
(b) Raman polariton branches at resonance ($\overline{\omega}_c = \omega_R/2$) as a function of the coupling $g$. In both panels, the dashed black lines correspond to an analytical solution using a Gaussian approximation theory.} 
\label{fig:polaritonbands}
\end{figure} 

{\em Parametric Raman polaritons.-} In Fig.~\ref{fig:polaritonbands} (a) we show the Raman spectra $n_s(\omega)$ for different cavity frequencies and a fixed $g$ where $\omega= \omega_p - \omega_s$ is the Raman shift. Away from resonance we see only one peak at $\omega \approx  \omega_R$, which corresponds to the Stokes peak~\cite{Raman_scattering_Reference1,Raman_scattering_Reference2} of the Raman mode~\cite{Note1}. Near the resonance at $\omega_c = \omega_R/2$ a second peak appears showing an avoided crossing, which signals the existence of a Raman polariton. To gain insight into the two polariton branches found numerically using the TWA method, we employ a Gaussian approximation~\cite{SupInfo}. Within this method, we find that the two polariton branches arise from resonant coupling between the Raman mode oscillating at $\omega_R$ and Gaussian squeezing oscillations of the photon, oscillating at $2 \omega_c$ leading to a new Raman polariton. We have computed analytically the dispersion of the lower and upper Raman polariton branches which are plotted with black dashed lines in Fig.~\ref{fig:polaritonbands}(a), showing good agreement with the two numerical peaks in the Raman spectrum (indicated by $\epsilon_{LR
P}$ and $\epsilon_{URP}$). The exact position of the avoided crossing is shifted to the left compared to the condition $\omega_c=\omega_R/2$, due to the renormalization of the cavity frequency by nonlinear interactions. Within the Gaussian approximation the effective cavity frequency is found to be $\bar{\omega}_c= \omega_c-12 g^2/\omega_R+3g_4$ so the improved estimate of the resonance condition is $\bar{\omega}_c=\omega_R/2$.

To quantify the strength of the Raman-cavity coupling, we define the Raman Rabi splitting as the difference between the upper and lower Raman polaritons on resonance, $2 \delta = \epsilon_{URP}(\bar{\omega}_c=\omega_R/2) -\epsilon_{LRP}(\bar{\omega}_c=\omega_R/2)$. In Fig.~\ref{fig:polaritonbands}(b) we plot the dependence of the Raman polariton branches on resonance $\epsilon^{r}_{URP}=\epsilon_{URP}(\bar{\omega}_c=\omega_R/2)$ and $\epsilon^{r}_{LRP}=\epsilon_{LRP}(\bar{\omega}_c=\omega_R/2)$ on the coupling strength $g$, and overlay the analytical prediction. The Rabi splitting grows linearly with $g$ and is given by the expression:
\begin{equation}
\begin{split}  \delta =  \sqrt{2 \langle \hat{x}^2 \rangle \omega_R} \left(1  - 27 g_4 \langle \hat{x}^2 \rangle^{3}/2\right) g + \mathcal{O}(g^3).
\end{split}
\label{eq:Rabi}
\end{equation}
Interestingly, the Gaussian theory suggests that the Rabi splitting could be parametrically enhanced by the cavity quantum fluctuations $\langle \hat{x}^2 \rangle$, where $\hat{x} = \frac{a + a^\dag}{\sqrt{2 \omega_c}}$. Perturbatively, $\langle \hat{x}^2 \rangle = \frac{1}{2\omega_c} + \mathcal{O}(g^2)$, which is the value we use in Eq.~\eqref{eq:Rabi} to plot the dashed lines in Fig.\ref{fig:polaritonbands}(b).

{\em Equilibrium parametric amplification-} 
While Raman spectroscopy provides experimental evidence for strong Raman-cavity coupling, we now expand the discussion to the equilibrium properties of the Raman polariton which exhibit equilibrium parametric amplification. This phenomenology corresponds to the amplification of photon fluctuations accompanied by a localization of Raman mode fluctuations, i.e. suppression of fluctuations, depicted schematically in Fig.~\ref{fig:cartoon} (c).

To illustrate the modification of each subsystem due to the coupling, we determine the deviation of the Raman and cavity fluctuations $\delta Q^2$ and $\delta x^2$ from the uncoupled case as
\begin{equation}
 \delta Q^2=\frac{\left<\hat{Q}^2\right>-\left<\hat{Q}^2\right>_0}{\left<\hat{Q}^2\right>_0} , 
 \\
 \delta x^2=\frac{\left<\hat{x}^2\right>-\left<\hat{x}^2\right>_0}{\left<\hat{x}^2\right>_0} 
 \label{eq:parameters}
\end{equation}
where $\left< ...\right>$ denotes the expectation value for a finite coupling $g$ and $\left< ...\right>_0$ the expectation value in the absence of coupling and cavity nonlinearities ($g=g_4=0$). As in the previous section, $\hat{x}$ is the cavity coordinate which is related to the electric field, $\hat{E}_{cav} = \sqrt{2 \omega_c} E_0 \hat{x} $,
where $E_0$ is the electric field amplitude of the noise of the cavity mode.
Note that also $\delta x^2 = \frac{\langle \hat{E}_{cav}^2 \rangle - \langle \hat{E}_{cav}^2 \rangle_0 }{\langle \hat{E}_{cav}^2 \rangle_0} $ is the variation of the quantum fluctuations in the electric field. To compute these Raman and cavity fluctuations we set the probe field $E_p (t)$ in Eqs.~(\ref{eq:motion1})-(\ref{eq:motion3}) to zero and average over steady states of the Langevin equations of motion~\cite{SupInfo}.

\begin{figure}[!tb]
\hspace*{-0.3cm} 
\includegraphics[width=0.5\textwidth]{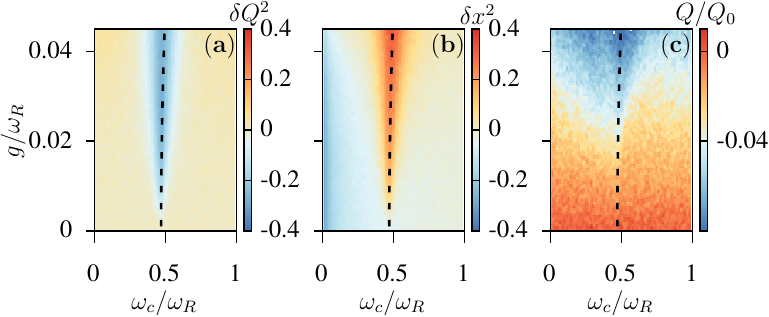}
\caption{ (a), (b) Variation of Raman and cavity fluctuations compared to the uncoupled case $g=0$ (see text) for different cavity frequencies and coupling strengths $g$. (c) Raman displacement in units of the quantum oscillator length associated to the Raman mode $Q_0=\sqrt{\hbar/M\omega_R}$. The dashed lines represent the parametric resonant condition $\omega_R=2\bar{\omega}_c$. On resonance, the main features of parametric Raman polaritons appear: localization (a) and shift (c) of the Raman mode in favor of cavity field amplification (b). The decay rates and nonlinear interaction $g_4$ are the same as in Fig.~\ref{fig:polaritonbands}.} 

\label{fig:Ramanpolaritonproperties}
\end{figure}

In Fig.~\ref{fig:Ramanpolaritonproperties} we show $\delta Q^2$ and $\delta x^2$ as well as the Raman coordinate $Q=\left<\hat{Q}\right>$ for different cavity frequencies and coupling strengths $g$. In all cases, a clear resonance can be seen around $\omega_c\approx \omega_R /2$ indicating a resonant regime in which both the Raman mode and cavity fluctuations are strongly modified. 
In this regime, the Raman fluctuations are suppressed by the cavity, $\delta Q^2<0$, so the Raman mode is localized while the cavity fluctuations are amplified by the Raman mode $\delta x^2 > 0$. Outside this resonant region the Raman fluctuations are unaffected and remain the same as in free space ($\delta Q^2\approx 0$). In a similar way, for off-resonant cavity frequencies, quantum vacuum fluctuations remain practically unchanged~\cite{note} meaning that the Raman medium barely perturbs the photon field ($\delta x^2\approx 0$). This observation justifies our choice to consider the coupling of a single cavity mode with a single Raman mode: due to the resonant character of the interaction, we expect that other off-resonant modes do not contribute.

For the parameters used in Fig.~\ref{fig:Ramanpolaritonproperties}, on resonance and close to the instability $g\approx\sqrt{g_4 \omega_R}/2$, the Raman mode is strongly localized by $\sim 40\%$ compared to the case of Raman fluctuations in free space while the photon field increases by the same amount with respect to the empty cavity case, even though the coupling is only $g \sim 4\% \omega_R $. We emphasize that the values $g$ used here are of the same order of magnitude as the decay rates $\gamma$ and $\kappa$. Therefore the system is between the weak and strong coupling regime but not in the ultrastrong coupling situation ($g>\omega_c$) where these resonant effects may be more pronounced~\cite{Frisk2019}.

In Fig.~\ref{fig:Ramanpolaritonproperties} (c) the expectation value of the Raman coordinate $Q$ is shown as a function of coupling strength $g$ and cavity frequency $\omega_c$. A clear shift of the Raman coordinate is observed for large values of $g$ which represents another form of control of the Raman mode by the cavity field. This is reminiscent of displacive excitation of coherent phonons~\cite{Dresselhaus,Merlin97,Giorgianni22}. However in the scenario considered here, the phonon shift occurs in equilibrium. Thus, the Raman mode is not only localized but also its coordinate is shifted by the vacuum fluctuations. This shift is an off-resonant process and therefore depends only weakly on the parametric resonance compared to the fluctuations in Figs.~\ref{fig:Ramanpolaritonproperties}(a)-~\ref{fig:Ramanpolaritonproperties}(b). 

We have checked that these parametric resonances in $\delta Q^2$, $\delta x^2$ and shift in $Q$ also occur for thermal states and survive for stronger nonlinearties $g_4$ and larger decay rates~\cite{SupInfo}. Interestingly, thermal fluctuations are less effective in localizing the Raman mode and amplifying
cavity fluctuations, but create a larger shift in the Raman coordinate.

{\em Experimental platforms.-} Our mechanism can be realized in materials hosting Raman modes, coupled resonantly with a cavity in the THz range. Interestingly, the strong coupling regime between infrared phonons and a tunable THz cavity has been experimentally demonstrated~\cite{jarc_tunable_2022} opening the door to the study of Raman-active materials in cavities using Raman spectroscopy [Fig. ~\ref{fig:cartoon} (a)]. Possible material candidates might be functionalized graphene nanoribbons with Raman activity around 6 THz~\cite{verzhbitskiy_raman_2016}, twisted bilayer graphene with low Raman modes $\lesssim$ 3 THz~\cite{he_observation_2013} or transition metal dichalcogenides (TMDs) with ultralow modes below 1 THz. All these examples lie in the experimental frequency range of up to $\sim$8 THz in different types of cavities~\cite{jarc_tunable_2022, appugliese_breakdown_2022,Valmorra2013,Maissen2014}.

The Raman-cavity coupling between the cavity mode and the zero-momentum Raman mode can be computed from first principles through the Raman tensor~\cite{SupInfo} and is given by
\begin{equation}
\frac{g}{\omega_R} = \frac{\sqrt{N}}{2}\frac{\epsilon_0 E_0^2 V_{cell}}{\omega_R} \tilde{R}
\end{equation}
where $\epsilon_0$ is the vacuum permittivity, $V_{cell}$ is the volume of the unit cell of the material hosting the phonon mode and $N = \frac{V_{samp}}{V_{cell}}$ is the total number of unit cells in the sample of volume $V_{samp}$. The dimensionless Raman coupling is given by $\tilde{R} \propto \vec{e}_c \cdot \partial_Q \underline{\underline{\epsilon}}(Q) \cdot \vec{e}_c$, which measures the change in the electric permittivity $\underline{\underline{\epsilon}}$ as a function of a shift of the Raman coordinate $Q$ per unit cell, and $\vec{e}_c$ is the polarization vector of the cavity. The electric field noise of a cavity is given by, $E_0 \sim \sqrt{\frac{\omega_c}{\epsilon_0 V_{eff}}}$~\cite{appugliese_breakdown_2022} and therefore, on parametric resonance $\omega_R = 2 \omega_c$, we estimate that $\frac{g}{\omega_R} \sim \frac{\sqrt{V_{samp} V_{cell}}}{V_{eff}} \tilde{R}
$. 

To circumvent the possible limitation of weak $g$ in Fabry-P\'{e}rot cavities, due to the small value of $V_{cell}/V_{eff}$, split-ring resonators (SRRs) cavities could be a solution where large cavity mode volume compression has been experimentally demonstrated. Typically, SRRs are build with cavity frequencies between 0.5-1 THz~\cite{Valmorra2013,Maissen2014,Scalari2012} which matches the range of breathing Raman modes of the order of 1 THz in twisted-TMDs like MoSe$_2$ or WSe$_2$~\cite{Lin2021,Alex2016}. Thus the condition $2\omega_c=\omega_R$ can be satisfied. From the above expression we may expect that also the coupling strength will be particularly increased for these twisted bilayer system of triangular lattices for twist angles around 0° or 60°, where the unit cell becomes very large. Considering 10nm x 10nm for the area of the unit cell, 1 $\mu m^2$ for the effective cavity area of SRRs, $g/\omega_R \sim 0.01$ assuming $\tilde{R}\sim 1$ as estimated for twisted TMDs using density functional theory~\cite{Alex2016}. For this $g$ we estimate that in the quantum noise limit $2\delta\sim 0.03$THz, $Q\sim-0.02Q_0$ (typically below 1 picometer), $\delta Q^2\sim -0.2$ and $\delta x^2\sim 0.2$. 

{\em Conclusions.-} 
We have presented how parametric resonances in Raman-cavity hybrids can be exploited to amplify photon quantum noise and localize Raman modes {\em at equilibrium}. Our study represents a proof of principle of how this nonlinear type of hybridization between Raman modes and photons, at the quantum fluctuation level, gives rise to equilibrium parametric amplification that can be leveraged to control quantum materials. In particular, the cavity control of Raman-active phonons demonstrated here is a crucial step towards cavity-material engineering in more complex systems leading to a phase transition as opposed to similar optomechanical few-body realizations. Strongly coupled Raman phonons are responsible for superconductivity in K$_3$C$_{60}$~\cite{Gunnarsson_97}, and statically shifting one of these modes was proposed as a mechanism for photoinduced superconductivity~\cite{eckhardt_theory_2023, chattopadhyay_mechanisms_2023}. More broadly, Raman phonons can change lattice symmetries, lift electronic orbital degeneracies~\cite{Goodenough98}, gap out gapless electronic systems~\cite{Liu_10} and manipulate spin-spin interactions~\cite{Emil_23}. 
Our work paves the way to new studies on all of these topics and the search of similar equilibrium parametric amplification in different scenarios such as Higgs-light hybrids in superconducting systems and in the quantum information realm using the recent  three-photon quantum-optics development~\cite{Chang2020,Minganti2023}. We further note that the process described here represents the opposite process to the parametric down conversion from one photon to two phonons~\cite{Cartella18,Teitelbaum18,Juraschek20}, and analogous conclusions could be extended to these examples. Finally, extending our analysis to multiple modes is a promising route to explore sum-frequency excitations of the Raman mode~\cite{Maehrlein17,Juraschek18,Johnson19} in equilibrium.

\ {\em Acknowledgments.-} 
We thank S. Felicetti, J. G. Cosme, L. Broers, J. Lorenzana, E. Demler, E. V. Bostr{\"o}m, C. Eckhardt, and F. Schlawin for useful discussions.  M.H.M. acknowledges financial support from the Alex von Humdoldt Foundation. H.P.O.C., J.S.. and L.M. acknowledge funding by the Deutsche Forschungsgemeinschaft (DFG, German Research Foundation) “SFB-925” Project No. 170620586 and the Cluster of Excellence “Advanced Imaging of Matter” (EXC 2056), Project No. 390715994.

\bibliography{references}

\begin{thebibliography}{93}%
\makeatletter
\providecommand \@ifxundefined [1]{%
 \@ifx{#1\undefined}
}%
\providecommand \@ifnum [1]{%
 \ifnum #1\expandafter \@firstoftwo
 \else \expandafter \@secondoftwo
 \fi
}%
\providecommand \@ifx [1]{%
 \ifx #1\expandafter \@firstoftwo
 \else \expandafter \@secondoftwo
 \fi
}%
\providecommand \natexlab [1]{#1}%
\providecommand \enquote  [1]{``#1''}%
\providecommand \bibnamefont  [1]{#1}%
\providecommand \bibfnamefont [1]{#1}%
\providecommand \citenamefont [1]{#1}%
\providecommand \href@noop [0]{\@secondoftwo}%
\providecommand \href [0]{\begingroup \@sanitize@url \@href}%
\providecommand \@href[1]{\@@startlink{#1}\@@href}%
\providecommand \@@href[1]{\endgroup#1\@@endlink}%
\providecommand \@sanitize@url [0]{\catcode `\\12\catcode `\$12\catcode `\&12\catcode `\#12\catcode `\^12\catcode `\_12\catcode `\%12\relax}%
\providecommand \@@startlink[1]{}%
\providecommand \@@endlink[0]{}%
\providecommand \url  [0]{\begingroup\@sanitize@url \@url }%
\providecommand \@url [1]{\endgroup\@href {#1}{\urlprefix }}%
\providecommand \urlprefix  [0]{URL }%
\providecommand \Eprint [0]{\href }%
\providecommand \doibase [0]{http://dx.doi.org/}%
\providecommand \selectlanguage [0]{\@gobble}%
\providecommand \bibinfo  [0]{\@secondoftwo}%
\providecommand \bibfield  [0]{\@secondoftwo}%
\providecommand \translation [1]{[#1]}%
\providecommand \BibitemOpen [0]{}%
\providecommand \bibitemStop [0]{}%
\providecommand \bibitemNoStop [0]{.\EOS\space}%
\providecommand \EOS [0]{\spacefactor3000\relax}%
\providecommand \BibitemShut  [1]{\csname bibitem#1\endcsname}%
\let\auto@bib@innerbib\@empty
\bibitem [{\citenamefont {{A. de la Torre, D. M. Kennes, M. Claassen, S. Gerber, J. W. {McIver}, and M. A. Sentef}}(2021)}]{de_la_torre_colloquium_2021}%
  \BibitemOpen
  \bibfield  {author} {\bibinfo {author} {\bibnamefont {{A. de la Torre, D. M. Kennes, M. Claassen, S. Gerber, J. W. {McIver}, and M. A. Sentef}}},\ }\bibfield  {title} {\enquote {\bibinfo {title} {Colloquium: Nonthermal pathways to ultrafast control in quantum materials},}\ }\href {\doibase 10.1103/RevModPhys.93.041002} {\bibfield  {journal} {\bibinfo  {journal} {Rev. Mod. Phys.}\ }\textbf {\bibinfo {volume} {93}},\ \bibinfo {pages} {041002} (\bibinfo {year} {2021})}\BibitemShut {NoStop}%
\bibitem [{\citenamefont {Basov}\ \emph {et~al.}(2017)\citenamefont {Basov}, \citenamefont {Averitt},\ and\ \citenamefont {Hsieh}}]{basov_towards_2017}%
  \BibitemOpen
  \bibfield  {author} {\bibinfo {author} {\bibfnamefont {D.~N.}\ \bibnamefont {Basov}}, \bibinfo {author} {\bibfnamefont {R.~D.}\ \bibnamefont {Averitt}}, \ and\ \bibinfo {author} {\bibfnamefont {D.}~\bibnamefont {Hsieh}},\ }\bibfield  {title} {\enquote {\bibinfo {title} {Towards properties on demand in quantum materials},}\ }\href {\doibase 10.1038/nmat5017} {\bibfield  {journal} {\bibinfo  {journal} {Nature Mater}\ }\textbf {\bibinfo {volume} {16}},\ \bibinfo {pages} {1077--1088} (\bibinfo {year} {2017})}\BibitemShut {NoStop}%
\bibitem [{\citenamefont {Bloch}\ \emph {et~al.}(2022)\citenamefont {Bloch}, \citenamefont {Cavalleri}, \citenamefont {Galitski}, \citenamefont {Hafezi},\ and\ \citenamefont {Rubio}}]{Bloch_22}%
  \BibitemOpen
  \bibfield  {author} {\bibinfo {author} {\bibfnamefont {J.}~\bibnamefont {Bloch}}, \bibinfo {author} {\bibfnamefont {A.}~\bibnamefont {Cavalleri}}, \bibinfo {author} {\bibfnamefont {V.}~\bibnamefont {Galitski}}, \bibinfo {author} {\bibfnamefont {M.}~\bibnamefont {Hafezi}}, \ and\ \bibinfo {author} {\bibfnamefont {A.}~\bibnamefont {Rubio}},\ }\bibfield  {title} {\enquote {\bibinfo {title} {Strongly correlated electron--photon systems},}\ }\href {\doibase 10.1038/s41586-022-04726-w} {\bibfield  {journal} {\bibinfo  {journal} {Nature}\ }\textbf {\bibinfo {volume} {606}},\ \bibinfo {pages} {41} (\bibinfo {year} {2022})}\BibitemShut {NoStop}%
\bibitem [{\citenamefont {Kennes}\ and\ \citenamefont {Rubio}(2023)}]{Kennes2023}%
  \BibitemOpen
  \bibfield  {author} {\bibinfo {author} {\bibfnamefont {D.~M.}\ \bibnamefont {Kennes}}\ and\ \bibinfo {author} {\bibfnamefont {A.}~\bibnamefont {Rubio}},\ }\enquote {\bibinfo {title} {{A new era of quantum materials mastery and quantum simulators in and out of equilibrium}},}\ in\ \href {\doibase 10.1007/978-3-031-32469-7_1} {\emph {\bibinfo {booktitle} {Sketches of Physics: The Celebration Collection}}}\ (\bibinfo  {publisher} {Springer International Publishing},\ \bibinfo {address} {New York},\ \bibinfo {year} {2023})\ pp.\ \bibinfo {pages} {1--39}\BibitemShut {NoStop}%
\bibitem [{\citenamefont {Nova}\ \emph {et~al.}(2019)\citenamefont {Nova}, \citenamefont {Disa}, \citenamefont {Fechner},\ and\ \citenamefont {Cavalleri}}]{Nova19}%
  \BibitemOpen
  \bibfield  {author} {\bibinfo {author} {\bibfnamefont {T.~F.}\ \bibnamefont {Nova}}, \bibinfo {author} {\bibfnamefont {A.~S.}\ \bibnamefont {Disa}}, \bibinfo {author} {\bibfnamefont {M.}~\bibnamefont {Fechner}}, \ and\ \bibinfo {author} {\bibfnamefont {A.}~\bibnamefont {Cavalleri}},\ }\bibfield  {title} {\enquote {\bibinfo {title} {Metastable ferroelectricity in optically strained {S}r{T}i{O}$_3$},}\ }\href {\doibase 10.1126/science.aaw4911} {\bibfield  {journal} {\bibinfo  {journal} {Science}\ }\textbf {\bibinfo {volume} {364}},\ \bibinfo {pages} {1075} (\bibinfo {year} {2019})}\BibitemShut {NoStop}%
\bibitem [{\citenamefont {Disa}\ \emph {et~al.}(2023)\citenamefont {Disa}, \citenamefont {Curtis}, \citenamefont {Fechner}, \citenamefont {Liu}, \citenamefont {von Hoegen}, \citenamefont {Först}, \citenamefont {Nova}, \citenamefont {Narang}, \citenamefont {Maljuk}, \citenamefont {Boris}, \citenamefont {Keimer},\ and\ \citenamefont {Cavalleri}}]{disa_photo-induced_2023}%
  \BibitemOpen
  \bibfield  {author} {\bibinfo {author} {\bibfnamefont {A.~S.}\ \bibnamefont {Disa}}, \bibinfo {author} {\bibfnamefont {J.}~\bibnamefont {Curtis}}, \bibinfo {author} {\bibfnamefont {M.}~\bibnamefont {Fechner}}, \bibinfo {author} {\bibfnamefont {A.}~\bibnamefont {Liu}}, \bibinfo {author} {\bibfnamefont {A.}~\bibnamefont {von Hoegen}}, \bibinfo {author} {\bibfnamefont {M.}~\bibnamefont {Först}}, \bibinfo {author} {\bibfnamefont {T.~F.}\ \bibnamefont {Nova}}, \bibinfo {author} {\bibfnamefont {P.}~\bibnamefont {Narang}}, \bibinfo {author} {\bibfnamefont {A.}~\bibnamefont {Maljuk}}, \bibinfo {author} {\bibfnamefont {A.~V.}\ \bibnamefont {Boris}}, \bibinfo {author} {\bibfnamefont {B.}~\bibnamefont {Keimer}}, \ and\ \bibinfo {author} {\bibfnamefont {A.}~\bibnamefont {Cavalleri}},\ }\bibfield  {title} {\enquote {\bibinfo {title} {Photo-induced high-temperature ferromagnetism in {YTiO}$_3$},}\ }\href {\doibase 10.1038/s41586-023-05853-8} {\bibfield  {journal} {\bibinfo  {journal} {Nature}\ }\textbf {\bibinfo
  {volume} {617}},\ \bibinfo {pages} {73} (\bibinfo {year} {2023})}\BibitemShut {NoStop}%
\bibitem [{\citenamefont {Siegrist}\ \emph {et~al.}(2019)\citenamefont {Siegrist}, \citenamefont {Gessner}, \citenamefont {Ossiander}, \citenamefont {Denker}, \citenamefont {Chang}, \citenamefont {Schröder}, \citenamefont {Guggenmos}, \citenamefont {Cui}, \citenamefont {Walowski}, \citenamefont {Martens}, \citenamefont {Dewhurst}, \citenamefont {Kleineberg}, \citenamefont {Münzenberg}, \citenamefont {Sharma},\ and\ \citenamefont {Schultze}}]{siegrist_light-wave_2019}%
  \BibitemOpen
  \bibfield  {author} {\bibinfo {author} {\bibfnamefont {F.}~\bibnamefont {Siegrist}}, \bibinfo {author} {\bibfnamefont {J.~A.}\ \bibnamefont {Gessner}}, \bibinfo {author} {\bibfnamefont {M.}~\bibnamefont {Ossiander}}, \bibinfo {author} {\bibfnamefont {C.}~\bibnamefont {Denker}}, \bibinfo {author} {\bibfnamefont {Y.~P.}\ \bibnamefont {Chang}}, \bibinfo {author} {\bibfnamefont {M.~C.}\ \bibnamefont {Schröder}}, \bibinfo {author} {\bibfnamefont {A.}~\bibnamefont {Guggenmos}}, \bibinfo {author} {\bibfnamefont {Y.}~\bibnamefont {Cui}}, \bibinfo {author} {\bibfnamefont {J.}~\bibnamefont {Walowski}}, \bibinfo {author} {\bibfnamefont {U.}~\bibnamefont {Martens}}, \bibinfo {author} {\bibfnamefont {J.~K.}\ \bibnamefont {Dewhurst}}, \bibinfo {author} {\bibfnamefont {U.}~\bibnamefont {Kleineberg}}, \bibinfo {author} {\bibfnamefont {M.}~\bibnamefont {Münzenberg}}, \bibinfo {author} {\bibfnamefont {S.}~\bibnamefont {Sharma}}, \ and\ \bibinfo {author} {\bibfnamefont {M.}~\bibnamefont {Schultze}},\ }\bibfield  {title}
  {\enquote {\bibinfo {title} {Light-wave dynamic control of magnetism},}\ }\href {\doibase 10.1038/s41586-019-1333-x} {\bibfield  {journal} {\bibinfo  {journal} {Nature}\ }\textbf {\bibinfo {volume} {571}},\ \bibinfo {pages} {240} (\bibinfo {year} {2019})}\BibitemShut {NoStop}%
\bibitem [{\citenamefont {Beaulieu}\ \emph {et~al.}(2021)\citenamefont {Beaulieu}, \citenamefont {Dong}, \citenamefont {Tancogne-Dejean}, \citenamefont {Dendzik}, \citenamefont {Pincelli}, \citenamefont {Maklar}, \citenamefont {Xian}, \citenamefont {Sentef}, \citenamefont {Wolf}, \citenamefont {Rubio}, \citenamefont {Rettig},\ and\ \citenamefont {Ernstorfer}}]{Samuel_21}%
  \BibitemOpen
  \bibfield  {author} {\bibinfo {author} {\bibfnamefont {S.}~\bibnamefont {Beaulieu}}, \bibinfo {author} {\bibfnamefont {S.}~\bibnamefont {Dong}}, \bibinfo {author} {\bibfnamefont {N.}~\bibnamefont {Tancogne-Dejean}}, \bibinfo {author} {\bibfnamefont {M.}~\bibnamefont {Dendzik}}, \bibinfo {author} {\bibfnamefont {T.}~\bibnamefont {Pincelli}}, \bibinfo {author} {\bibfnamefont {J.}~\bibnamefont {Maklar}}, \bibinfo {author} {\bibfnamefont {R.~Patrick}\ \bibnamefont {Xian}}, \bibinfo {author} {\bibfnamefont {M.~A.}\ \bibnamefont {Sentef}}, \bibinfo {author} {\bibfnamefont {M.}~\bibnamefont {Wolf}}, \bibinfo {author} {\bibfnamefont {A.}~\bibnamefont {Rubio}}, \bibinfo {author} {\bibfnamefont {L.}~\bibnamefont {Rettig}}, \ and\ \bibinfo {author} {\bibfnamefont {R.}~\bibnamefont {Ernstorfer}},\ }\bibfield  {title} {\enquote {\bibinfo {title} {Ultrafast dynamical {L}ifshitz transition},}\ }\href {\doibase 10.1126/sciadv.abd9275} {\bibfield  {journal} {\bibinfo  {journal} {Science Advances}\ }\textbf {\bibinfo
  {volume} {7}},\ \bibinfo {pages} {eabd9275} (\bibinfo {year} {2021})}\BibitemShut {NoStop}%
\bibitem [{\citenamefont {Boström}\ \emph {et~al.}(2020)\citenamefont {Boström}, \citenamefont {Claassen}, \citenamefont {McIver}, \citenamefont {Jotzu}, \citenamefont {Rubio},\ and\ \citenamefont {Sentef}}]{Emil_20}%
  \BibitemOpen
  \bibfield  {author} {\bibinfo {author} {\bibfnamefont {E.~V.}\ \bibnamefont {Boström}}, \bibinfo {author} {\bibfnamefont {M.}~\bibnamefont {Claassen}}, \bibinfo {author} {\bibfnamefont {J.~W.}\ \bibnamefont {McIver}}, \bibinfo {author} {\bibfnamefont {G.}~\bibnamefont {Jotzu}}, \bibinfo {author} {\bibfnamefont {A.}~\bibnamefont {Rubio}}, \ and\ \bibinfo {author} {\bibfnamefont {M.~A.}\ \bibnamefont {Sentef}},\ }\bibfield  {title} {\enquote {\bibinfo {title} {{Light-induced topological magnons in two-dimensional van der {W}aals magnets}},}\ }\href {\doibase 10.21468/SciPostPhys.9.4.061} {\bibfield  {journal} {\bibinfo  {journal} {SciPost Phys.}\ }\textbf {\bibinfo {volume} {9}},\ \bibinfo {pages} {061} (\bibinfo {year} {2020})}\BibitemShut {NoStop}%
\bibitem [{\citenamefont {Mitrano}\ \emph {et~al.}(2016)\citenamefont {Mitrano}, \citenamefont {Cantaluppi}, \citenamefont {Nicoletti}, \citenamefont {Kaiser}, \citenamefont {Perucchi}, \citenamefont {Lupi}, \citenamefont {Di~Pietro}, \citenamefont {Pontiroli}, \citenamefont {Riccò}, \citenamefont {Clark}, \citenamefont {Jaksch},\ and\ \citenamefont {Cavalleri}}]{mitrano_possible_2016}%
  \BibitemOpen
  \bibfield  {author} {\bibinfo {author} {\bibfnamefont {M.}~\bibnamefont {Mitrano}}, \bibinfo {author} {\bibfnamefont {A.}~\bibnamefont {Cantaluppi}}, \bibinfo {author} {\bibfnamefont {D.}~\bibnamefont {Nicoletti}}, \bibinfo {author} {\bibfnamefont {S.}~\bibnamefont {Kaiser}}, \bibinfo {author} {\bibfnamefont {A.}~\bibnamefont {Perucchi}}, \bibinfo {author} {\bibfnamefont {S.}~\bibnamefont {Lupi}}, \bibinfo {author} {\bibfnamefont {P.}~\bibnamefont {Di~Pietro}}, \bibinfo {author} {\bibfnamefont {D.}~\bibnamefont {Pontiroli}}, \bibinfo {author} {\bibfnamefont {M.}~\bibnamefont {Riccò}}, \bibinfo {author} {\bibfnamefont {S.~R.}\ \bibnamefont {Clark}}, \bibinfo {author} {\bibfnamefont {D.}~\bibnamefont {Jaksch}}, \ and\ \bibinfo {author} {\bibfnamefont {A.}~\bibnamefont {Cavalleri}},\ }\bibfield  {title} {\enquote {\bibinfo {title} {Possible light-induced superconductivity in {K}$_3${C}$_{60}$ at high temperature},}\ }\href {\doibase 10.1038/nature16522} {\bibfield  {journal} {\bibinfo  {journal} {Nature}\
  }\textbf {\bibinfo {volume} {530}},\ \bibinfo {pages} {461} (\bibinfo {year} {2016})}\BibitemShut {NoStop}%
\bibitem [{\citenamefont {Budden}\ \emph {et~al.}(2021)\citenamefont {Budden}, \citenamefont {Gebert}, \citenamefont {Buzzi}, \citenamefont {Jotzu}, \citenamefont {Wang}, \citenamefont {Matsuyama}, \citenamefont {Meier}, \citenamefont {Laplace}, \citenamefont {Pontiroli}, \citenamefont {Riccò}, \citenamefont {Schlawin}, \citenamefont {Jaksch},\ and\ \citenamefont {Cavalleri}}]{budden_evidence_2021}%
  \BibitemOpen
  \bibfield  {author} {\bibinfo {author} {\bibfnamefont {M.}~\bibnamefont {Budden}}, \bibinfo {author} {\bibfnamefont {T.}~\bibnamefont {Gebert}}, \bibinfo {author} {\bibfnamefont {M.}~\bibnamefont {Buzzi}}, \bibinfo {author} {\bibfnamefont {G.}~\bibnamefont {Jotzu}}, \bibinfo {author} {\bibfnamefont {E.}~\bibnamefont {Wang}}, \bibinfo {author} {\bibfnamefont {T.}~\bibnamefont {Matsuyama}}, \bibinfo {author} {\bibfnamefont {G.}~\bibnamefont {Meier}}, \bibinfo {author} {\bibfnamefont {Y.}~\bibnamefont {Laplace}}, \bibinfo {author} {\bibfnamefont {D.}~\bibnamefont {Pontiroli}}, \bibinfo {author} {\bibfnamefont {M.}~\bibnamefont {Riccò}}, \bibinfo {author} {\bibfnamefont {F.}~\bibnamefont {Schlawin}}, \bibinfo {author} {\bibfnamefont {D.}~\bibnamefont {Jaksch}}, \ and\ \bibinfo {author} {\bibfnamefont {A.}~\bibnamefont {Cavalleri}},\ }\bibfield  {title} {\enquote {\bibinfo {title} {Evidence for metastable photo-induced superconductivity in {K}$_3${C}$_{60}$},}\ }\href {\doibase 10.1038/s41567-020-01148-1}
  {\bibfield  {journal} {\bibinfo  {journal} {Nat. Phys.}\ }\textbf {\bibinfo {volume} {17}},\ \bibinfo {pages} {611} (\bibinfo {year} {2021})}\BibitemShut {NoStop}%
\bibitem [{\citenamefont {Rowe}\ \emph {et~al.}(2023)\citenamefont {Rowe}, \citenamefont {Yuan}, \citenamefont {Buzzi}, \citenamefont {Jotzu}, \citenamefont {Zhu}, \citenamefont {Fechner}, \citenamefont {Först}, \citenamefont {Liu}, \citenamefont {Pontiroli}, \citenamefont {Riccò},\ and\ \citenamefont {Cavalleri}}]{rowe_giant_2023}%
  \BibitemOpen
  \bibfield  {author} {\bibinfo {author} {\bibfnamefont {E.}~\bibnamefont {Rowe}}, \bibinfo {author} {\bibfnamefont {B.}~\bibnamefont {Yuan}}, \bibinfo {author} {\bibfnamefont {M.}~\bibnamefont {Buzzi}}, \bibinfo {author} {\bibfnamefont {G.}~\bibnamefont {Jotzu}}, \bibinfo {author} {\bibfnamefont {Y.}~\bibnamefont {Zhu}}, \bibinfo {author} {\bibfnamefont {M.}~\bibnamefont {Fechner}}, \bibinfo {author} {\bibfnamefont {M.}~\bibnamefont {Först}}, \bibinfo {author} {\bibfnamefont {B.}~\bibnamefont {Liu}}, \bibinfo {author} {\bibfnamefont {D.}~\bibnamefont {Pontiroli}}, \bibinfo {author} {\bibfnamefont {M.}~\bibnamefont {Riccò}}, \ and\ \bibinfo {author} {\bibfnamefont {A.}~\bibnamefont {Cavalleri}},\ }\href@noop {} {\enquote {\bibinfo {title} {Giant resonant enhancement for photo-induced superconductivity in {K}$_3${C}$_{60}$},}\ } (\bibinfo {year} {2023}),\ \Eprint {http://arxiv.org/abs/2301.08633 [cond-mat]} {2301.08633 [cond-mat]} \BibitemShut {NoStop}%
\bibitem [{\citenamefont {Chattopadhyay}\ \emph {et~al.}(2023)\citenamefont {Chattopadhyay}, \citenamefont {Eckhardt}, \citenamefont {Kennes}, \citenamefont {Sentef}, \citenamefont {Shin}, \citenamefont {Rubio}, \citenamefont {Cavalleri}, \citenamefont {Demler},\ and\ \citenamefont {Michael}}]{chattopadhyay_mechanisms_2023}%
  \BibitemOpen
  \bibfield  {author} {\bibinfo {author} {\bibfnamefont {S.}~\bibnamefont {Chattopadhyay}}, \bibinfo {author} {\bibfnamefont {C.~J.}\ \bibnamefont {Eckhardt}}, \bibinfo {author} {\bibfnamefont {D.~M.}\ \bibnamefont {Kennes}}, \bibinfo {author} {\bibfnamefont {M.~A.}\ \bibnamefont {Sentef}}, \bibinfo {author} {\bibfnamefont {D.}~\bibnamefont {Shin}}, \bibinfo {author} {\bibfnamefont {A.}~\bibnamefont {Rubio}}, \bibinfo {author} {\bibfnamefont {A.}~\bibnamefont {Cavalleri}}, \bibinfo {author} {\bibfnamefont {E.~A.}\ \bibnamefont {Demler}}, \ and\ \bibinfo {author} {\bibfnamefont {M.~H.}\ \bibnamefont {Michael}},\ }\href@noop {} {\enquote {\bibinfo {title} {Mechanisms for {L}ong-{L}ived, {P}hoto-{I}nduced {S}uperconductivity},}\ } (\bibinfo {year} {2023}),\ \Eprint {http://arxiv.org/abs/2303.15355 [cond-mat, physics:physics]} {2303.15355 [cond-mat, physics:physics]} \BibitemShut {NoStop}%
\bibitem [{\citenamefont {von Hoegen}\ \emph {et~al.}(2022)\citenamefont {von Hoegen}, \citenamefont {Fechner}, \citenamefont {Först}, \citenamefont {Taherian}, \citenamefont {Rowe}, \citenamefont {Ribak}, \citenamefont {Porras}, \citenamefont {Keimer}, \citenamefont {Michael}, \citenamefont {Demler},\ and\ \citenamefont {Cavalleri}}]{von_hoegen_amplification_2022}%
  \BibitemOpen
  \bibfield  {author} {\bibinfo {author} {\bibfnamefont {A.}~\bibnamefont {von Hoegen}}, \bibinfo {author} {\bibfnamefont {M.}~\bibnamefont {Fechner}}, \bibinfo {author} {\bibfnamefont {M.}~\bibnamefont {Först}}, \bibinfo {author} {\bibfnamefont {N.}~\bibnamefont {Taherian}}, \bibinfo {author} {\bibfnamefont {E.}~\bibnamefont {Rowe}}, \bibinfo {author} {\bibfnamefont {A.}~\bibnamefont {Ribak}}, \bibinfo {author} {\bibfnamefont {J.}~\bibnamefont {Porras}}, \bibinfo {author} {\bibfnamefont {B.}~\bibnamefont {Keimer}}, \bibinfo {author} {\bibfnamefont {M.}~\bibnamefont {Michael}}, \bibinfo {author} {\bibfnamefont {E.}~\bibnamefont {Demler}}, \ and\ \bibinfo {author} {\bibfnamefont {A.}~\bibnamefont {Cavalleri}},\ }\bibfield  {title} {\enquote {\bibinfo {title} {Amplification of {S}uperconducting {F}luctuations in {D}riven {YB}a$_2$ {C}u$_3$ {O}$_{6+x}$},}\ }\href {\doibase 10.1103/PhysRevX.12.031008} {\bibfield  {journal} {\bibinfo  {journal} {Phys. Rev. X}\ }\textbf {\bibinfo {volume} {12}},\ \bibinfo {pages}
  {031008} (\bibinfo {year} {2022})}\BibitemShut {NoStop}%
\bibitem [{\citenamefont {Michael}\ \emph {et~al.}(2020)\citenamefont {Michael}, \citenamefont {von Hoegen}, \citenamefont {Fechner}, \citenamefont {Först}, \citenamefont {Cavalleri},\ and\ \citenamefont {Demler}}]{michael_parametric_2020}%
  \BibitemOpen
  \bibfield  {author} {\bibinfo {author} {\bibfnamefont {M.~H.}\ \bibnamefont {Michael}}, \bibinfo {author} {\bibfnamefont {A.}~\bibnamefont {von Hoegen}}, \bibinfo {author} {\bibfnamefont {M.}~\bibnamefont {Fechner}}, \bibinfo {author} {\bibfnamefont {M.}~\bibnamefont {Först}}, \bibinfo {author} {\bibfnamefont {A.}~\bibnamefont {Cavalleri}}, \ and\ \bibinfo {author} {\bibfnamefont {E.}~\bibnamefont {Demler}},\ }\bibfield  {title} {\enquote {\bibinfo {title} {Parametric resonance of {J}osephson plasma waves: A theory for optically amplified interlayer superconductivity in {YB}a$_2$ {C}u$_3$ {O}$_{6+x}$},}\ }\href {\doibase 10.1103/PhysRevB.102.174505} {\bibfield  {journal} {\bibinfo  {journal} {Phys. Rev. B}\ }\textbf {\bibinfo {volume} {102}},\ \bibinfo {pages} {174505} (\bibinfo {year} {2020})}\BibitemShut {NoStop}%
\bibitem [{\citenamefont {{McIver}}\ \emph {et~al.}(2020)\citenamefont {{McIver}}, \citenamefont {Schulte}, \citenamefont {Stein}, \citenamefont {Matsuyama}, \citenamefont {Jotzu}, \citenamefont {Meier},\ and\ \citenamefont {Cavalleri}}]{mciver_light-induced_2020}%
  \BibitemOpen
  \bibfield  {author} {\bibinfo {author} {\bibfnamefont {J.~W.}\ \bibnamefont {{McIver}}}, \bibinfo {author} {\bibfnamefont {B.}~\bibnamefont {Schulte}}, \bibinfo {author} {\bibfnamefont {F.~U.}\ \bibnamefont {Stein}}, \bibinfo {author} {\bibfnamefont {T.}~\bibnamefont {Matsuyama}}, \bibinfo {author} {\bibfnamefont {G.}~\bibnamefont {Jotzu}}, \bibinfo {author} {\bibfnamefont {G.}~\bibnamefont {Meier}}, \ and\ \bibinfo {author} {\bibfnamefont {A.}~\bibnamefont {Cavalleri}},\ }\bibfield  {title} {\enquote {\bibinfo {title} {Light-induced anomalous {H}all effect in graphene},}\ }\href {\doibase 10.1038/s41567-019-0698-y} {\bibfield  {journal} {\bibinfo  {journal} {Nat. Phys.}\ }\textbf {\bibinfo {volume} {16}},\ \bibinfo {pages} {38} (\bibinfo {year} {2020})}\BibitemShut {NoStop}%
\bibitem [{\citenamefont {Kogar}\ \emph {et~al.}(2020)\citenamefont {Kogar}, \citenamefont {Zong}, \citenamefont {Dolgirev}, \citenamefont {Shen}, \citenamefont {Straquadine}, \citenamefont {Bie}, \citenamefont {Wang}, \citenamefont {Rohwer}, \citenamefont {Tung}, \citenamefont {Yang}, \citenamefont {Li}, \citenamefont {Yang}, \citenamefont {Weathersby}, \citenamefont {Park}, \citenamefont {Kozina}, \citenamefont {Sie}, \citenamefont {Wen}, \citenamefont {Jarillo-Herrero}, \citenamefont {Fisher}, \citenamefont {Wang},\ and\ \citenamefont {Gedik}}]{kogar_light-induced_2020}%
  \BibitemOpen
  \bibfield  {author} {\bibinfo {author} {\bibfnamefont {A.}~\bibnamefont {Kogar}}, \bibinfo {author} {\bibfnamefont {A.}~\bibnamefont {Zong}}, \bibinfo {author} {\bibfnamefont {P.~E.}\ \bibnamefont {Dolgirev}}, \bibinfo {author} {\bibfnamefont {X.}~\bibnamefont {Shen}}, \bibinfo {author} {\bibfnamefont {J.}~\bibnamefont {Straquadine}}, \bibinfo {author} {\bibfnamefont {Y.Q.}\ \bibnamefont {Bie}}, \bibinfo {author} {\bibfnamefont {X.}~\bibnamefont {Wang}}, \bibinfo {author} {\bibfnamefont {T.}~\bibnamefont {Rohwer}}, \bibinfo {author} {\bibfnamefont {I.-C.}\ \bibnamefont {Tung}}, \bibinfo {author} {\bibfnamefont {Y.}~\bibnamefont {Yang}}, \bibinfo {author} {\bibfnamefont {R.}~\bibnamefont {Li}}, \bibinfo {author} {\bibfnamefont {J.}~\bibnamefont {Yang}}, \bibinfo {author} {\bibfnamefont {S.}~\bibnamefont {Weathersby}}, \bibinfo {author} {\bibfnamefont {S.}~\bibnamefont {Park}}, \bibinfo {author} {\bibfnamefont {M.~E.}\ \bibnamefont {Kozina}}, \bibinfo {author} {\bibfnamefont {E.~J.}\ \bibnamefont {Sie}},
  \bibinfo {author} {\bibfnamefont {H.}~\bibnamefont {Wen}}, \bibinfo {author} {\bibfnamefont {P.}~\bibnamefont {Jarillo-Herrero}}, \bibinfo {author} {\bibfnamefont {I.~R.}\ \bibnamefont {Fisher}}, \bibinfo {author} {\bibfnamefont {X.}~\bibnamefont {Wang}}, \ and\ \bibinfo {author} {\bibfnamefont {N.}~\bibnamefont {Gedik}},\ }\bibfield  {title} {\enquote {\bibinfo {title} {Light-induced charge density wave in {LaTe}$_3$},}\ }\href {\doibase 10.1038/s41567-019-0705-3} {\bibfield  {journal} {\bibinfo  {journal} {Nat. Phys.}\ }\textbf {\bibinfo {volume} {16}},\ \bibinfo {pages} {159} (\bibinfo {year} {2020})}\BibitemShut {NoStop}%
\bibitem [{\citenamefont {Dolgirev}\ \emph {et~al.}(2023)\citenamefont {Dolgirev}, \citenamefont {Michael}, \citenamefont {Zong}, \citenamefont {Gedik},\ and\ \citenamefont {Demler}}]{dolgirev_self-similar_2020}%
  \BibitemOpen
  \bibfield  {author} {\bibinfo {author} {\bibfnamefont {P.~E.}\ \bibnamefont {Dolgirev}}, \bibinfo {author} {\bibfnamefont {M.~H.}\ \bibnamefont {Michael}}, \bibinfo {author} {\bibfnamefont {A.}~\bibnamefont {Zong}}, \bibinfo {author} {\bibfnamefont {N.}~\bibnamefont {Gedik}}, \ and\ \bibinfo {author} {\bibfnamefont {E.}~\bibnamefont {Demler}},\ }\bibfield  {title} {\enquote {\bibinfo {title} {Self-similar dynamics of order parameter fluctuations in pump-probe experiments},}\ }\href {\doibase 10.1103/PhysRevB.101.174306} {\bibfield  {journal} {\bibinfo  {journal} {Phys. Rev. B}\ }\textbf {\bibinfo {volume} {101}},\ \bibinfo {pages} {174306} (\bibinfo {year} {2023})}\BibitemShut {NoStop}%
\bibitem [{\citenamefont {Lyubarov}\ \emph {et~al.}(2022)\citenamefont {Lyubarov}, \citenamefont {Lumer}, \citenamefont {Dikopoltsev}, \citenamefont {Lustig}, \citenamefont {Sharabi},\ and\ \citenamefont {Segev}}]{Lyubarov_2022}%
  \BibitemOpen
  \bibfield  {author} {\bibinfo {author} {\bibfnamefont {M.}~\bibnamefont {Lyubarov}}, \bibinfo {author} {\bibfnamefont {Y.}~\bibnamefont {Lumer}}, \bibinfo {author} {\bibfnamefont {A.}~\bibnamefont {Dikopoltsev}}, \bibinfo {author} {\bibfnamefont {E.}~\bibnamefont {Lustig}}, \bibinfo {author} {\bibfnamefont {Y.}~\bibnamefont {Sharabi}}, \ and\ \bibinfo {author} {\bibfnamefont {M.}~\bibnamefont {Segev}},\ }\bibfield  {title} {\enquote {\bibinfo {title} {Amplified emission and lasing in photonic time crystals},}\ }\href {\doibase 10.1126/science.abo3324} {\bibfield  {journal} {\bibinfo  {journal} {Science}\ }\textbf {\bibinfo {volume} {377}},\ \bibinfo {pages} {425} (\bibinfo {year} {2022})}\BibitemShut {NoStop}%
\bibitem [{\citenamefont {Michael}\ \emph {et~al.}(2022)\citenamefont {Michael}, \citenamefont {Först}, \citenamefont {Nicoletti}, \citenamefont {Haque}, \citenamefont {Zhang}, \citenamefont {Cavalleri}, \citenamefont {Averitt}, \citenamefont {Podolsky},\ and\ \citenamefont {Demler}}]{michael_generalized_2022}%
  \BibitemOpen
  \bibfield  {author} {\bibinfo {author} {\bibfnamefont {M.~H.}\ \bibnamefont {Michael}}, \bibinfo {author} {\bibfnamefont {M.}~\bibnamefont {Först}}, \bibinfo {author} {\bibfnamefont {D.}~\bibnamefont {Nicoletti}}, \bibinfo {author} {\bibfnamefont {S.~R.~Ul}\ \bibnamefont {Haque}}, \bibinfo {author} {\bibfnamefont {Y.}~\bibnamefont {Zhang}}, \bibinfo {author} {\bibfnamefont {A.}~\bibnamefont {Cavalleri}}, \bibinfo {author} {\bibfnamefont {R.~D.}\ \bibnamefont {Averitt}}, \bibinfo {author} {\bibfnamefont {D.}~\bibnamefont {Podolsky}}, \ and\ \bibinfo {author} {\bibfnamefont {E.}~\bibnamefont {Demler}},\ }\bibfield  {title} {\enquote {\bibinfo {title} {Generalized {F}resnel-{F}loquet equations for driven quantum materials},}\ }\href {\doibase 10.1103/PhysRevB.105.174301} {\bibfield  {journal} {\bibinfo  {journal} {Phys. Rev. B}\ }\textbf {\bibinfo {volume} {105}},\ \bibinfo {pages} {174301} (\bibinfo {year} {2022})}\BibitemShut {NoStop}%
\bibitem [{\citenamefont {Michael}\ \emph {et~al.}(2023)\citenamefont {Michael}, \citenamefont {Haque}, \citenamefont {Windgaetter}, \citenamefont {Latini}, \citenamefont {Zhang}, \citenamefont {Rubio}, \citenamefont {Averitt},\ and\ \citenamefont {Demler}}]{michael2023theory}%
  \BibitemOpen
  \bibfield  {author} {\bibinfo {author} {\bibfnamefont {M.~H.}\ \bibnamefont {Michael}}, \bibinfo {author} {\bibfnamefont {S.~R.~Ul}\ \bibnamefont {Haque}}, \bibinfo {author} {\bibfnamefont {L.}~\bibnamefont {Windgaetter}}, \bibinfo {author} {\bibfnamefont {S.}~\bibnamefont {Latini}}, \bibinfo {author} {\bibfnamefont {Y.}~\bibnamefont {Zhang}}, \bibinfo {author} {\bibfnamefont {A.}~\bibnamefont {Rubio}}, \bibinfo {author} {\bibfnamefont {R.~D.}\ \bibnamefont {Averitt}}, \ and\ \bibinfo {author} {\bibfnamefont {E.}~\bibnamefont {Demler}},\ }\href@noop {} {\enquote {\bibinfo {title} {Theory of time-crystalline behaviour mediated by phonon squeezing in {Ta$_2$NiSe$_5$}},}\ } (\bibinfo {year} {2023}),\ \Eprint {http://arxiv.org/abs/2207.08851} {arXiv:2207.08851 [cond-mat.str-el]} \BibitemShut {NoStop}%
\bibitem [{\citenamefont {Haque}\ \emph {et~al.}(2023)\citenamefont {Haque}, \citenamefont {Michael}, \citenamefont {Zhu}, \citenamefont {Zhang}, \citenamefont {Windgätter}, \citenamefont {Latini}, \citenamefont {Wakefield}, \citenamefont {Zhang}, \citenamefont {Zhang}, \citenamefont {Rubio}, \citenamefont {Checkelsky}, \citenamefont {Demler},\ and\ \citenamefont {Averitt}}]{haque2023terahertz}%
  \BibitemOpen
  \bibfield  {author} {\bibinfo {author} {\bibfnamefont {S.~R.~Ul}\ \bibnamefont {Haque}}, \bibinfo {author} {\bibfnamefont {M.~H.}\ \bibnamefont {Michael}}, \bibinfo {author} {\bibfnamefont {J.}~\bibnamefont {Zhu}}, \bibinfo {author} {\bibfnamefont {Y.}~\bibnamefont {Zhang}}, \bibinfo {author} {\bibfnamefont {L.}~\bibnamefont {Windgätter}}, \bibinfo {author} {\bibfnamefont {S.}~\bibnamefont {Latini}}, \bibinfo {author} {\bibfnamefont {J.~P.}\ \bibnamefont {Wakefield}}, \bibinfo {author} {\bibfnamefont {G.~F.}\ \bibnamefont {Zhang}}, \bibinfo {author} {\bibfnamefont {J.}~\bibnamefont {Zhang}}, \bibinfo {author} {\bibfnamefont {A.}~\bibnamefont {Rubio}}, \bibinfo {author} {\bibfnamefont {J.~G.}\ \bibnamefont {Checkelsky}}, \bibinfo {author} {\bibfnamefont {E.}~\bibnamefont {Demler}}, \ and\ \bibinfo {author} {\bibfnamefont {R.~D.}\ \bibnamefont {Averitt}},\ }\href@noop {} {\enquote {\bibinfo {title} {Terahertz parametric amplification as a reporter of exciton condensate dynamics},}\ } (\bibinfo {year}
  {2023}),\ \Eprint {http://arxiv.org/abs/2304.09249} {arXiv:2304.09249 [cond-mat.str-el]} \BibitemShut {NoStop}%
\bibitem [{\citenamefont {Dolgirev}\ \emph {et~al.}(2022)\citenamefont {Dolgirev}, \citenamefont {Zong}, \citenamefont {Michael}, \citenamefont {Curtis}, \citenamefont {Podolsky}, \citenamefont {Cavalleri},\ and\ \citenamefont {Demler}}]{dolgirev_periodic_2022}%
  \BibitemOpen
  \bibfield  {author} {\bibinfo {author} {\bibfnamefont {P.~E.}\ \bibnamefont {Dolgirev}}, \bibinfo {author} {\bibfnamefont {A.}~\bibnamefont {Zong}}, \bibinfo {author} {\bibfnamefont {M.~H.}\ \bibnamefont {Michael}}, \bibinfo {author} {\bibfnamefont {J.~B.}\ \bibnamefont {Curtis}}, \bibinfo {author} {\bibfnamefont {D.}~\bibnamefont {Podolsky}}, \bibinfo {author} {\bibfnamefont {A.}~\bibnamefont {Cavalleri}}, \ and\ \bibinfo {author} {\bibfnamefont {E.}~\bibnamefont {Demler}},\ }\bibfield  {title} {\enquote {\bibinfo {title} {Periodic dynamics in superconductors induced by an impulsive optical quench},}\ }\href {\doibase 10.1038/s42005-022-01007-w} {\bibfield  {journal} {\bibinfo  {journal} {Commun Phys}\ }\textbf {\bibinfo {volume} {5}},\ \bibinfo {pages} {1} (\bibinfo {year} {2022})}\BibitemShut {NoStop}%
\bibitem [{\citenamefont {Else}\ \emph {et~al.}(2020)\citenamefont {Else}, \citenamefont {Monroe}, \citenamefont {Nayak},\ and\ \citenamefont {Yao}}]{Else2020}%
  \BibitemOpen
  \bibfield  {author} {\bibinfo {author} {\bibfnamefont {D.~V.}\ \bibnamefont {Else}}, \bibinfo {author} {\bibfnamefont {C.}~\bibnamefont {Monroe}}, \bibinfo {author} {\bibfnamefont {C.}~\bibnamefont {Nayak}}, \ and\ \bibinfo {author} {\bibfnamefont {N.~Y.}\ \bibnamefont {Yao}},\ }\bibfield  {title} {\enquote {\bibinfo {title} {Discrete {T}ime {C}rystals},}\ }\href {\doibase 10.1146/annurev-conmatphys-031119-050658} {\bibfield  {journal} {\bibinfo  {journal} {Annual Review of Condensed Matter Physics}\ }\textbf {\bibinfo {volume} {11}},\ \bibinfo {pages} {467} (\bibinfo {year} {2020})}\BibitemShut {NoStop}%
\bibitem [{\citenamefont {Zhang}\ \emph {et~al.}(2017)\citenamefont {Zhang}, \citenamefont {Hess}, \citenamefont {Kyprianidis}, \citenamefont {Becker}, \citenamefont {Lee}, \citenamefont {Smith}, \citenamefont {Pagano}, \citenamefont {Potirniche}, \citenamefont {Potter}, \citenamefont {Vishwanath}, \citenamefont {Yao},\ and\ \citenamefont {Monroe}}]{Zhang2017}%
  \BibitemOpen
  \bibfield  {author} {\bibinfo {author} {\bibfnamefont {J.}~\bibnamefont {Zhang}}, \bibinfo {author} {\bibfnamefont {P.~W.}\ \bibnamefont {Hess}}, \bibinfo {author} {\bibfnamefont {A.}~\bibnamefont {Kyprianidis}}, \bibinfo {author} {\bibfnamefont {P.}~\bibnamefont {Becker}}, \bibinfo {author} {\bibfnamefont {A.}~\bibnamefont {Lee}}, \bibinfo {author} {\bibfnamefont {J.}~\bibnamefont {Smith}}, \bibinfo {author} {\bibfnamefont {G.}~\bibnamefont {Pagano}}, \bibinfo {author} {\bibfnamefont {I.~D.}\ \bibnamefont {Potirniche}}, \bibinfo {author} {\bibfnamefont {A.~C.}\ \bibnamefont {Potter}}, \bibinfo {author} {\bibfnamefont {A.}~\bibnamefont {Vishwanath}}, \bibinfo {author} {\bibfnamefont {N.~Y.}\ \bibnamefont {Yao}}, \ and\ \bibinfo {author} {\bibfnamefont {C.}~\bibnamefont {Monroe}},\ }\bibfield  {title} {\enquote {\bibinfo {title} {Observation of a discrete time crystal},}\ }\href {https://www.nature.com/articles/nature21413} {\bibfield  {journal} {\bibinfo  {journal} {Nature}\ }\textbf {\bibinfo {volume}
  {543}},\ \bibinfo {pages} {217} (\bibinfo {year} {2017})}\BibitemShut {NoStop}%
\bibitem [{\citenamefont {Choi}\ \emph {et~al.}(2017)\citenamefont {Choi}, \citenamefont {Choi}, \citenamefont {Landig}, \citenamefont {Kucsko}, \citenamefont {Zhou}, \citenamefont {Isoya}, \citenamefont {Jelezko}, \citenamefont {Onoda}, \citenamefont {Sumiya}, \citenamefont {Khemani}, \citenamefont {von Keyserlingk}, \citenamefont {Yao}, \citenamefont {Demler},\ and\ \citenamefont {Lukin}}]{Choi2017}%
  \BibitemOpen
  \bibfield  {author} {\bibinfo {author} {\bibfnamefont {S.}~\bibnamefont {Choi}}, \bibinfo {author} {\bibfnamefont {J.}~\bibnamefont {Choi}}, \bibinfo {author} {\bibfnamefont {R.}~\bibnamefont {Landig}}, \bibinfo {author} {\bibfnamefont {G.}~\bibnamefont {Kucsko}}, \bibinfo {author} {\bibfnamefont {H.}~\bibnamefont {Zhou}}, \bibinfo {author} {\bibfnamefont {J.}~\bibnamefont {Isoya}}, \bibinfo {author} {\bibfnamefont {F.}~\bibnamefont {Jelezko}}, \bibinfo {author} {\bibfnamefont {S.}~\bibnamefont {Onoda}}, \bibinfo {author} {\bibfnamefont {H.}~\bibnamefont {Sumiya}}, \bibinfo {author} {\bibfnamefont {V.}~\bibnamefont {Khemani}}, \bibinfo {author} {\bibfnamefont {C.}~\bibnamefont {von Keyserlingk}}, \bibinfo {author} {\bibfnamefont {N.~Y.}\ \bibnamefont {Yao}}, \bibinfo {author} {\bibfnamefont {E.}~\bibnamefont {Demler}}, \ and\ \bibinfo {author} {\bibfnamefont {M.~D.}\ \bibnamefont {Lukin}},\ }\bibfield  {title} {\enquote {\bibinfo {title} {Observation of discrete time-crystalline order in a disordered
  dipolar many-body system},}\ }\href {https://www.nature.com/articles/nature21426} {\bibfield  {journal} {\bibinfo  {journal} {Nature}\ }\textbf {\bibinfo {volume} {543}},\ \bibinfo {pages} {221} (\bibinfo {year} {2017})}\BibitemShut {NoStop}%
\bibitem [{\citenamefont {Ke\ss{}ler}\ \emph {et~al.}(2021)\citenamefont {Ke\ss{}ler}, \citenamefont {Kongkhambut}, \citenamefont {Georges}, \citenamefont {Mathey}, \citenamefont {Cosme},\ and\ \citenamefont {Hemmerich}}]{Kessler2021}%
  \BibitemOpen
  \bibfield  {author} {\bibinfo {author} {\bibfnamefont {H.}~\bibnamefont {Ke\ss{}ler}}, \bibinfo {author} {\bibfnamefont {P.}~\bibnamefont {Kongkhambut}}, \bibinfo {author} {\bibfnamefont {C.}~\bibnamefont {Georges}}, \bibinfo {author} {\bibfnamefont {L.}~\bibnamefont {Mathey}}, \bibinfo {author} {\bibfnamefont {J.~G.}\ \bibnamefont {Cosme}}, \ and\ \bibinfo {author} {\bibfnamefont {A.}~\bibnamefont {Hemmerich}},\ }\bibfield  {title} {\enquote {\bibinfo {title} {Observation of a {D}issipative {T}ime {C}rystal},}\ }\href {\doibase 10.1103/PhysRevLett.127.043602} {\bibfield  {journal} {\bibinfo  {journal} {Phys. Rev. Lett.}\ }\textbf {\bibinfo {volume} {127}},\ \bibinfo {pages} {043602} (\bibinfo {year} {2021})}\BibitemShut {NoStop}%
\bibitem [{\citenamefont {Kongkhambut}\ \emph {et~al.}(2021)\citenamefont {Kongkhambut}, \citenamefont {Ke\ss{}ler}, \citenamefont {Skulte}, \citenamefont {Mathey}, \citenamefont {Cosme},\ and\ \citenamefont {Hemmerich}}]{Kongkhambut2021}%
  \BibitemOpen
  \bibfield  {author} {\bibinfo {author} {\bibfnamefont {P.}~\bibnamefont {Kongkhambut}}, \bibinfo {author} {\bibfnamefont {H.}~\bibnamefont {Ke\ss{}ler}}, \bibinfo {author} {\bibfnamefont {J.}~\bibnamefont {Skulte}}, \bibinfo {author} {\bibfnamefont {L.}~\bibnamefont {Mathey}}, \bibinfo {author} {\bibfnamefont {J.~G.}\ \bibnamefont {Cosme}}, \ and\ \bibinfo {author} {\bibfnamefont {A.}~\bibnamefont {Hemmerich}},\ }\bibfield  {title} {\enquote {\bibinfo {title} {Realization of a {P}eriodically {D}riven {O}pen {T}hree-{L}evel {D}icke {M}odel},}\ }\href {\doibase 10.1103/PhysRevLett.127.253601} {\bibfield  {journal} {\bibinfo  {journal} {Phys. Rev. Lett.}\ }\textbf {\bibinfo {volume} {127}},\ \bibinfo {pages} {253601} (\bibinfo {year} {2021})}\BibitemShut {NoStop}%
\bibitem [{\citenamefont {Taheri}\ \emph {et~al.}(2022)\citenamefont {Taheri}, \citenamefont {Matsko}, \citenamefont {Maleki},\ and\ \citenamefont {Sacha}}]{Taheri2022}%
  \BibitemOpen
  \bibfield  {author} {\bibinfo {author} {\bibfnamefont {H.}~\bibnamefont {Taheri}}, \bibinfo {author} {\bibfnamefont {A.~B.}\ \bibnamefont {Matsko}}, \bibinfo {author} {\bibfnamefont {L.}~\bibnamefont {Maleki}}, \ and\ \bibinfo {author} {\bibfnamefont {K.}~\bibnamefont {Sacha}},\ }\bibfield  {title} {\enquote {\bibinfo {title} {All-optical dissipative discrete time crystals},}\ }\href {https://www.nature.com/articles/s41467-022-28462-x} {\bibfield  {journal} {\bibinfo  {journal} {Nature Communications}\ }\textbf {\bibinfo {volume} {13}},\ \bibinfo {pages} {848} (\bibinfo {year} {2022})}\BibitemShut {NoStop}%
\bibitem [{\citenamefont {Zaletel}\ \emph {et~al.}(2023)\citenamefont {Zaletel}, \citenamefont {Lukin}, \citenamefont {Monroe}, \citenamefont {Nayak}, \citenamefont {Wilczek},\ and\ \citenamefont {Yao}}]{Zaletel2023}%
  \BibitemOpen
  \bibfield  {author} {\bibinfo {author} {\bibfnamefont {M.~P.}\ \bibnamefont {Zaletel}}, \bibinfo {author} {\bibfnamefont {M.}~\bibnamefont {Lukin}}, \bibinfo {author} {\bibfnamefont {C.}~\bibnamefont {Monroe}}, \bibinfo {author} {\bibfnamefont {C.}~\bibnamefont {Nayak}}, \bibinfo {author} {\bibfnamefont {F.}~\bibnamefont {Wilczek}}, \ and\ \bibinfo {author} {\bibfnamefont {N.~Y.}\ \bibnamefont {Yao}},\ }\bibfield  {title} {\enquote {\bibinfo {title} {Colloquium: Quantum and classical discrete time crystals},}\ }\href {\doibase 10.1103/RevModPhys.95.031001} {\bibfield  {journal} {\bibinfo  {journal} {Rev. Mod. Phys.}\ }\textbf {\bibinfo {volume} {95}},\ \bibinfo {pages} {031001} (\bibinfo {year} {2023})}\BibitemShut {NoStop}%
\bibitem [{\citenamefont {Ojeda~Collado}\ \emph {et~al.}(2021)\citenamefont {Ojeda~Collado}, \citenamefont {Usaj}, \citenamefont {Balseiro}, \citenamefont {Zanette},\ and\ \citenamefont {Lorenzana}}]{HP_2021}%
  \BibitemOpen
  \bibfield  {author} {\bibinfo {author} {\bibfnamefont {H.~P.}\ \bibnamefont {Ojeda~Collado}}, \bibinfo {author} {\bibfnamefont {G.}~\bibnamefont {Usaj}}, \bibinfo {author} {\bibfnamefont {C.~A.}\ \bibnamefont {Balseiro}}, \bibinfo {author} {\bibfnamefont {D.~H.}\ \bibnamefont {Zanette}}, \ and\ \bibinfo {author} {\bibfnamefont {J.}~\bibnamefont {Lorenzana}},\ }\bibfield  {title} {\enquote {\bibinfo {title} {Emergent parametric resonances and time-crystal phases in driven {B}ardeen-{C}ooper-{S}chrieffer systems},}\ }\href {\doibase 10.1103/PhysRevResearch.3.L042023} {\bibfield  {journal} {\bibinfo  {journal} {Phys. Rev. Res.}\ }\textbf {\bibinfo {volume} {3}},\ \bibinfo {pages} {L042023} (\bibinfo {year} {2021})}\BibitemShut {NoStop}%
\bibitem [{\citenamefont {Collado}\ \emph {et~al.}(2023)\citenamefont {Collado}, \citenamefont {Usaj}, \citenamefont {Balseiro}, \citenamefont {Zanette},\ and\ \citenamefont {Lorenzana}}]{HP_2023}%
  \BibitemOpen
  \bibfield  {author} {\bibinfo {author} {\bibfnamefont {H.~P.~Ojeda}\ \bibnamefont {Collado}}, \bibinfo {author} {\bibfnamefont {G.}~\bibnamefont {Usaj}}, \bibinfo {author} {\bibfnamefont {C.~A.}\ \bibnamefont {Balseiro}}, \bibinfo {author} {\bibfnamefont {D.~H.}\ \bibnamefont {Zanette}}, \ and\ \bibinfo {author} {\bibfnamefont {J.}~\bibnamefont {Lorenzana}},\ }\bibfield  {title} {\enquote {\bibinfo {title} {Dynamical phase transitions in periodically driven {B}ardeen-{C}ooper-{S}chrieffer systems},}\ }\href {\doibase 10.1103/PhysRevResearch.5.023014} {\bibfield  {journal} {\bibinfo  {journal} {Phys. Rev. Res.}\ }\textbf {\bibinfo {volume} {5}},\ \bibinfo {pages} {023014} (\bibinfo {year} {2023})}\BibitemShut {NoStop}%
\bibitem [{\citenamefont {Schlawin}\ \emph {et~al.}(2022)\citenamefont {Schlawin}, \citenamefont {Kennes},\ and\ \citenamefont {Sentef}}]{schlawin_cavity_2022}%
  \BibitemOpen
  \bibfield  {author} {\bibinfo {author} {\bibfnamefont {F.}~\bibnamefont {Schlawin}}, \bibinfo {author} {\bibfnamefont {D.~M.}\ \bibnamefont {Kennes}}, \ and\ \bibinfo {author} {\bibfnamefont {M.~A.}\ \bibnamefont {Sentef}},\ }\bibfield  {title} {\enquote {\bibinfo {title} {Cavity quantum materials},}\ }\href {https://pubs.aip.org/aip/apr/article/9/1/011312/2835409} {\bibfield  {journal} {\bibinfo  {journal} {Applied Physics Reviews}\ }\textbf {\bibinfo {volume} {9}},\ \bibinfo {pages} {011312} (\bibinfo {year} {2022})}\BibitemShut {NoStop}%
\bibitem [{\citenamefont {Curtis}\ \emph {et~al.}(2023)\citenamefont {Curtis}, \citenamefont {Michael},\ and\ \citenamefont {Demler}}]{curtis2023local}%
  \BibitemOpen
  \bibfield  {author} {\bibinfo {author} {\bibfnamefont {J.~B.}\ \bibnamefont {Curtis}}, \bibinfo {author} {\bibfnamefont {M.~H.}\ \bibnamefont {Michael}}, \ and\ \bibinfo {author} {\bibfnamefont {E.}~\bibnamefont {Demler}},\ }\bibfield  {title} {\enquote {\bibinfo {title} {Local fluctuations in cavity control of ferroelectricity},}\ }\href {\doibase 10.1103/PhysRevResearch.5.043118} {\bibfield  {journal} {\bibinfo  {journal} {Phys. Rev. Res.}\ }\textbf {\bibinfo {volume} {5}},\ \bibinfo {pages} {043118} (\bibinfo {year} {2023})}\BibitemShut {NoStop}%
\bibitem [{\citenamefont {Ruggenthaler}\ \emph {et~al.}(2018)\citenamefont {Ruggenthaler}, \citenamefont {Tancogne-Dejean}, \citenamefont {Flick}, \citenamefont {Appel},\ and\ \citenamefont {Rubio}}]{Ruggi_18}%
  \BibitemOpen
  \bibfield  {author} {\bibinfo {author} {\bibfnamefont {M.}~\bibnamefont {Ruggenthaler}}, \bibinfo {author} {\bibfnamefont {N.}~\bibnamefont {Tancogne-Dejean}}, \bibinfo {author} {\bibfnamefont {J.}~\bibnamefont {Flick}}, \bibinfo {author} {\bibfnamefont {H.}~\bibnamefont {Appel}}, \ and\ \bibinfo {author} {\bibfnamefont {A.}~\bibnamefont {Rubio}},\ }\bibfield  {title} {\enquote {\bibinfo {title} {From a quantum-electrodynamical light--matter description to novel spectroscopies},}\ }\href {\doibase 10.1038/s41570-018-0118} {\bibfield  {journal} {\bibinfo  {journal} {Nature Reviews Chemistry}\ }\textbf {\bibinfo {volume} {2}},\ \bibinfo {pages} {0118} (\bibinfo {year} {2018})}\BibitemShut {NoStop}%
\bibitem [{\citenamefont {Jarc}\ \emph {et~al.}(2022)\citenamefont {Jarc}, \citenamefont {Mathengattil}, \citenamefont {Giusti}, \citenamefont {Barnaba}, \citenamefont {Singh}, \citenamefont {Montanaro}, \citenamefont {Glerean}, \citenamefont {Rigoni}, \citenamefont {Zilio}, \citenamefont {Winnerl},\ and\ \citenamefont {Fausti}}]{jarc_tunable_2022}%
  \BibitemOpen
  \bibfield  {author} {\bibinfo {author} {\bibfnamefont {G.}~\bibnamefont {Jarc}}, \bibinfo {author} {\bibfnamefont {S.~Y.}\ \bibnamefont {Mathengattil}}, \bibinfo {author} {\bibfnamefont {F.}~\bibnamefont {Giusti}}, \bibinfo {author} {\bibfnamefont {M.}~\bibnamefont {Barnaba}}, \bibinfo {author} {\bibfnamefont {A.}~\bibnamefont {Singh}}, \bibinfo {author} {\bibfnamefont {A.}~\bibnamefont {Montanaro}}, \bibinfo {author} {\bibfnamefont {F.}~\bibnamefont {Glerean}}, \bibinfo {author} {\bibfnamefont {E.~M.}\ \bibnamefont {Rigoni}}, \bibinfo {author} {\bibfnamefont {S.~D.}\ \bibnamefont {Zilio}}, \bibinfo {author} {\bibfnamefont {S.}~\bibnamefont {Winnerl}}, \ and\ \bibinfo {author} {\bibfnamefont {D.}~\bibnamefont {Fausti}},\ }\bibfield  {title} {\enquote {\bibinfo {title} {Tunable cryogenic {THz} cavity for strong light-matter coupling in complex materials},}\ }\href {\doibase 10.1063/5.0080045} {\bibfield  {journal} {\bibinfo  {journal} {Review of Scientific Instruments}\ }\textbf {\bibinfo {volume} {93}},\
  \bibinfo {pages} {033102} (\bibinfo {year} {2022})}\BibitemShut {NoStop}%
\bibitem [{\citenamefont {Appugliese}\ \emph {et~al.}(2022)\citenamefont {Appugliese}, \citenamefont {Enkner}, \citenamefont {Paravicini-Bagliani}, \citenamefont {Beck}, \citenamefont {Reichl}, \citenamefont {Wegscheider}, \citenamefont {Scalari}, \citenamefont {Ciuti},\ and\ \citenamefont {Faist}}]{appugliese_breakdown_2022}%
  \BibitemOpen
  \bibfield  {author} {\bibinfo {author} {\bibfnamefont {F.}~\bibnamefont {Appugliese}}, \bibinfo {author} {\bibfnamefont {J.}~\bibnamefont {Enkner}}, \bibinfo {author} {\bibfnamefont {G.~L.}\ \bibnamefont {Paravicini-Bagliani}}, \bibinfo {author} {\bibfnamefont {M.}~\bibnamefont {Beck}}, \bibinfo {author} {\bibfnamefont {C.}~\bibnamefont {Reichl}}, \bibinfo {author} {\bibfnamefont {W.}~\bibnamefont {Wegscheider}}, \bibinfo {author} {\bibfnamefont {G.}~\bibnamefont {Scalari}}, \bibinfo {author} {\bibfnamefont {C.}~\bibnamefont {Ciuti}}, \ and\ \bibinfo {author} {\bibfnamefont {J.}~\bibnamefont {Faist}},\ }\bibfield  {title} {\enquote {\bibinfo {title} {Breakdown of topological protection by cavity vacuum fields in the integer quantum {H}all effect},}\ }\href {\doibase 10.1126/science.abl5818} {\bibfield  {journal} {\bibinfo  {journal} {Science}\ }\textbf {\bibinfo {volume} {375}},\ \bibinfo {pages} {1030--1034} (\bibinfo {year} {2022})}\BibitemShut {NoStop}%
\bibitem [{\citenamefont {Lenk}\ \emph {et~al.}(2022)\citenamefont {Lenk}, \citenamefont {Li}, \citenamefont {Werner},\ and\ \citenamefont {Eckstein}}]{Lenk_2022}%
  \BibitemOpen
  \bibfield  {author} {\bibinfo {author} {\bibfnamefont {K.}~\bibnamefont {Lenk}}, \bibinfo {author} {\bibfnamefont {J.}~\bibnamefont {Li}}, \bibinfo {author} {\bibfnamefont {P.}~\bibnamefont {Werner}}, \ and\ \bibinfo {author} {\bibfnamefont {M.}~\bibnamefont {Eckstein}},\ }\bibfield  {title} {\enquote {\bibinfo {title} {Dynamical mean-field study of a photon-mediated ferroelectric phase transition},}\ }\href {https://doi.org/10.1103%2Fphysrevb.106.245124} {\bibfield  {journal} {\bibinfo  {journal} {Phys. Rev. B}\ }\textbf {\bibinfo {volume} {106}},\ \bibinfo {pages} {245124} (\bibinfo {year} {2022})}\BibitemShut {NoStop}%
\bibitem [{\citenamefont {Eckhardt}\ \emph {et~al.}(2023)\citenamefont {Eckhardt}, \citenamefont {Chattopadhyay}, \citenamefont {Kennes}, \citenamefont {Demler}, \citenamefont {Sentef},\ and\ \citenamefont {Michael}}]{eckhardt_theory_2023}%
  \BibitemOpen
  \bibfield  {author} {\bibinfo {author} {\bibfnamefont {C.~J.}\ \bibnamefont {Eckhardt}}, \bibinfo {author} {\bibfnamefont {S.}~\bibnamefont {Chattopadhyay}}, \bibinfo {author} {\bibfnamefont {D.~M.}\ \bibnamefont {Kennes}}, \bibinfo {author} {\bibfnamefont {E.~A.}\ \bibnamefont {Demler}}, \bibinfo {author} {\bibfnamefont {M.~A.}\ \bibnamefont {Sentef}}, \ and\ \bibinfo {author} {\bibfnamefont {M.~H.}\ \bibnamefont {Michael}},\ }\href@noop {} {\enquote {\bibinfo {title} {Theory of resonantly enhanced photo-induced superconductivity},}\ } (\bibinfo {year} {2023}),\ \Eprint {http://arxiv.org/abs/2303.02176 [cond-mat]} {2303.02176 [cond-mat]} \BibitemShut {NoStop}%
\bibitem [{\citenamefont {Baydin}\ \emph {et~al.}(2023)\citenamefont {Baydin}, \citenamefont {Manjappa}, \citenamefont {Mishra}, \citenamefont {Xu}, \citenamefont {Tay}, \citenamefont {Kim}, \citenamefont {Hernandez}, \citenamefont {Rappl}, \citenamefont {Abramof}, \citenamefont {Singh},\ and\ \citenamefont {Kono}}]{Baydin_23}%
  \BibitemOpen
  \bibfield  {author} {\bibinfo {author} {\bibfnamefont {A.}~\bibnamefont {Baydin}}, \bibinfo {author} {\bibfnamefont {M.}~\bibnamefont {Manjappa}}, \bibinfo {author} {\bibfnamefont {S.~Subhra}\ \bibnamefont {Mishra}}, \bibinfo {author} {\bibfnamefont {H.}~\bibnamefont {Xu}}, \bibinfo {author} {\bibfnamefont {F.}~\bibnamefont {Tay}}, \bibinfo {author} {\bibfnamefont {D.}~\bibnamefont {Kim}}, \bibinfo {author} {\bibfnamefont {F.~G.~G.}\ \bibnamefont {Hernandez}}, \bibinfo {author} {\bibfnamefont {P.~H.~O.}\ \bibnamefont {Rappl}}, \bibinfo {author} {\bibfnamefont {E.}~\bibnamefont {Abramof}}, \bibinfo {author} {\bibfnamefont {R.}~\bibnamefont {Singh}}, \ and\ \bibinfo {author} {\bibfnamefont {J.}~\bibnamefont {Kono}},\ }\bibfield  {title} {\enquote {\bibinfo {title} {Deep-strong coupling between cavity photons and terahertz to phonons in pbte},}\ }in\ \href {\doibase 10.1364/CLEO_FS.2023.FF3D.2} {\emph {\bibinfo {booktitle} {CLEO 2023}}}\ (\bibinfo  {publisher} {Optica Publishing Group},\ \bibinfo {year}
  {2023})\ p.\ \bibinfo {pages} {FF3D.2}\BibitemShut {NoStop}%
\bibitem [{\citenamefont {Orgiu}\ \emph {et~al.}(2015)\citenamefont {Orgiu}, \citenamefont {George}, \citenamefont {Hutchison}, \citenamefont {Devaux}, \citenamefont {Dayen}, \citenamefont {Doudin}, \citenamefont {Stellacci}, \citenamefont {Genet}, \citenamefont {Schachenmayer}, \citenamefont {Genes}, \citenamefont {Pupillo}, \citenamefont {Samorì},\ and\ \citenamefont {Ebbesen}}]{orgiu_conductivity_2015}%
  \BibitemOpen
  \bibfield  {author} {\bibinfo {author} {\bibfnamefont {E.}~\bibnamefont {Orgiu}}, \bibinfo {author} {\bibfnamefont {J.}~\bibnamefont {George}}, \bibinfo {author} {\bibfnamefont {J.~A.}\ \bibnamefont {Hutchison}}, \bibinfo {author} {\bibfnamefont {E.}~\bibnamefont {Devaux}}, \bibinfo {author} {\bibfnamefont {J.~F.}\ \bibnamefont {Dayen}}, \bibinfo {author} {\bibfnamefont {B.}~\bibnamefont {Doudin}}, \bibinfo {author} {\bibfnamefont {F.}~\bibnamefont {Stellacci}}, \bibinfo {author} {\bibfnamefont {C.}~\bibnamefont {Genet}}, \bibinfo {author} {\bibfnamefont {J.}~\bibnamefont {Schachenmayer}}, \bibinfo {author} {\bibfnamefont {C.}~\bibnamefont {Genes}}, \bibinfo {author} {\bibfnamefont {G.}~\bibnamefont {Pupillo}}, \bibinfo {author} {\bibfnamefont {P.}~\bibnamefont {Samorì}}, \ and\ \bibinfo {author} {\bibfnamefont {T.~W.}\ \bibnamefont {Ebbesen}},\ }\bibfield  {title} {\enquote {\bibinfo {title} {Conductivity in organic semiconductors hybridized with the vacuum field},}\ }\href {\doibase 10.1038/nmat4392}
  {\bibfield  {journal} {\bibinfo  {journal} {Nat Mater}\ }\textbf {\bibinfo {volume} {14}},\ \bibinfo {pages} {1123--1129} (\bibinfo {year} {2015})}\BibitemShut {NoStop}%
\bibitem [{\citenamefont {{A. Thomas, E. Devaux, K. Nagarajan, T. Chervy, M. Seidel, D. Hagenmüller, S. Schütz, J. Schachenmayer, C. Genet, G. Pupillo, and T. W. Ebbesen}}(2019)}]{thomas_exploring_2019}%
  \BibitemOpen
  \bibfield  {author} {\bibinfo {author} {\bibnamefont {{A. Thomas, E. Devaux, K. Nagarajan, T. Chervy, M. Seidel, D. Hagenmüller, S. Schütz, J. Schachenmayer, C. Genet, G. Pupillo, and T. W. Ebbesen}}},\ }\href@noop {} {\enquote {\bibinfo {title} {Exploring superconductivity under strong coupling with the vacuum electromagnetic field},}\ } (\bibinfo {year} {2019}),\ \Eprint {http://arxiv.org/abs/1911.01459 [cond-mat, physics:quant-ph]} {1911.01459 [cond-mat, physics:quant-ph]} \BibitemShut {NoStop}%
\bibitem [{\citenamefont {{A. Thomas, E. Devaux, K. Nagarajan, G. Rogez, M. Seidel, F. Richard, C. Genet, M. Drillon, and T. W. Ebbesen}}(2021)}]{thomas_large_2021}%
  \BibitemOpen
  \bibfield  {author} {\bibinfo {author} {\bibnamefont {{A. Thomas, E. Devaux, K. Nagarajan, G. Rogez, M. Seidel, F. Richard, C. Genet, M. Drillon, and T. W. Ebbesen}}},\ }\bibfield  {title} {\enquote {\bibinfo {title} {Large enhancement of ferro-magnetism under collective strong coupling of {YBCO} nanoparticles},}\ }\href {\doibase 10.1021/acs.nanolett.1c00973} {\bibfield  {journal} {\bibinfo  {journal} {Nano Lett.}\ }\textbf {\bibinfo {volume} {21}},\ \bibinfo {pages} {4365--4370} (\bibinfo {year} {2021})}\BibitemShut {NoStop}%
\bibitem [{\citenamefont {Thomas}\ \emph {et~al.}(2019)\citenamefont {Thomas}, \citenamefont {Lethuillier-Karl}, \citenamefont {Nagarajan}, \citenamefont {Vergauwe}, \citenamefont {George}, \citenamefont {Chervy}, \citenamefont {Shalabney}, \citenamefont {Devaux}, \citenamefont {Genet}, \citenamefont {Moran},\ and\ \citenamefont {Ebbesen}}]{thomas_tilting_2019}%
  \BibitemOpen
  \bibfield  {author} {\bibinfo {author} {\bibfnamefont {A.}~\bibnamefont {Thomas}}, \bibinfo {author} {\bibfnamefont {L.}~\bibnamefont {Lethuillier-Karl}}, \bibinfo {author} {\bibfnamefont {K.}~\bibnamefont {Nagarajan}}, \bibinfo {author} {\bibfnamefont {R.~M.~A.}\ \bibnamefont {Vergauwe}}, \bibinfo {author} {\bibfnamefont {J.}~\bibnamefont {George}}, \bibinfo {author} {\bibfnamefont {T.}~\bibnamefont {Chervy}}, \bibinfo {author} {\bibfnamefont {A.}~\bibnamefont {Shalabney}}, \bibinfo {author} {\bibfnamefont {E.}~\bibnamefont {Devaux}}, \bibinfo {author} {\bibfnamefont {C.}~\bibnamefont {Genet}}, \bibinfo {author} {\bibfnamefont {J.}~\bibnamefont {Moran}}, \ and\ \bibinfo {author} {\bibfnamefont {T.~W.}\ \bibnamefont {Ebbesen}},\ }\bibfield  {title} {\enquote {\bibinfo {title} {Tilting a ground-state reactivity landscape by vibrational strong coupling},}\ }\href {\doibase 10.1126/science.aau7742} {\bibfield  {journal} {\bibinfo  {journal} {Science}\ }\textbf {\bibinfo {volume} {363}},\ \bibinfo {pages}
  {615--619} (\bibinfo {year} {2019})}\BibitemShut {NoStop}%
\bibitem [{\citenamefont {Nagarajan}\ \emph {et~al.}(2021)\citenamefont {Nagarajan}, \citenamefont {Thomas},\ and\ \citenamefont {Ebbesen}}]{nagarajan_chemistry_2021}%
  \BibitemOpen
  \bibfield  {author} {\bibinfo {author} {\bibfnamefont {K.}~\bibnamefont {Nagarajan}}, \bibinfo {author} {\bibfnamefont {A.}~\bibnamefont {Thomas}}, \ and\ \bibinfo {author} {\bibfnamefont {T.~W.}\ \bibnamefont {Ebbesen}},\ }\bibfield  {title} {\enquote {\bibinfo {title} {Chemistry under vibrational strong coupling},}\ }\href {\doibase 10.1021/jacs.1c07420} {\bibfield  {journal} {\bibinfo  {journal} {J. Am. Chem. Soc.}\ }\textbf {\bibinfo {volume} {143}},\ \bibinfo {pages} {16877--16889} (\bibinfo {year} {2021})}\BibitemShut {NoStop}%
\bibitem [{\citenamefont {Sch{\"a}fer}\ \emph {et~al.}(2022)\citenamefont {Sch{\"a}fer}, \citenamefont {Flick}, \citenamefont {Ronca}, \citenamefont {Narang},\ and\ \citenamefont {Rubio}}]{Sclaefer_19}%
  \BibitemOpen
  \bibfield  {author} {\bibinfo {author} {\bibfnamefont {C.}~\bibnamefont {Sch{\"a}fer}}, \bibinfo {author} {\bibfnamefont {J.}~\bibnamefont {Flick}}, \bibinfo {author} {\bibfnamefont {E.}~\bibnamefont {Ronca}}, \bibinfo {author} {\bibfnamefont {P.}~\bibnamefont {Narang}}, \ and\ \bibinfo {author} {\bibfnamefont {A.}~\bibnamefont {Rubio}},\ }\bibfield  {title} {\enquote {\bibinfo {title} {Shining light on the microscopic resonant mechanism responsible for cavity-mediated chemical reactivity},}\ }\href {\doibase 10.1038/s41467-022-35363-6} {\bibfield  {journal} {\bibinfo  {journal} {Nature Communications}\ }\textbf {\bibinfo {volume} {13}},\ \bibinfo {pages} {7817} (\bibinfo {year} {2022})}\BibitemShut {NoStop}%
\bibitem [{\citenamefont {Fainstein}\ \emph {et~al.}(1995)\citenamefont {Fainstein}, \citenamefont {Jusserand},\ and\ \citenamefont {Thierry-Mieg}}]{Fainstein95}%
  \BibitemOpen
  \bibfield  {author} {\bibinfo {author} {\bibfnamefont {A.}~\bibnamefont {Fainstein}}, \bibinfo {author} {\bibfnamefont {B.}~\bibnamefont {Jusserand}}, \ and\ \bibinfo {author} {\bibfnamefont {V.}~\bibnamefont {Thierry-Mieg}},\ }\bibfield  {title} {\enquote {\bibinfo {title} {Raman scattering enhancement by optical confinement in a semiconductor planar microcavity},}\ }\href {\doibase 10.1103/PhysRevLett.75.3764} {\bibfield  {journal} {\bibinfo  {journal} {Phys. Rev. Lett.}\ }\textbf {\bibinfo {volume} {75}},\ \bibinfo {pages} {3764} (\bibinfo {year} {1995})}\BibitemShut {NoStop}%
\bibitem [{\citenamefont {Sulzer}\ \emph {et~al.}(2022)\citenamefont {Sulzer}, \citenamefont {Högner}, \citenamefont {Raab}, \citenamefont {Fürst}, \citenamefont {Fill}, \citenamefont {Gerz}, \citenamefont {Hofer}, \citenamefont {Voronina},\ and\ \citenamefont {Pupeza}}]{Sulzer22}%
  \BibitemOpen
  \bibfield  {author} {\bibinfo {author} {\bibfnamefont {P.}~\bibnamefont {Sulzer}}, \bibinfo {author} {\bibfnamefont {M.}~\bibnamefont {Högner}}, \bibinfo {author} {\bibfnamefont {A.~K.}\ \bibnamefont {Raab}}, \bibinfo {author} {\bibfnamefont {L.}~\bibnamefont {Fürst}}, \bibinfo {author} {\bibfnamefont {E.}~\bibnamefont {Fill}}, \bibinfo {author} {\bibfnamefont {D.}~\bibnamefont {Gerz}}, \bibinfo {author} {\bibfnamefont {C.}~\bibnamefont {Hofer}}, \bibinfo {author} {\bibfnamefont {L.}~\bibnamefont {Voronina}}, \ and\ \bibinfo {author} {\bibfnamefont {I.}~\bibnamefont {Pupeza}},\ }\bibfield  {title} {\enquote {\bibinfo {title} {Cavity-enhanced field-resolved spectroscopy},}\ }\href {\doibase 10.1038/s41566-022-01057-0} {\bibfield  {journal} {\bibinfo  {journal} {Nat. Photonics}\ }\textbf {\bibinfo {volume} {16}},\ \bibinfo {pages} {692} (\bibinfo {year} {2022})}\BibitemShut {NoStop}%
\bibitem [{\citenamefont {Reber}\ \emph {et~al.}(2016)\citenamefont {Reber}, \citenamefont {Chen},\ and\ \citenamefont {Allison}}]{Melanie16}%
  \BibitemOpen
  \bibfield  {author} {\bibinfo {author} {\bibfnamefont {M.~A.~R.}\ \bibnamefont {Reber}}, \bibinfo {author} {\bibfnamefont {Y.}~\bibnamefont {Chen}}, \ and\ \bibinfo {author} {\bibfnamefont {T.~K.}\ \bibnamefont {Allison}},\ }\bibfield  {title} {\enquote {\bibinfo {title} {Cavity-enhanced ultrafast spectroscopy: Ultrafast meets ultrasensitive},}\ }\href {\doibase 10.1364/OPTICA.3.000311} {\bibfield  {journal} {\bibinfo  {journal} {Optica}\ }\textbf {\bibinfo {volume} {3}},\ \bibinfo {pages} {311} (\bibinfo {year} {2016})}\BibitemShut {NoStop}%
\bibitem [{\citenamefont {Gagliardi}\ and\ \citenamefont {Loock}(2013)}]{Gagliardi13}%
  \BibitemOpen
  \bibfield  {author} {\bibinfo {author} {\bibfnamefont {G.}~\bibnamefont {Gagliardi}}\ and\ \bibinfo {author} {\bibfnamefont {H.~P.}\ \bibnamefont {Loock}},\ }\bibfield  {title} {\enquote {\bibinfo {title} {Cavity enhanced spectroscopy and sensing},}\ }\href {\doibase 10.1007/978-3-642-40003-2} {\bibfield  {journal} {\bibinfo  {journal} {Springer, New York}\ } (\bibinfo {year} {2013}),\ 10.1007/978-3-642-40003-2}\BibitemShut {NoStop}%
\bibitem [{\citenamefont {Först}\ \emph {et~al.}(2011)\citenamefont {Först}, \citenamefont {Manzoni}, \citenamefont {Kaiser}, \citenamefont {Tomioka}, \citenamefont {Tokura}, \citenamefont {Merlin},\ and\ \citenamefont {Cavalleri}}]{forst_nonlinear_2011}%
  \BibitemOpen
  \bibfield  {author} {\bibinfo {author} {\bibfnamefont {M.}~\bibnamefont {Först}}, \bibinfo {author} {\bibfnamefont {C.}~\bibnamefont {Manzoni}}, \bibinfo {author} {\bibfnamefont {S.}~\bibnamefont {Kaiser}}, \bibinfo {author} {\bibfnamefont {Y.}~\bibnamefont {Tomioka}}, \bibinfo {author} {\bibfnamefont {Y.}~\bibnamefont {Tokura}}, \bibinfo {author} {\bibfnamefont {R.}~\bibnamefont {Merlin}}, \ and\ \bibinfo {author} {\bibfnamefont {A.}~\bibnamefont {Cavalleri}},\ }\bibfield  {title} {\enquote {\bibinfo {title} {Nonlinear phononics as an ultrafast route to lattice control},}\ }\href {\doibase 10.1038/nphys2055} {\bibfield  {journal} {\bibinfo  {journal} {Nature Phys}\ }\textbf {\bibinfo {volume} {7}},\ \bibinfo {pages} {854--856} (\bibinfo {year} {2011})}\BibitemShut {NoStop}%
\bibitem [{\citenamefont {Schmidt}\ \emph {et~al.}(2016)\citenamefont {Schmidt}, \citenamefont {Esteban}, \citenamefont {González-Tudela}, \citenamefont {Giedke},\ and\ \citenamefont {Aizpurua}}]{Schmidt}%
  \BibitemOpen
  \bibfield  {author} {\bibinfo {author} {\bibfnamefont {M.~K.}\ \bibnamefont {Schmidt}}, \bibinfo {author} {\bibfnamefont {R.}~\bibnamefont {Esteban}}, \bibinfo {author} {\bibfnamefont {A.}~\bibnamefont {González-Tudela}}, \bibinfo {author} {\bibfnamefont {G.}~\bibnamefont {Giedke}}, \ and\ \bibinfo {author} {\bibfnamefont {J.}~\bibnamefont {Aizpurua}},\ }\bibfield  {title} {\enquote {\bibinfo {title} {Quantum mechanical description of raman scattering from molecules in plasmonic cavities},}\ }\href {\doibase 10.1021/acsnano.6b02484} {\bibfield  {journal} {\bibinfo  {journal} {ACS Nano}\ }\textbf {\bibinfo {volume} {10}},\ \bibinfo {pages} {6291} (\bibinfo {year} {2016})}\BibitemShut {NoStop}%
\bibitem [{\citenamefont {Timsina}\ \emph {et~al.}(2024)\citenamefont {Timsina}, \citenamefont {Hammadia}, \citenamefont {Milani}, \citenamefont {Brolo},\ and\ \citenamefont {de~Sousa}}]{Sanker}%
  \BibitemOpen
  \bibfield  {author} {\bibinfo {author} {\bibfnamefont {S.}~\bibnamefont {Timsina}}, \bibinfo {author} {\bibfnamefont {T.}~\bibnamefont {Hammadia}}, \bibinfo {author} {\bibfnamefont {F.~S}\ \bibnamefont {Milani}, \bibfnamefont {S.~G. de Aguiar~J\'unior}}, \bibinfo {author} {\bibfnamefont {A.}~\bibnamefont {Brolo}}, \ and\ \bibinfo {author} {\bibfnamefont {R.}~\bibnamefont {de~Sousa}},\ }\bibfield  {title} {\enquote {\bibinfo {title} {Resonant squeezed light from photonic cooper pairs},}\ }\href {\doibase 10.1103/PhysRevResearch.6.033067} {\bibfield  {journal} {\bibinfo  {journal} {Phys. Rev. Res.}\ }\textbf {\bibinfo {volume} {6}},\ \bibinfo {pages} {033067} (\bibinfo {year} {2024})}\BibitemShut {NoStop}%
\bibitem [{\citenamefont {{R. Matsunaga, N. Tsuji, H. Fujita, A. Sugioka, K. Makise, Y. Uzawa, H. Terai, Z. Wang, H. Aoki and R. Shimano}}(2014)}]{Matsunaga2014}%
  \BibitemOpen
  \bibfield  {author} {\bibinfo {author} {\bibnamefont {{R. Matsunaga, N. Tsuji, H. Fujita, A. Sugioka, K. Makise, Y. Uzawa, H. Terai, Z. Wang, H. Aoki and R. Shimano}}},\ }\bibfield  {title} {\enquote {\bibinfo {title} {Light-induced collective pseudospin precession resonating with {H}iggs mode in a superconductor},}\ }\href {\doibase 10.1126/science.1254697} {\bibfield  {journal} {\bibinfo  {journal} {Science}\ }\textbf {\bibinfo {volume} {345}},\ \bibinfo {pages} {1145--1149} (\bibinfo {year} {2014})}\BibitemShut {NoStop}%
\bibitem [{\citenamefont {Buzzi}\ \emph {et~al.}(2021)\citenamefont {Buzzi}, \citenamefont {Jotzu}, \citenamefont {Cavalleri}, \citenamefont {Cirac}, \citenamefont {Demler}, \citenamefont {Halperin}, \citenamefont {Lukin}, \citenamefont {Shi}, \citenamefont {Wang},\ and\ \citenamefont {Podolsky}}]{Buzzi21}%
  \BibitemOpen
  \bibfield  {author} {\bibinfo {author} {\bibfnamefont {M.}~\bibnamefont {Buzzi}}, \bibinfo {author} {\bibfnamefont {G.}~\bibnamefont {Jotzu}}, \bibinfo {author} {\bibfnamefont {A.}~\bibnamefont {Cavalleri}}, \bibinfo {author} {\bibfnamefont {J.~I.}\ \bibnamefont {Cirac}}, \bibinfo {author} {\bibfnamefont {E.~A.}\ \bibnamefont {Demler}}, \bibinfo {author} {\bibfnamefont {B.~I.}\ \bibnamefont {Halperin}}, \bibinfo {author} {\bibfnamefont {M.~D.}\ \bibnamefont {Lukin}}, \bibinfo {author} {\bibfnamefont {T.}~\bibnamefont {Shi}}, \bibinfo {author} {\bibfnamefont {Y.}~\bibnamefont {Wang}}, \ and\ \bibinfo {author} {\bibfnamefont {D.}~\bibnamefont {Podolsky}},\ }\bibfield  {title} {\enquote {\bibinfo {title} {Higgs-{M}ediated {O}ptical {A}mplification in a {N}onequilibrium {S}uperconductor},}\ }\href {\doibase 10.1103/PhysRevX.11.011055} {\bibfield  {journal} {\bibinfo  {journal} {Phys. Rev. X}\ }\textbf {\bibinfo {volume} {11}},\ \bibinfo {pages} {011055} (\bibinfo {year} {2021})}\BibitemShut {NoStop}%
\bibitem [{\citenamefont {Collado}\ \emph {et~al.}(2018)\citenamefont {Collado}, \citenamefont {Lorenzana}, \citenamefont {Usaj},\ and\ \citenamefont {Balseiro}}]{HP_2018}%
  \BibitemOpen
  \bibfield  {author} {\bibinfo {author} {\bibfnamefont {H.~P.~Ojeda}\ \bibnamefont {Collado}}, \bibinfo {author} {\bibfnamefont {J.}~\bibnamefont {Lorenzana}}, \bibinfo {author} {\bibfnamefont {G.}~\bibnamefont {Usaj}}, \ and\ \bibinfo {author} {\bibfnamefont {C.~A.}\ \bibnamefont {Balseiro}},\ }\bibfield  {title} {\enquote {\bibinfo {title} {Population inversion and dynamical phase transitions in a driven superconductor},}\ }\href {\doibase 10.1103/PhysRevB.98.214519} {\bibfield  {journal} {\bibinfo  {journal} {Phys. Rev. B}\ }\textbf {\bibinfo {volume} {98}},\ \bibinfo {pages} {214519} (\bibinfo {year} {2018})}\BibitemShut {NoStop}%
\bibitem [{\citenamefont {Juraschek}\ \emph {et~al.}(2021)\citenamefont {Juraschek}, \citenamefont {Neuman}, \citenamefont {Flick},\ and\ \citenamefont {Narang}}]{juraschek_cavity_2021}%
  \BibitemOpen
  \bibfield  {author} {\bibinfo {author} {\bibfnamefont {D.~M.}\ \bibnamefont {Juraschek}}, \bibinfo {author} {\bibfnamefont {T.}~\bibnamefont {Neuman}}, \bibinfo {author} {\bibfnamefont {J.}~\bibnamefont {Flick}}, \ and\ \bibinfo {author} {\bibfnamefont {P.}~\bibnamefont {Narang}},\ }\bibfield  {title} {\enquote {\bibinfo {title} {Cavity control of nonlinear phononics},}\ }\href {\doibase 10.1103/PhysRevResearch.3.L032046} {\bibfield  {journal} {\bibinfo  {journal} {Phys. Rev. Res.}\ }\textbf {\bibinfo {volume} {3}},\ \bibinfo {pages} {L032046} (\bibinfo {year} {2021})}\BibitemShut {NoStop}%
\bibitem [{\citenamefont {Polkovnikov}(2010)}]{polkovnikov_phase_2010}%
  \BibitemOpen
  \bibfield  {author} {\bibinfo {author} {\bibfnamefont {A.}~\bibnamefont {Polkovnikov}},\ }\bibfield  {title} {\enquote {\bibinfo {title} {Phase space representation of quantum dynamics},}\ }\href {\doibase 10.1016/j.aop.2010.02.006} {\bibfield  {journal} {\bibinfo  {journal} {Annals of Physics}\ }\textbf {\bibinfo {volume} {325}},\ \bibinfo {pages} {1790} (\bibinfo {year} {2010})}\BibitemShut {NoStop}%
\bibitem [{\citenamefont {Cosme}\ \emph {et~al.}(2019)\citenamefont {Cosme}, \citenamefont {Skulte},\ and\ \citenamefont {Mathey}}]{cosme_time_2019}%
  \BibitemOpen
  \bibfield  {author} {\bibinfo {author} {\bibfnamefont {J.~G.}\ \bibnamefont {Cosme}}, \bibinfo {author} {\bibfnamefont {J.}~\bibnamefont {Skulte}}, \ and\ \bibinfo {author} {\bibfnamefont {L.}~\bibnamefont {Mathey}},\ }\bibfield  {title} {\enquote {\bibinfo {title} {Time crystals in a shaken atom-cavity system},}\ }\href {\doibase 10.1103/PhysRevA.100.053615} {\bibfield  {journal} {\bibinfo  {journal} {Phys. Rev. A}\ }\textbf {\bibinfo {volume} {100}},\ \bibinfo {pages} {053615} (\bibinfo {year} {2019})}\BibitemShut {NoStop}%
\bibitem [{\citenamefont {Skulte}\ \emph {et~al.}(2023)\citenamefont {Skulte}, \citenamefont {Kongkhambut}, \citenamefont {Rao}, \citenamefont {Mathey}, \citenamefont {Ke\ss{}ler}, \citenamefont {Hemmerich},\ and\ \citenamefont {Cosme}}]{Skulte2023}%
  \BibitemOpen
  \bibfield  {author} {\bibinfo {author} {\bibfnamefont {J.}~\bibnamefont {Skulte}}, \bibinfo {author} {\bibfnamefont {P.}~\bibnamefont {Kongkhambut}}, \bibinfo {author} {\bibfnamefont {S.}~\bibnamefont {Rao}}, \bibinfo {author} {\bibfnamefont {L.}~\bibnamefont {Mathey}}, \bibinfo {author} {\bibfnamefont {H.}~\bibnamefont {Ke\ss{}ler}}, \bibinfo {author} {\bibfnamefont {A.}~\bibnamefont {Hemmerich}}, \ and\ \bibinfo {author} {\bibfnamefont {J.~G.}\ \bibnamefont {Cosme}},\ }\bibfield  {title} {\enquote {\bibinfo {title} {{Condensate Formation in a Dark State of a Driven Atom-Cavity System}},}\ }\href {\doibase 10.1103/PhysRevLett.130.163603} {\bibfield  {journal} {\bibinfo  {journal} {Phys. Rev. Lett.}\ }\textbf {\bibinfo {volume} {130}},\ \bibinfo {pages} {163603} (\bibinfo {year} {2023})}\BibitemShut {NoStop}%
\bibitem [{\citenamefont {Basov}\ \emph {et~al.}(2021)\citenamefont {Basov}, \citenamefont {Asenjo-Garcia}, \citenamefont {Schuck}, \citenamefont {Zhu},\ and\ \citenamefont {Rubio}}]{Polaritonpanorama}%
  \BibitemOpen
  \bibfield  {author} {\bibinfo {author} {\bibfnamefont {D.~N.}\ \bibnamefont {Basov}}, \bibinfo {author} {\bibfnamefont {A.}~\bibnamefont {Asenjo-Garcia}}, \bibinfo {author} {\bibfnamefont {P.~J.}\ \bibnamefont {Schuck}}, \bibinfo {author} {\bibfnamefont {X.}~\bibnamefont {Zhu}}, \ and\ \bibinfo {author} {\bibfnamefont {A.}~\bibnamefont {Rubio}},\ }\bibfield  {title} {\enquote {\bibinfo {title} {Polariton panorama},}\ }\href {\doibase doi:10.1515/nanoph-2020-0449} {\bibfield  {journal} {\bibinfo  {journal} {Nanophotonics}\ }\textbf {\bibinfo {volume} {10}},\ \bibinfo {pages} {549} (\bibinfo {year} {2021})}\BibitemShut {NoStop}%
\bibitem [{Sup()}]{SupInfo}%
  \BibitemOpen
  \bibfield  {title} {\enquote {\bibinfo {title} {See {S}upplemental {M}aterial at [url] for additional information about the gaussian approximation, detailed discussion of {TWA} simulations, the effects of nonlinearities, strong dissipation and thermal effects, as well as a discussion on the {R}aman-cavity coupling strength. {T}he {S}upplemental {M}aterial includes {R}efs.~\cite{BreuerPetruccione,Landau,Alex2016,Liang_14}},}\ }\href@noop {} {\ }\BibitemShut {NoStop}%
\bibitem [{\citenamefont {Kongkhambut}\ \emph {et~al.}(2022)\citenamefont {Kongkhambut}, \citenamefont {Skulte}, \citenamefont {Mathey}, \citenamefont {Cosme}, \citenamefont {Hemmerich},\ and\ \citenamefont {Keßler}}]{kongkhambut_observation_2022}%
  \BibitemOpen
  \bibfield  {author} {\bibinfo {author} {\bibfnamefont {P.}~\bibnamefont {Kongkhambut}}, \bibinfo {author} {\bibfnamefont {J.}~\bibnamefont {Skulte}}, \bibinfo {author} {\bibfnamefont {L.}~\bibnamefont {Mathey}}, \bibinfo {author} {\bibfnamefont {J.~G.}\ \bibnamefont {Cosme}}, \bibinfo {author} {\bibfnamefont {A.}~\bibnamefont {Hemmerich}}, \ and\ \bibinfo {author} {\bibfnamefont {H.}~\bibnamefont {Keßler}},\ }\bibfield  {title} {\enquote {\bibinfo {title} {Observation of a continuous time crystal},}\ }\href {\doibase 10.1126/science.abo3382} {\bibfield  {journal} {\bibinfo  {journal} {Science}\ }\textbf {\bibinfo {volume} {377}},\ \bibinfo {pages} {670--673} (\bibinfo {year} {2022})}\BibitemShut {NoStop}%
\bibitem [{\citenamefont {Cardona}\ and\ \citenamefont {G.Güntherodt}(1982)}]{Raman_scattering_Reference1}%
  \BibitemOpen
  \bibfield  {author} {\bibinfo {author} {\bibfnamefont {M.}~\bibnamefont {Cardona}}\ and\ \bibinfo {author} {\bibnamefont {G.Güntherodt}},\ }\bibfield  {title} {\enquote {\bibinfo {title} {Light {S}cattering in {S}olids {II}},}\ }\href@noop {} {\bibfield  {journal} {\bibinfo  {journal} {Springer-Verlag, Berlin}\ } (\bibinfo {year} {1982})}\BibitemShut {NoStop}%
\bibitem [{\citenamefont {Brüesch}(1986)}]{Raman_scattering_Reference2}%
  \BibitemOpen
  \bibfield  {author} {\bibinfo {author} {\bibfnamefont {P.}~\bibnamefont {Brüesch}},\ }\bibfield  {title} {\enquote {\bibinfo {title} {Phonons: {T}heory and {E}xperiments {II}},}\ }\href@noop {} {\bibfield  {journal} {\bibinfo  {journal} {Springer-Verlag, Berlin}\ } (\bibinfo {year} {1986})}\BibitemShut {NoStop}%
\bibitem [{Not()}]{Note1}%
  \BibitemOpen
  \bibfield  {title} {\enquote {\bibinfo {title} {Our calculations correspond to the quantum noise limit at zero temperature where only {S}tokes peaks show up. {I}n the presence of thermal fluctuations, at finite temperature, both {S}tokes and anti-{S}tokes peaks appear in the spectrum that we present in the {S}upplemental {M}aterial~\cite{SupInfo} for completeness.}}\ }\href@noop {} {\ }\BibitemShut {NoStop}%
\bibitem [{not()}]{note}%
  \BibitemOpen
  \bibfield  {title} {\enquote {\bibinfo {title} {In the limit of small cavity frequencies, the nonlinearities of strength $g_4$ become relevant which tends to suppress the cavity fluctuations ($\delta x^2<0$) weakly with respect to the case of $g=g_4=0$.}}\ }\href@noop {} {\ }\BibitemShut {NoStop}%
\bibitem [{\citenamefont {Frisk~Kockum}\ \emph {et~al.}(2019)\citenamefont {Frisk~Kockum}, \citenamefont {Miranowicz}, \citenamefont {De~Liberato}, \citenamefont {Savasta},\ and\ \citenamefont {Nori}}]{Frisk2019}%
  \BibitemOpen
  \bibfield  {author} {\bibinfo {author} {\bibfnamefont {A.}~\bibnamefont {Frisk~Kockum}}, \bibinfo {author} {\bibfnamefont {A.}~\bibnamefont {Miranowicz}}, \bibinfo {author} {\bibfnamefont {S.}~\bibnamefont {De~Liberato}}, \bibinfo {author} {\bibfnamefont {S.}~\bibnamefont {Savasta}}, \ and\ \bibinfo {author} {\bibfnamefont {F.}~\bibnamefont {Nori}},\ }\bibfield  {title} {\enquote {\bibinfo {title} {Ultrastrong coupling between light and matter},}\ }\href {\doibase 10.1038/s42254-018-0006-2} {\bibfield  {journal} {\bibinfo  {journal} {Nature Reviews Physics}\ }\textbf {\bibinfo {volume} {1}},\ \bibinfo {pages} {19} (\bibinfo {year} {2019})}\BibitemShut {NoStop}%
\bibitem [{\citenamefont {Zeiger}\ \emph {et~al.}(1992)\citenamefont {Zeiger}, \citenamefont {Vidal}, \citenamefont {Cheng}, \citenamefont {Ippen}, \citenamefont {Dresselhaus},\ and\ \citenamefont {Dresselhaus}}]{Dresselhaus}%
  \BibitemOpen
  \bibfield  {author} {\bibinfo {author} {\bibfnamefont {H.~J}\ \bibnamefont {Zeiger}}, \bibinfo {author} {\bibfnamefont {J.}~\bibnamefont {Vidal}}, \bibinfo {author} {\bibfnamefont {T.~K.}\ \bibnamefont {Cheng}}, \bibinfo {author} {\bibfnamefont {E.~P.}\ \bibnamefont {Ippen}}, \bibinfo {author} {\bibfnamefont {G.}~\bibnamefont {Dresselhaus}}, \ and\ \bibinfo {author} {\bibfnamefont {M.~S.}\ \bibnamefont {Dresselhaus}},\ }\bibfield  {title} {\enquote {\bibinfo {title} {Theory for displacive excitation of coherent phonons},}\ }\href {\doibase 10.1103/PhysRevB.45.768} {\bibfield  {journal} {\bibinfo  {journal} {Phys. Rev. B}\ }\textbf {\bibinfo {volume} {45}},\ \bibinfo {pages} {768} (\bibinfo {year} {1992})}\BibitemShut {NoStop}%
\bibitem [{\citenamefont {Merlin}(1997)}]{Merlin97}%
  \BibitemOpen
  \bibfield  {author} {\bibinfo {author} {\bibfnamefont {R.}~\bibnamefont {Merlin}},\ }\bibfield  {title} {\enquote {\bibinfo {title} {Generating coherent thz phonons with light pulses},}\ }\href {\doibase 10.1016/S0038-1098(96)00721-1} {\bibfield  {journal} {\bibinfo  {journal} {Solid State Commun.}\ }\textbf {\bibinfo {volume} {102}},\ \bibinfo {pages} {207} (\bibinfo {year} {1997})}\BibitemShut {NoStop}%
\bibitem [{\citenamefont {Giorgianni}\ \emph {et~al.}(2022)\citenamefont {Giorgianni}, \citenamefont {Udina}, \citenamefont {Cea}, \citenamefont {Paris}, \citenamefont {Caputo}, \citenamefont {Radovic}, \citenamefont {Boie}, \citenamefont {Sakai}, \citenamefont {Schneider},\ and\ \citenamefont {Johnson}}]{Giorgianni22}%
  \BibitemOpen
  \bibfield  {author} {\bibinfo {author} {\bibfnamefont {F.}~\bibnamefont {Giorgianni}}, \bibinfo {author} {\bibfnamefont {M.}~\bibnamefont {Udina}}, \bibinfo {author} {\bibfnamefont {T.}~\bibnamefont {Cea}}, \bibinfo {author} {\bibfnamefont {E.}~\bibnamefont {Paris}}, \bibinfo {author} {\bibfnamefont {M.}~\bibnamefont {Caputo}}, \bibinfo {author} {\bibfnamefont {M.}~\bibnamefont {Radovic}}, \bibinfo {author} {\bibfnamefont {L.}~\bibnamefont {Boie}}, \bibinfo {author} {\bibfnamefont {J.}~\bibnamefont {Sakai}}, \bibinfo {author} {\bibfnamefont {C.~W.}\ \bibnamefont {Schneider}}, \ and\ \bibinfo {author} {\bibfnamefont {S.~L.}\ \bibnamefont {Johnson}},\ }\bibfield  {title} {\enquote {\bibinfo {title} {Terahertz displacive excitation of a coherent raman-active phonon in v$_2$o$_3$},}\ }\href {\doibase 10.1038/s42005-022-00882-7} {\bibfield  {journal} {\bibinfo  {journal} {Commun. Phys.}\ }\textbf {\bibinfo {volume} {5}},\ \bibinfo {pages} {103} (\bibinfo {year} {2022})}\BibitemShut {NoStop}%
\bibitem [{\citenamefont {Verzhbitskiy}\ \emph {et~al.}(2016)\citenamefont {Verzhbitskiy}, \citenamefont {De~Corato}, \citenamefont {Ruini}, \citenamefont {Molinari}, \citenamefont {Narita}, \citenamefont {Hu}, \citenamefont {Schwab}, \citenamefont {Bruna}, \citenamefont {Yoon}, \citenamefont {Milana}, \citenamefont {Feng}, \citenamefont {Müllen}, \citenamefont {Ferrari}, \citenamefont {Casiraghi},\ and\ \citenamefont {Prezzi}}]{verzhbitskiy_raman_2016}%
  \BibitemOpen
  \bibfield  {author} {\bibinfo {author} {\bibfnamefont {I.~A.}\ \bibnamefont {Verzhbitskiy}}, \bibinfo {author} {\bibfnamefont {M.}~\bibnamefont {De~Corato}}, \bibinfo {author} {\bibfnamefont {A.}~\bibnamefont {Ruini}}, \bibinfo {author} {\bibfnamefont {E.}~\bibnamefont {Molinari}}, \bibinfo {author} {\bibfnamefont {A.}~\bibnamefont {Narita}}, \bibinfo {author} {\bibfnamefont {Y.}~\bibnamefont {Hu}}, \bibinfo {author} {\bibfnamefont {M.~G.}\ \bibnamefont {Schwab}}, \bibinfo {author} {\bibfnamefont {M.}~\bibnamefont {Bruna}}, \bibinfo {author} {\bibfnamefont {D.}~\bibnamefont {Yoon}}, \bibinfo {author} {\bibfnamefont {S.}~\bibnamefont {Milana}}, \bibinfo {author} {\bibfnamefont {X.}~\bibnamefont {Feng}}, \bibinfo {author} {\bibfnamefont {K.}~\bibnamefont {Müllen}}, \bibinfo {author} {\bibfnamefont {A.~C.}\ \bibnamefont {Ferrari}}, \bibinfo {author} {\bibfnamefont {C.}~\bibnamefont {Casiraghi}}, \ and\ \bibinfo {author} {\bibfnamefont {D.}~\bibnamefont {Prezzi}},\ }\bibfield  {title} {\enquote {\bibinfo
  {title} {Raman fingerprints of atomically precise graphene nanoribbons},}\ }\href {\doibase 10.1021/acs.nanolett.5b04183} {\bibfield  {journal} {\bibinfo  {journal} {Nano Lett.}\ }\textbf {\bibinfo {volume} {16}},\ \bibinfo {pages} {3442} (\bibinfo {year} {2016})}\BibitemShut {NoStop}%
\bibitem [{\citenamefont {He}\ \emph {et~al.}(2013)\citenamefont {He}, \citenamefont {Chung}, \citenamefont {Delaney}, \citenamefont {Keiser}, \citenamefont {Jauregui}, \citenamefont {Shand}, \citenamefont {Chancey}, \citenamefont {Wang}, \citenamefont {Bao},\ and\ \citenamefont {Chen}}]{he_observation_2013}%
  \BibitemOpen
  \bibfield  {author} {\bibinfo {author} {\bibfnamefont {R.}~\bibnamefont {He}}, \bibinfo {author} {\bibfnamefont {T.-F.}\ \bibnamefont {Chung}}, \bibinfo {author} {\bibfnamefont {C.}~\bibnamefont {Delaney}}, \bibinfo {author} {\bibfnamefont {C.}~\bibnamefont {Keiser}}, \bibinfo {author} {\bibfnamefont {L.~A.}\ \bibnamefont {Jauregui}}, \bibinfo {author} {\bibfnamefont {P.~M.}\ \bibnamefont {Shand}}, \bibinfo {author} {\bibfnamefont {C.~C.}\ \bibnamefont {Chancey}}, \bibinfo {author} {\bibfnamefont {Y.}~\bibnamefont {Wang}}, \bibinfo {author} {\bibfnamefont {J.}~\bibnamefont {Bao}}, \ and\ \bibinfo {author} {\bibfnamefont {Y.~P.}\ \bibnamefont {Chen}},\ }\bibfield  {title} {\enquote {\bibinfo {title} {Observation of low energy {R}aman modes in twisted bilayer graphene},}\ }\href {\doibase 10.1021/nl4013387} {\bibfield  {journal} {\bibinfo  {journal} {Nano Lett}\ }\textbf {\bibinfo {volume} {13}},\ \bibinfo {pages} {3594} (\bibinfo {year} {2013})}\BibitemShut {NoStop}%
\bibitem [{\citenamefont {Valmorra}\ \emph {et~al.}(2013)\citenamefont {Valmorra}, \citenamefont {Scalari}, \citenamefont {Maissen}, \citenamefont {Fu}, \citenamefont {Schönenberger}, \citenamefont {Choi}, \citenamefont {Park}, \citenamefont {Beck},\ and\ \citenamefont {Faist}}]{Valmorra2013}%
  \BibitemOpen
  \bibfield  {author} {\bibinfo {author} {\bibfnamefont {F.}~\bibnamefont {Valmorra}}, \bibinfo {author} {\bibfnamefont {G.}~\bibnamefont {Scalari}}, \bibinfo {author} {\bibfnamefont {C.}~\bibnamefont {Maissen}}, \bibinfo {author} {\bibfnamefont {W.}~\bibnamefont {Fu}}, \bibinfo {author} {\bibfnamefont {C.}~\bibnamefont {Schönenberger}}, \bibinfo {author} {\bibfnamefont {J.~W.}\ \bibnamefont {Choi}}, \bibinfo {author} {\bibfnamefont {H.~G.}\ \bibnamefont {Park}}, \bibinfo {author} {\bibfnamefont {M.}~\bibnamefont {Beck}}, \ and\ \bibinfo {author} {\bibfnamefont {J.}~\bibnamefont {Faist}},\ }\bibfield  {title} {\enquote {\bibinfo {title} {Low-{B}ias {A}ctive {C}ontrol of {T}erahertz {W}aves by {C}oupling {L}arge-{A}rea {CVD} {G}raphene to a {T}erahertz metamaterial},}\ }\href {\doibase 10.1021/nl4012547} {\bibfield  {journal} {\bibinfo  {journal} {Nano Letters}\ }\textbf {\bibinfo {volume} {13}},\ \bibinfo {pages} {3193} (\bibinfo {year} {2013})}\BibitemShut {NoStop}%
\bibitem [{\citenamefont {Maissen}\ \emph {et~al.}(2014)\citenamefont {Maissen}, \citenamefont {Scalari}, \citenamefont {Valmorra}, \citenamefont {Beck}, \citenamefont {Faist}, \citenamefont {Cibella}, \citenamefont {Leoni}, \citenamefont {Reichl}, \citenamefont {Charpentier},\ and\ \citenamefont {Wegscheider}}]{Maissen2014}%
  \BibitemOpen
  \bibfield  {author} {\bibinfo {author} {\bibfnamefont {C.}~\bibnamefont {Maissen}}, \bibinfo {author} {\bibfnamefont {G.}~\bibnamefont {Scalari}}, \bibinfo {author} {\bibfnamefont {F.}~\bibnamefont {Valmorra}}, \bibinfo {author} {\bibfnamefont {M.}~\bibnamefont {Beck}}, \bibinfo {author} {\bibfnamefont {J.}~\bibnamefont {Faist}}, \bibinfo {author} {\bibfnamefont {S.}~\bibnamefont {Cibella}}, \bibinfo {author} {\bibfnamefont {R.}~\bibnamefont {Leoni}}, \bibinfo {author} {\bibfnamefont {C.}~\bibnamefont {Reichl}}, \bibinfo {author} {\bibfnamefont {C.}~\bibnamefont {Charpentier}}, \ and\ \bibinfo {author} {\bibfnamefont {W.}~\bibnamefont {Wegscheider}},\ }\bibfield  {title} {\enquote {\bibinfo {title} {Ultrastrong coupling in the near field of complementary split-ring resonators},}\ }\href {\doibase 10.1103/PhysRevB.90.205309} {\bibfield  {journal} {\bibinfo  {journal} {Phys. Rev. B}\ }\textbf {\bibinfo {volume} {90}},\ \bibinfo {pages} {205309} (\bibinfo {year} {2014})}\BibitemShut {NoStop}%
\bibitem [{\citenamefont {Scalari}\ \emph {et~al.}(2012)\citenamefont {Scalari}, \citenamefont {Maissen}, \citenamefont {Turčinková}, \citenamefont {Hagenmüller}, \citenamefont {Liberato}, \citenamefont {Ciuti}, \citenamefont {Reichl}, \citenamefont {Schuh}, \citenamefont {Wegscheider}, \citenamefont {Beck},\ and\ \citenamefont {Faist}}]{Scalari2012}%
  \BibitemOpen
  \bibfield  {author} {\bibinfo {author} {\bibfnamefont {G.}~\bibnamefont {Scalari}}, \bibinfo {author} {\bibfnamefont {C.}~\bibnamefont {Maissen}}, \bibinfo {author} {\bibfnamefont {D.}~\bibnamefont {Turčinková}}, \bibinfo {author} {\bibfnamefont {D.}~\bibnamefont {Hagenmüller}}, \bibinfo {author} {\bibfnamefont {S.~De}\ \bibnamefont {Liberato}}, \bibinfo {author} {\bibfnamefont {C.}~\bibnamefont {Ciuti}}, \bibinfo {author} {\bibfnamefont {C.}~\bibnamefont {Reichl}}, \bibinfo {author} {\bibfnamefont {D.}~\bibnamefont {Schuh}}, \bibinfo {author} {\bibfnamefont {W.}~\bibnamefont {Wegscheider}}, \bibinfo {author} {\bibfnamefont {M.}~\bibnamefont {Beck}}, \ and\ \bibinfo {author} {\bibfnamefont {J.}~\bibnamefont {Faist}},\ }\bibfield  {title} {\enquote {\bibinfo {title} {Ultrastrong coupling of the cyclotron transition of a 2d electron gas to a {THz} metamaterial},}\ }\href {\doibase 10.1126/science.1216022} {\bibfield  {journal} {\bibinfo  {journal} {Science}\ }\textbf {\bibinfo {volume} {335}},\ \bibinfo
  {pages} {1323} (\bibinfo {year} {2012})}\BibitemShut {NoStop}%
\bibitem [{\citenamefont {Lin}\ \emph {et~al.}(2021)\citenamefont {Lin}, \citenamefont {Holler}, \citenamefont {Bauer}, \citenamefont {Parzefall}, \citenamefont {Scheuck}, \citenamefont {Peng}, \citenamefont {Korn}, \citenamefont {Bange}, \citenamefont {Lupton},\ and\ \citenamefont {Schüller}}]{Lin2021}%
  \BibitemOpen
  \bibfield  {author} {\bibinfo {author} {\bibfnamefont {K.-Q.}\ \bibnamefont {Lin}}, \bibinfo {author} {\bibfnamefont {J.}~\bibnamefont {Holler}}, \bibinfo {author} {\bibfnamefont {J.~M.}\ \bibnamefont {Bauer}}, \bibinfo {author} {\bibfnamefont {P.}~\bibnamefont {Parzefall}}, \bibinfo {author} {\bibfnamefont {M.}~\bibnamefont {Scheuck}}, \bibinfo {author} {\bibfnamefont {B.}~\bibnamefont {Peng}}, \bibinfo {author} {\bibfnamefont {T.}~\bibnamefont {Korn}}, \bibinfo {author} {\bibfnamefont {S.}~\bibnamefont {Bange}}, \bibinfo {author} {\bibfnamefont {J.~M.}\ \bibnamefont {Lupton}}, \ and\ \bibinfo {author} {\bibfnamefont {C.}~\bibnamefont {Schüller}},\ }\bibfield  {title} {\enquote {\bibinfo {title} {{Large-Scale Mapping of Moiré Superlattices by Hyperspectral Raman Imaging}},}\ }\href {\doibase https://doi.org/10.1002/adma.202008333} {\bibfield  {journal} {\bibinfo  {journal} {Advanced Materials}\ }\textbf {\bibinfo {volume} {33}},\ \bibinfo {pages} {2008333} (\bibinfo {year} {2021})}\BibitemShut {NoStop}%
\bibitem [{\citenamefont {Puretzky}\ \emph {et~al.}(2016)\citenamefont {Puretzky}, \citenamefont {Liang}, \citenamefont {Li}, \citenamefont {Xiao}, \citenamefont {Sumpter}, \citenamefont {Meunier},\ and\ \citenamefont {Geohegan}}]{Alex2016}%
  \BibitemOpen
  \bibfield  {author} {\bibinfo {author} {\bibfnamefont {A.~A.}\ \bibnamefont {Puretzky}}, \bibinfo {author} {\bibfnamefont {L.}~\bibnamefont {Liang}}, \bibinfo {author} {\bibfnamefont {X.}~\bibnamefont {Li}}, \bibinfo {author} {\bibfnamefont {K.}~\bibnamefont {Xiao}}, \bibinfo {author} {\bibfnamefont {B.~G.}\ \bibnamefont {Sumpter}}, \bibinfo {author} {\bibfnamefont {V.}~\bibnamefont {Meunier}}, \ and\ \bibinfo {author} {\bibfnamefont {D.~B.}\ \bibnamefont {Geohegan}},\ }\bibfield  {title} {\enquote {\bibinfo {title} {{Twisted MoSe2 Bilayers with Variable Local Stacking and Interlayer Coupling Revealed by Low-Frequency Raman Spectroscopy}},}\ }\href {\doibase 10.1021/acsnano.5b07807} {\bibfield  {journal} {\bibinfo  {journal} {ACS Nano}\ }\textbf {\bibinfo {volume} {10}},\ \bibinfo {pages} {2736} (\bibinfo {year} {2016})}\BibitemShut {NoStop}%
\bibitem [{\citenamefont {Gunnarsson}(1997)}]{Gunnarsson_97}%
  \BibitemOpen
  \bibfield  {author} {\bibinfo {author} {\bibfnamefont {O.}~\bibnamefont {Gunnarsson}},\ }\bibfield  {title} {\enquote {\bibinfo {title} {Superconductivity in fullerides},}\ }\href {\doibase 10.1103/RevModPhys.69.575} {\bibfield  {journal} {\bibinfo  {journal} {Rev. Mod. Phys.}\ }\textbf {\bibinfo {volume} {69}},\ \bibinfo {pages} {575--606} (\bibinfo {year} {1997})}\BibitemShut {NoStop}%
\bibitem [{\citenamefont {Goodenough}(1998)}]{Goodenough98}%
  \BibitemOpen
  \bibfield  {author} {\bibinfo {author} {\bibfnamefont {J.~B.}\ \bibnamefont {Goodenough}},\ }\bibfield  {title} {\enquote {\bibinfo {title} {{Jahn-Teller Phenomena In Solids}},}\ }\href {\doibase 10.1146/annurev.matsci.28.1.1} {\bibfield  {journal} {\bibinfo  {journal} {Annual Review of Materials Science}\ }\textbf {\bibinfo {volume} {28}},\ \bibinfo {pages} {1} (\bibinfo {year} {1998})}\BibitemShut {NoStop}%
\bibitem [{\citenamefont {Liu}\ \emph {et~al.}(2010)\citenamefont {Liu}, \citenamefont {Zhang}, \citenamefont {Brinkley}, \citenamefont {Bian}, \citenamefont {Miller},\ and\ \citenamefont {Chiang}}]{Liu_10}%
  \BibitemOpen
  \bibfield  {author} {\bibinfo {author} {\bibfnamefont {Y.}~\bibnamefont {Liu}}, \bibinfo {author} {\bibfnamefont {Longxiang}\ \bibnamefont {Zhang}}, \bibinfo {author} {\bibfnamefont {M.~K.}\ \bibnamefont {Brinkley}}, \bibinfo {author} {\bibfnamefont {G.}~\bibnamefont {Bian}}, \bibinfo {author} {\bibfnamefont {T.}~\bibnamefont {Miller}}, \ and\ \bibinfo {author} {\bibfnamefont {T.-C.}\ \bibnamefont {Chiang}},\ }\bibfield  {title} {\enquote {\bibinfo {title} {{Phonon-Induced Gaps in Graphene and Graphite Observed by Angle-Resolved Photoemission}},}\ }\href {\doibase 10.1103/PhysRevLett.105.136804} {\bibfield  {journal} {\bibinfo  {journal} {Phys. Rev. Lett.}\ }\textbf {\bibinfo {volume} {105}},\ \bibinfo {pages} {136804} (\bibinfo {year} {2010})}\BibitemShut {NoStop}%
\bibitem [{\citenamefont {V.~Bostr{\"o}m}\ \emph {et~al.}(2023)\citenamefont {V.~Bostr{\"o}m}, \citenamefont {Sriram}, \citenamefont {Claassen},\ and\ \citenamefont {Rubio}}]{Emil_23}%
  \BibitemOpen
  \bibfield  {author} {\bibinfo {author} {\bibfnamefont {E.}~\bibnamefont {V.~Bostr{\"o}m}}, \bibinfo {author} {\bibfnamefont {A.}~\bibnamefont {Sriram}}, \bibinfo {author} {\bibfnamefont {M.}~\bibnamefont {Claassen}}, \ and\ \bibinfo {author} {\bibfnamefont {A.}~\bibnamefont {Rubio}},\ }\bibfield  {title} {\enquote {\bibinfo {title} {Controlling the magnetic state of the proximate quantum spin liquid $\alpha$-{R}u{C}l$_3$ with an optical cavity},}\ }\href {\doibase 10.1038/s41524-023-01158-6} {\bibfield  {journal} {\bibinfo  {journal} {npj Computational Materials}\ }\textbf {\bibinfo {volume} {9}},\ \bibinfo {pages} {202} (\bibinfo {year} {2023})}\BibitemShut {NoStop}%
\bibitem [{\citenamefont {Sandbo~Chang}\ \emph {et~al.}(2020)\citenamefont {Sandbo~Chang}, \citenamefont {Sabín}, \citenamefont {Forn-Díaz}, \citenamefont {Quijandría}, \citenamefont {Vadiraj}, \citenamefont {Nsanzineza}, \citenamefont {Johansson},\ and\ \citenamefont {Wilson}}]{Chang2020}%
  \BibitemOpen
  \bibfield  {author} {\bibinfo {author} {\bibfnamefont {C.~W.}\ \bibnamefont {Sandbo~Chang}}, \bibinfo {author} {\bibfnamefont {C.}~\bibnamefont {Sabín}}, \bibinfo {author} {\bibfnamefont {P.}~\bibnamefont {Forn-Díaz}}, \bibinfo {author} {\bibfnamefont {F.}~\bibnamefont {Quijandría}}, \bibinfo {author} {\bibfnamefont {A.~M.}\ \bibnamefont {Vadiraj}}, \bibinfo {author} {\bibfnamefont {I.}~\bibnamefont {Nsanzineza}}, \bibinfo {author} {\bibfnamefont {G.}~\bibnamefont {Johansson}}, \ and\ \bibinfo {author} {\bibfnamefont {C.~M.}\ \bibnamefont {Wilson}},\ }\bibfield  {title} {\enquote {\bibinfo {title} {{Observation of Three-Photon Spontaneous Parametric Down-Conversion in a Superconducting Parametric Cavity}},}\ }\href {\doibase 10.1103/PhysRevX.10.011011} {\bibfield  {journal} {\bibinfo  {journal} {Phys. Rev. X}\ }\textbf {\bibinfo {volume} {10}},\ \bibinfo {pages} {011011} (\bibinfo {year} {2020})}\BibitemShut {NoStop}%
\bibitem [{\citenamefont {Minganti}\ \emph {et~al.}(2023)\citenamefont {Minganti}, \citenamefont {Garbe}, \citenamefont {Le~Boit\'e},\ and\ \citenamefont {Felicetti}}]{Minganti2023}%
  \BibitemOpen
  \bibfield  {author} {\bibinfo {author} {\bibfnamefont {F.}~\bibnamefont {Minganti}}, \bibinfo {author} {\bibfnamefont {L.}~\bibnamefont {Garbe}}, \bibinfo {author} {\bibfnamefont {A.}~\bibnamefont {Le~Boit\'e}}, \ and\ \bibinfo {author} {\bibfnamefont {S.}~\bibnamefont {Felicetti}},\ }\bibfield  {title} {\enquote {\bibinfo {title} {Non-{G}aussian superradiant transition via three-body ultrastrong coupling},}\ }\href {\doibase 10.1103/PhysRevA.107.013715} {\bibfield  {journal} {\bibinfo  {journal} {Phys. Rev. A}\ }\textbf {\bibinfo {volume} {107}},\ \bibinfo {pages} {013715} (\bibinfo {year} {2023})}\BibitemShut {NoStop}%
\bibitem [{\citenamefont {Cartella}\ \emph {et~al.}(2018)\citenamefont {Cartella}, \citenamefont {Nova}, \citenamefont {Fechner}, \citenamefont {Merlin},\ and\ \citenamefont {Cavalleri}}]{Cartella18}%
  \BibitemOpen
  \bibfield  {author} {\bibinfo {author} {\bibfnamefont {A.}~\bibnamefont {Cartella}}, \bibinfo {author} {\bibfnamefont {T.~F.}\ \bibnamefont {Nova}}, \bibinfo {author} {\bibfnamefont {M.}~\bibnamefont {Fechner}}, \bibinfo {author} {\bibfnamefont {R.}~\bibnamefont {Merlin}}, \ and\ \bibinfo {author} {\bibfnamefont {A.}~\bibnamefont {Cavalleri}},\ }\bibfield  {title} {\enquote {\bibinfo {title} {Parametric amplification of optical phonons},}\ }\href {\doibase 10.1073/pnas.1809725115} {\bibfield  {journal} {\bibinfo  {journal} {Proc. Nat. Acad. Sci. U.S.A.}\ }\textbf {\bibinfo {volume} {115}},\ \bibinfo {pages} {12148} (\bibinfo {year} {2018})}\BibitemShut {NoStop}%
\bibitem [{\citenamefont {Teitelbaum}\ \emph {et~al.}(2018)\citenamefont {Teitelbaum}, \citenamefont {Henighan}, \citenamefont {Huang}, \citenamefont {Liu}, \citenamefont {Jiang}, \citenamefont {Zhu}, \citenamefont {Chollet}, \citenamefont {Sato}, \citenamefont {Murray}, \citenamefont {Fahy}, \citenamefont {O'Mahony}, \citenamefont {Bailey}, \citenamefont {Uher}, \citenamefont {Trigo},\ and\ \citenamefont {Reis}}]{Teitelbaum18}%
  \BibitemOpen
  \bibfield  {author} {\bibinfo {author} {\bibfnamefont {S.~W.}\ \bibnamefont {Teitelbaum}}, \bibinfo {author} {\bibfnamefont {T.}~\bibnamefont {Henighan}}, \bibinfo {author} {\bibfnamefont {Y.}~\bibnamefont {Huang}}, \bibinfo {author} {\bibfnamefont {H.}~\bibnamefont {Liu}}, \bibinfo {author} {\bibfnamefont {M.~P.}\ \bibnamefont {Jiang}}, \bibinfo {author} {\bibfnamefont {D.}~\bibnamefont {Zhu}}, \bibinfo {author} {\bibfnamefont {M.}~\bibnamefont {Chollet}}, \bibinfo {author} {\bibfnamefont {T.}~\bibnamefont {Sato}}, \bibinfo {author} {\bibfnamefont {E.~D.}\ \bibnamefont {Murray}}, \bibinfo {author} {\bibfnamefont {S.}~\bibnamefont {Fahy}}, \bibinfo {author} {\bibfnamefont {S.}~\bibnamefont {O'Mahony}}, \bibinfo {author} {\bibfnamefont {T.~P.}\ \bibnamefont {Bailey}}, \bibinfo {author} {\bibfnamefont {C.}~\bibnamefont {Uher}}, \bibinfo {author} {\bibfnamefont {M.}~\bibnamefont {Trigo}}, \ and\ \bibinfo {author} {\bibfnamefont {D.~A.}\ \bibnamefont {Reis}},\ }\bibfield  {title} {\enquote {\bibinfo {title}
  {Direct measurement of anharmonic decay channels of a coherent phonon},}\ }\href {\doibase 10.1103/PhysRevLett.121.125901} {\bibfield  {journal} {\bibinfo  {journal} {Phys. Rev. Lett.}\ }\textbf {\bibinfo {volume} {121}},\ \bibinfo {pages} {125901} (\bibinfo {year} {2018})}\BibitemShut {NoStop}%
\bibitem [{\citenamefont {Juraschek}\ \emph {et~al.}(2020)\citenamefont {Juraschek}, \citenamefont {Meier},\ and\ \citenamefont {Narang}}]{Juraschek20}%
  \BibitemOpen
  \bibfield  {author} {\bibinfo {author} {\bibfnamefont {D.~M.}\ \bibnamefont {Juraschek}}, \bibinfo {author} {\bibfnamefont {Q.~N.}\ \bibnamefont {Meier}}, \ and\ \bibinfo {author} {\bibfnamefont {P.}~\bibnamefont {Narang}},\ }\bibfield  {title} {\enquote {\bibinfo {title} {Parametric excitation of an optically silent goldstone-like phonon mode},}\ }\href {\doibase 10.1103/PhysRevLett.124.117401} {\bibfield  {journal} {\bibinfo  {journal} {Phys. Rev. Lett.}\ }\textbf {\bibinfo {volume} {124}},\ \bibinfo {pages} {117401} (\bibinfo {year} {2020})}\BibitemShut {NoStop}%
\bibitem [{\citenamefont {Maehrlein}\ \emph {et~al.}(2017)\citenamefont {Maehrlein}, \citenamefont {Paarmann}, \citenamefont {Wolf},\ and\ \citenamefont {Kampfrath}}]{Maehrlein17}%
  \BibitemOpen
  \bibfield  {author} {\bibinfo {author} {\bibfnamefont {S.}~\bibnamefont {Maehrlein}}, \bibinfo {author} {\bibfnamefont {A.}~\bibnamefont {Paarmann}}, \bibinfo {author} {\bibfnamefont {M.}~\bibnamefont {Wolf}}, \ and\ \bibinfo {author} {\bibfnamefont {T.}~\bibnamefont {Kampfrath}},\ }\bibfield  {title} {\enquote {\bibinfo {title} {Terahertz sum-frequency excitation of a {R}aman-active phonon},}\ }\href {\doibase 10.1103/PhysRevLett.119.127402} {\bibfield  {journal} {\bibinfo  {journal} {Phys. Rev. Lett.}\ }\textbf {\bibinfo {volume} {119}},\ \bibinfo {pages} {127402} (\bibinfo {year} {2017})}\BibitemShut {NoStop}%
\bibitem [{\citenamefont {Juraschek}\ and\ \citenamefont {Maehrlein}(2018)}]{Juraschek18}%
  \BibitemOpen
  \bibfield  {author} {\bibinfo {author} {\bibfnamefont {D.~M.}\ \bibnamefont {Juraschek}}\ and\ \bibinfo {author} {\bibfnamefont {S.~F.}\ \bibnamefont {Maehrlein}},\ }\bibfield  {title} {\enquote {\bibinfo {title} {Sum-frequency ionic {R}aman scattering},}\ }\href {\doibase 10.1103/PhysRevB.97.174302} {\bibfield  {journal} {\bibinfo  {journal} {Phys. Rev. B}\ }\textbf {\bibinfo {volume} {97}},\ \bibinfo {pages} {174302} (\bibinfo {year} {2018})}\BibitemShut {NoStop}%
\bibitem [{\citenamefont {Johnson}\ \emph {et~al.}(2019)\citenamefont {Johnson}, \citenamefont {Knighton},\ and\ \citenamefont {Johnson}}]{Johnson19}%
  \BibitemOpen
  \bibfield  {author} {\bibinfo {author} {\bibfnamefont {C.~L.}\ \bibnamefont {Johnson}}, \bibinfo {author} {\bibfnamefont {B.~E.}\ \bibnamefont {Knighton}}, \ and\ \bibinfo {author} {\bibfnamefont {J.~A.}\ \bibnamefont {Johnson}},\ }\bibfield  {title} {\enquote {\bibinfo {title} {Distinguishing nonlinear terahertz excitation pathways with two-dimensional spectroscopy},}\ }\href {\doibase 10.1103/PhysRevLett.122.073901} {\bibfield  {journal} {\bibinfo  {journal} {Phys. Rev. Lett.}\ }\textbf {\bibinfo {volume} {122}},\ \bibinfo {pages} {073901} (\bibinfo {year} {2019})}\BibitemShut {NoStop}%
\bibitem [{\citenamefont {Breuer}\ and\ \citenamefont {Petruccione}(2002)}]{BreuerPetruccione}%
  \BibitemOpen
  \bibfield  {author} {\bibinfo {author} {\bibfnamefont {H.~P.}\ \bibnamefont {Breuer}}\ and\ \bibinfo {author} {\bibfnamefont {F.}~\bibnamefont {Petruccione}},\ }\bibfield  {title} {\enquote {\bibinfo {title} {{Open Quantum Systems}},}\ }\href@noop {} {\bibfield  {journal} {\bibinfo  {journal} {Cambridge University Press, Cambridge, England}\ } (\bibinfo {year} {2002})}\BibitemShut {NoStop}%
\bibitem [{\citenamefont {Landau}\ and\ \citenamefont {Lifshitz}(1976)}]{Landau}%
  \BibitemOpen
  \bibfield  {author} {\bibinfo {author} {\bibfnamefont {L.~D.}\ \bibnamefont {Landau}}\ and\ \bibinfo {author} {\bibfnamefont {E.~M.}\ \bibnamefont {Lifshitz}},\ }\bibfield  {title} {\enquote {\bibinfo {title} {{Mechanics, Course of Theoretical Physics}},}\ }\href@noop {} {\bibfield  {journal} {\bibinfo  {journal} {Butterworth-Heinenann, Oxford,}\ }\textbf {\bibinfo {volume} {1}} (\bibinfo {year} {1976})}\BibitemShut {NoStop}%
\bibitem [{\citenamefont {Liang}\ and\ \citenamefont {Meunier}(2014)}]{Liang_14}%
  \BibitemOpen
  \bibfield  {author} {\bibinfo {author} {\bibfnamefont {L.}~\bibnamefont {Liang}}\ and\ \bibinfo {author} {\bibfnamefont {V.}~\bibnamefont {Meunier}},\ }\bibfield  {title} {\enquote {\bibinfo {title} {First-principles {R}aman spectra of {M}o{S}$_2$, {WS}$_2$ and their heterostructures.}}\ }\href {https://api.semanticscholar.org/CorpusID:264680658} {\bibfield  {journal} {\bibinfo  {journal} {Nanoscale}\ }\textbf {\bibinfo {volume} {6}},\ \bibinfo {pages} {5394} (\bibinfo {year} {2014})}\BibitemShut {NoStop}%
\end{thebibliography}%

\pagebreak
\onecolumngrid

\begin{center}
\textbf{\large Supplementary Material for "Equilibrium Parametric Amplification in Raman-Cavity Hybrids"}
\end{center}

\setcounter{equation}{0}
\setcounter{figure}{0}
\setcounter{table}{0}
\setcounter{page}{1}
\makeatletter
\renewcommand{\theequation}{S\arabic{equation}}
\renewcommand{\thefigure}{S\arabic{figure}}


\section{Protocols and numerical implementation}

We solve the stochastic Heisenberg-Langevin equations of motion introduced in the main text using the truncated Wigner approximation (TWA) method~\cite{polkovnikov_phase_2010,cosme_time_2019}. The equations read
\begin{align}
\begin{split} \partial_{t}{a} =& -i\omega_c a - 2 i g (a+a^{*})(b+b^{*})- i g_4 (a+a^{*})^{3}- \kappa a + \xi_{a} , \label{eq:smotion1}
\end{split}
 \\ \label{eq:smotion2} \partial_{t}{b_r} =& \omega_R b_i,
  \\ \partial_{t}{b_i} =& -\omega_R b_r -  g (a+a^{*})^2 -2 g_s E_p (t) a_{s,r}- \gamma b_i + \xi_{b}, \label{eq:smotion2Im}
   \\ \partial_{t} a_{s,r} =& \omega_s a_{s,i}-\kappa_s a_{s,r} + \xi_{s,r},
 \\ \partial_t a_{s,i} =& -\omega_s a_{s,r}-2 g_s E_p (t) b_r -\kappa_s a_{s,i}+ \xi_{s,i}. \label{eq:smotion3}
\end{align}
where the subscripts $r,i$ denote the real and imaginary part of the field. To initialize the modes we sample from the corresponding Wigner distributions. We assume that all our modes have an expectation value of zero. Hence, the Wigner distribution from which we sample corresponds to a Gaussian distribution with mean zero and standard deviation of $1/2$. For each set of parameter we sample over $15000$ trajectories. We further include stochastic delta-correlated noise $\xi_{a}, \xi_{b}$ and $\xi_{s}$ satisfying $\left<\xi_a^{*}(t_1)\xi_a(t_2)\right>=\kappa\delta(t_1-t_2)$,  $\left<\xi_b(t_1)\xi_b(t_2)\right>=\gamma \delta(t_1-t_2)$ and $\left<\xi_s^{*}(t_1)\xi_s(t_2)\right>=\kappa_s\delta(t_1-t_2)$.
Initially we ramp up the Raman-cavity coupling $g$ from zero to its final value at time $t_0$ as:
\begin{equation}
    g(t)=g(\tanh((t-t_0)/\tau)+1)/2.
\end{equation}
We hold this coupling for the rest of the dynamics until the steady state is reached and turn on the probing field at time $t_p$ afterwards. We consider a probing field with an associated electric field  $E_p(t)=E^{0}_p (t) \sin({\omega_p t})$ with
\begin{equation}
    E^{0}_p(t)=E^{0}_p(\tanh((t-t_p)/\tau)+1)/2.
\end{equation}
Finally, to obtain the Raman spectra, we compute the number of scattered photons $n_s=a^{*}_s a_s=a_{s,r}^2+a_{s,i}^2$ at a fixed time $t^*\gg t_p$ by averaging over all the realizations.
We take $\omega_R t_0= 10$ and for the decay rates that we use in the main text $\omega_R t_p \approx 100$ is enough to be in the steady state. We use $\tau=1/\omega_R$ and checked that same results can be obtained for very different values $\tau=10/\omega_R$. We choose $\omega_R t^*=250$ which means the system is under the probing field during a time window $t^* - t_p=150/\omega_R$. With these parameters we obtain clear Raman spectra. 


To compute the modification of cavity and Raman fluctuations; $\delta x^2$ and $\delta Q^2$, as well as Raman shift $Q$ shown in Fig. 3 of the main text, we drop the probe field ($E_p (t)=0$) in the equation of motion and solve the dynamics to compute such observables at the steady state $\omega_R t_p \approx 100$.

\section{Raman spectra in the presence of thermal noise}

Here we present the Raman spectra for the combined system in the presence of thermal noise associated to both the cavity and Raman mode. In this case we solve the dynamics Eq.~(S1)-(S5) but now considering white noises satisfying 
\begin{equation}
\left<\xi_a^{*}(t_1)\xi_a(t_2)\right>=\kappa\coth (\omega_c/2k_B T)\delta(t_1-t_2),
\end{equation}
\begin{equation}
\left<\xi_b(t_1)\xi_b(t_2)\right>=\gamma\coth (\omega_R/2k_B T)\delta(t_1-t_2)
\end{equation}
 where $k_B$ is the Boltzmann constant. These autocorrelation relations guarantee the fluctuation-dissipation theorem hold for both subsystem (in the uncoupled case) assuming they are  connected to a reservoir at temperature $T$ and considering a Markovian approximation~\cite{BreuerPetruccione}.

 \begin{figure}[!tbh]
\hspace*{-0.7cm} 
\includegraphics[width=0.75\textwidth]{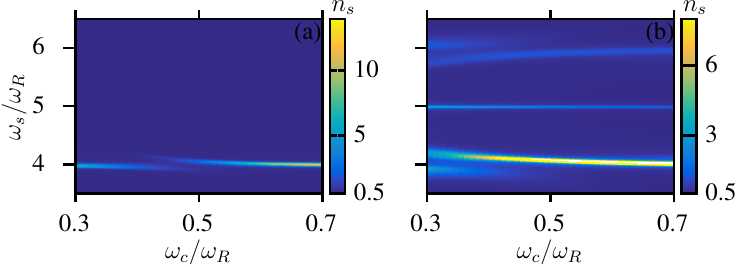}
\caption{(a) Raman spectra $n_s\left(\omega_s \right)$ for different cavity frequencies $\omega_c$ in the quantum noise limit. (b) Raman spectra including thermal noise with $k_B T=2.5\hbar\omega_R$. In both cases we consider a probe laser of strength $g_s E_p^0=0.04\omega_R$ and frequency $\omega_p=5\omega_R$ with $g=0.04\omega_R$, $g_4=\kappa=\gamma=\kappa_s=0.01\omega_R$.} 
\label{fig:spectrathermalnoise}
\end{figure} 

In Fig.~\ref{fig:spectrathermalnoise}(a) we plot the same Raman spectra shown in the main text (in the quantum noise limit) to be contrasted with the Raman spectra in the presence of thermal noise shown in Fig.~\ref{fig:spectrathermalnoise}(b).
In both cases we show the raw data $n_s(\omega_s)$ instead of $n_s(\omega)$ with $\omega= \omega_p - \omega_s$ being the Raman shift. It allows us to see Stoke and anti-Stoke contributions separately.
As discussed in the main text, in the quantum noise limit (Fig.~\ref{fig:spectrathermalnoise}(a)), only a Stoke peak appears in the spectra around $\omega_s=\omega_p -\omega_R=4\omega_R$. In contrast, if thermal noise is added anti-Stoke processes also occur and we find additional peaks around $\omega_s=\omega_p +\omega_R=6\omega_R$ (see Fig.~\ref{fig:spectrathermalnoise}(b)).

\section{Gaussian theory for Cavity fluctuations}

In this section, we complement the numerical Truncated Wigner simulations by analytical calculations using gaussian theory. The purpose of the gaussian theory is to derive analytical formulas for the parametric Raman phonon polariton branches, and show perturbatively that a parametric resonance is indeed present in the ground state. Apart from analytical intuition, the agreement between numerics (Raman spectra) and the analytical dispersion derived by the gaussian theory serves as a cross check on our results.

The semi-classical Langevin equations of motion, able to capture vacuum fluctuations for bosonic modes are given by the equations~(S1)-(S3) where the noise terms, $\xi_a$ and $\xi_b$ are delta-correlated noise, $\langle \xi_a^*(t_1) \xi_a(t_2) \rangle = \kappa \delta(t_1 - t_2) $, $\langle \xi_b(t_1) \xi_b(t_2)\rangle = \gamma \delta(t_1 - t_2) $ and $\langle \xi_a(t_1) \xi_b (t_2) \rangle =0$. The equations of motion in terms of the frequency Fourier components defined as: $a(\omega) = \int dt e^{i \omega t
} a(t)$ and $a(t) = \int \frac{d \omega}{2 \pi} e^{- i \omega t} a(\omega)$, are given for the Raman mode by combining equations (S2)-(S3) as: 
\begin{align}
    \left(- \omega^2 - i \gamma \omega + \omega^2_R  \right) b_r =&  - g \omega_R \int \frac{d \omega'}{2 \pi} (a + a^*)(\omega- \omega') (a + a^*)( \omega') + \omega_R \xi_b(\omega) .
\label{eq:EqOfbr}
\end{align}
Replacing this expression into Eq.~(S1) for the cavity we find:
\begin{equation}
\begin{split}
    - i \omega a(\omega) =& - i \omega_c a(\omega) - \kappa a(\omega) + \xi_a(\omega) \\
    &- 2 i g\int \frac{d \omega'}{2 \pi} (a(\omega - \omega') + a^*(\omega' - \omega)) \Bigg( \frac{2 \omega_R}{- (\omega')^2 - i \gamma \omega' + \omega_R^2} \xi_b(\omega') \\
    &- \frac{2 g \omega_R}{- (\omega')^2 - i \gamma \omega' + \omega_R^2}\int \frac{d \omega''}{2 \pi} (a(\omega'- \omega'') + a^*( \omega''- \omega')) (a(\omega'') + a^*(-\omega'')) \Bigg) \\
    &- i g_4  \int \frac{d \omega' d \omega''}{(2 \pi)^2} (a(\omega- \omega')  + a^*( \omega' - \omega)) ( a(\omega' - \omega'') + a^*(\omega''- \omega') (a(\omega'') + a^*(- \omega'')).
\end{split}
\label{eq:EqNonlinea}
\end{equation}
At this level, Eq.~(\ref{eq:EqNonlinea}) is an exact representation of our full Langevin dynamics in terms of nonlinear equation of motion of $a$ and the noise terms $\xi_a$ and $\xi_b$. To derive equilibrium properties, we use a Gaussian approximation in order to derive effective linear equations of motion perform using Wick's theorem. In particular, we assume: 
\begin{align}
    a(\omega_1) a(\omega_2) a(\omega_3) \approx& \langle a(\omega_1) a(\omega_2)\rangle a(\omega_3) + \langle a(\omega_1) a(\omega_3)\rangle a(\omega_2) + \langle a(\omega_2) a(\omega_3)\rangle a(\omega_1), \label{eq:Wicks}\\
    \langle \xi_b(\omega_1) a(\omega_2) \rangle \approx& \langle \xi_b(\omega_1) \rangle a(\omega_2) + \langle a(\omega_2) \rangle \xi_b(\omega_1) = 0 ,
\end{align}
where by definition $\langle \xi_b(\omega_1) \rangle = 0$ and in the ground state, inversion symmetry is conserved such that we have no coherent state of photons $\langle a(\omega_1) \rangle = 0$. This approximation is valid only as long as the connected four-point expectation value, $\langle a(\omega_1) a(\omega_2) a(\omega_3) a(\omega_4) \rangle_c $, and higher order connected correlators are small. This type of correlations are generated by nonlinearities therefore the gaussian theory should be applicable at least in the perturbative regime for small couplings, $g$ and $g_4$.  

In the absence of any interactions, $g = g_4 = 0$, Eq.~(\ref{eq:EqNonlinea}) recovers the expectation value of the vacuum fluctuations:  $\langle \frac{a^*(t) a(t)+ a(t) a^*(t)}{2}\rangle = \int \frac{d \omega d \omega'}{(2 \pi)^2} e^{ i t (\omega- \omega')} a^*(\omega) a(\omega' ) =  \int \frac{d \omega}{2 \pi} \frac{\kappa}{\kappa^2 + \left( \omega - \omega_c\right)^2} = \frac{1}{2} $. After using Wick's theorem (see Eq.~(\ref{eq:Wicks})) the equations can be simplified further by using time-translation symmetry in equilibrium which constraints gaussian expectations value to take the form $\langle a^*(\omega) a(\omega')\rangle = 2 \pi \delta(\omega- \omega') n(\omega)$, 
 $\langle a(\omega) a(\omega') \rangle = 2 \pi \delta(\omega + \omega') f(\omega) $and are expressed only in terms of a frequency dependent occupation number, $n(\omega) $ and a frequency dependent squeezing parameter, $f(\omega)$. Within this approximation, Eq.~(\ref{eq:EqNonlinea}) becomes:
 \begin{equation}
 \begin{split}
     - i \omega a(\omega) =& - i \omega_c a(\omega) - \kappa a(\omega) + \xi_a(\omega) \\
    &i \frac{4 g^2}{\omega_R} \left( a(\omega) + a^*(- \omega) \right)\left( \int \frac{d\omega'}{2\pi} \left( 2 n(\omega') + f(\omega') + f^*(\omega') \right) \right)  \\
    &i 8 g^2 \left(a(\omega) + a^*(- \omega) \right) \int \frac{d\omega'}{2\pi}\frac{\omega_R}{- (\omega')^2 - i \gamma \omega  + \omega_R^2}\left( 2 n(\omega - \omega') + f(\omega - \omega') + f^*(\omega - \omega') \right)   \\ 
     &-i 3 g_4   \left(a( \omega) +  a^*(- \omega) \right)\left( \int \frac{d\omega''}{2\pi} \left( 2 n(\omega') + f(\omega') + f^*(\omega') \right) \right).
\end{split}
\end{equation}
The above result can be compactly re-written in terms of an effective cavity frequency, $\bar{\omega}_{c}(\omega)$ and a squeezing force $\Delta(\omega)$:
\begin{equation}
    (-i\omega + \kappa)a(\omega) = - i \bar{\omega}_{c}(\omega) a(\omega) - i \Delta(\omega) a^*( - \omega) + \xi_a(\omega),
\end{equation}
where the effective parameters are given by:
\begin{align}
\begin{split}    
\bar{\omega}_{c}(\omega) =& \omega_c + \left(- \frac{4 g^2} {\omega_R} + 3 g_4\right) \left( \int \frac{d\omega'}{2\pi} \left( 2 n(\omega') + f(\omega') + f^*(\omega') \right) \right)\\
& - 8 g^2 \left( \int \frac{d\omega'}{2\pi} \frac{\omega_R}{-(\omega')^2 - i \gamma \omega' + \omega_R^2 }\left( 2 n(\omega - \omega') + f(\omega - \omega') + f^*(\omega - \omega') \right) \right)
\end{split} , \\
\begin{split}    \Delta(\omega) =& \left(- \frac{4 g^2} {\omega_R} + 3 g_4\right) \left( \int \frac{d\omega'}{2\pi} \left( 2 n(\omega') + f(\omega') + f^*(\omega') \right) \right)\\
& - 8 g^2 \left( \int \frac{d\omega'}{2\pi} \frac{\omega_R}{-(\omega')^2 - i \gamma \omega' + \omega_R^2 }\left( 2 n(\omega - \omega') + f(\omega - \omega') + f^*(\omega - \omega') \right) \right).
\end{split}
\end{align}
The fluctuations are determined through the dependence of $a(\omega)$, $a^*(-\omega)$ to the noise terms $\xi(\omega)$ and $\xi^*(-\omega)$ : 
\begin{align}
    &\begin{pmatrix} - i \omega + \kappa + i \bar{\omega}_{c}(\omega) && 
 i \Delta(\omega) 
 \\ - i \Delta(-\omega) && 
 - i \omega + \kappa - i \bar{\omega}_{c}(-\omega) \end{pmatrix} \cdot \begin{pmatrix} a(\omega) \\
 a^*( - \omega) \end{pmatrix}  = \begin{pmatrix} \xi_a(\omega) \\
 \xi^*_a(-\omega) \end{pmatrix} ,
\end{align}
which leads to:
\begin{equation}
    a(\omega) = \frac{(- i \omega + \kappa - i \bar{\omega}_{c}(-\omega)) \xi_a(\omega) - i \Delta(\omega) \xi^*_a(-\omega)}{
-\Delta(-\omega) \Delta(\omega) + (- i \omega + \kappa - i \bar{\omega}_{c}(-\omega)) (- i \omega + \kappa + i \bar{\omega}_{c}(\omega))}.
\end{equation}
Finally, the ground state fluctuations are found by solving the self consistent equations: $\langle a^*(\omega) a(\omega')\rangle = 2 \pi \delta(\omega- \omega') n(\omega)$, 
 $\langle a(\omega) a(\omega') \rangle = 2 \pi \delta(\omega + \omega') f(\omega) $.

\subsection{Renormalized cavity frequency}

We explore the normalized cavity frequency quoted in the main text analytically by using perturbation theory in $g^2$ and $g_4$. We express the fluctuations as a series expansion, $n(\omega) \approx n_0(\omega) + n_1(\omega)$ and $f(\omega) = f_0(\omega) + f_1(\omega)$, where in the absence of any coupling to the Raman mode, the fluctuations take the form:
\begin{align}
    n_0(\omega) =& \frac{\kappa}{\kappa^2 + (\omega- \omega_c)^2}, \\
    f_0(\omega) =& 0,
\end{align}
To leading order in the couplings, the renormalized frequency, $\bar{\omega}_{c}(\omega)$, and squeezing parameter, $\Delta(\omega)$ are given by:
\begin{align}
    \bar{\omega}_{c}(\omega) = \omega_c &- \frac{4 g^2} {\omega_R} + 3 g_4 - 8 g^2 \frac{\omega_R}{ \omega_R^2 - \left( \omega - \omega_c\right)^2 - (i \gamma + 2 i \kappa) ( \omega - \omega_c) + \gamma \kappa + \kappa^2  }, \\
    \Delta(\omega) = &- \frac{4 g^2} {\omega_R} + 3 g_4 - 8 g^2 \frac{\omega_R}{ \omega_R^2 - \left( \omega - \omega_c\right)^2 - (i \gamma + 2 i \kappa) ( \omega - \omega_c) + \gamma \kappa + \kappa^2  }
\end{align}
for small $\kappa$ and $\gamma$. Similarly, to leading order, the squeezing parameter is given by $\Delta = \Delta(\omega_c) = - \frac{12 g^2}{\omega_R} + 3 g_4$. Corrections in the frequency of the cavity mode due to the squeezing go as $|\Delta(\omega)|^2 \sim \mathcal{O}(g^4, g_4 g^2 , g_4^2)$, and corresponds to a higher order contribution. As a result, to leading order in the couplings the cavity response frequency is given by:
\begin{equation}
 \bar{\omega}_{c}(\omega)  = \bar{\omega}_{c}(\omega = \omega_c ) = \omega_c - \frac{12 g^2}{\omega_R} + 3 g_4. 
\end{equation}

\subsection{Parametric enhancement of cavity fluctuations}

To linear order in $g^2$ and $g_4$, the fluctuation functions take the form:
\begin{align}
    n(\omega) =& \frac{\kappa}{\kappa^2 + (\omega - \bar{\omega}_{c})^2}, \\
\begin{split} f(\omega) =& \frac{i \Delta(\omega)}{( i \omega + \kappa - i \omega_{c} )(- i \omega + \kappa - i \omega_{c} ) (- i \omega + \kappa + i \omega_{c})}+\\
&\frac{i \Delta(-\omega)}{( -i \omega + \kappa - i \omega_{c} )(i \omega + \kappa - i \omega_{c} ) (i \omega + \kappa + i \omega_{c})}
\end{split}
\end{align}

The squeezing term, $f(\omega)$, is resonantly amplified when $\omega_c \approx \omega_R/2$, showing that Gaussian theory can indeed capture the non-trivial equilibrium amplification process. On resonance, perturbation theory breaks down and one should self-consistently solve for $f(\omega)$ and $n(\omega)$. In this Letter, we instead rely on the numerically evaluated solution. 

\section{Raman-cavity polariton frequency}

To analytically extract the Raman polariton branches, we consider the linear dynamics of excitations on top of the equilibrium fluctuations discussed in the previous section.

As mentioned in the main text, the Raman coherent oscillations linearly hybridized with squeezing fluctuations of the cavity mode. Here for convenience we write the Hamiltonian in the alternative but equivalent form
\begin{equation}
    H = \frac{\omega_{R}^2}{2} \hat{Q}^2 + \frac{\hat{P}^2}{2} + g\times 2 \omega_c \sqrt{2 \omega_R } \hat{Q} \hat{x}^2 + \omega_c^2 \frac{\hat{x}^2}{2} + \frac{\hat{p}^2}{2} + g_4 \omega^2_c \hat{x}^4
    \label{eq:hamilt}
\end{equation}
where the Raman coordinate is defined as $\hat{Q} = \frac{\hat{b} + \hat{b}^\dag}{\sqrt{2 \omega_R} }$, the Raman conjugate momentum as $\hat{P} = \frac{i \sqrt{\omega_R} (\hat{b}^\dag - \hat{b})}{\sqrt{2}}$, the cavity coordinate as $\hat{x} = \frac{\hat{a} + \hat{a}^\dag}{\sqrt{2 \omega_c} } $ and the cavity conjugate momentum as $\hat{p} = \frac{i \sqrt{\omega_c} (\hat{a}^\dag - \hat{a})}{\sqrt{2}}$. In this basis, completing the square in Eq.~\eqref{eq:hamilt} the Hamiltonian reads:
\begin{equation}
    H = \frac{\omega_{R}^2}{2} \left( \hat{Q} + g \frac{2\sqrt{2 \omega_R} \omega_c}{\omega_R^2} \hat{x}^2 \right)^2  + \left( g_4 \omega^2_c  - \frac{4 g^2 \omega_c^2}{\omega_R} \right) \hat{x}^4 + \frac{\hat{P}^2}{2} + \omega_c^2 \frac{\hat{x}^2}{2} + \frac{\hat{p}^2}{2},
\end{equation}
which leads to the condition for stability quoted in the main text, $g_4 > \frac{4 g^2}{\omega_R}$.

The equations of motion for the Raman mode, $\hat{Q}$, and the fluctuations of the cavity mode, $\left ( \hat{x}^2, \{\hat{x} , \hat{p}\} ,\hat{p}^2 \right)$ are given by:
\begin{align}
    \frac{d \hat{Q}}{dt} =& \hat{P}, \\
    \frac{d \hat{P}}{dt} =&   - \omega_R^2 \hat{Q}  - g\times 2 \omega_c \sqrt{2 \omega_R } \hat{x}^2 , \\
    \frac{d \hat{x}^2}{dt} =& \{\hat{x}, \hat{p}\} ,\\
    \frac{d \{\hat{x}, \hat{p}\}}{dt} =& 2 \hat{p}^2 - 2 \omega_c^2 \hat{x}^2 - 4 g\times 2 \omega_c \sqrt{2 \omega_R } \hat{x}^2 \hat{Q} - 2 g_4 \times 4 \omega^2_c \hat{x}^4, \\
    \frac{d \hat{p}^2}{d t} =& - \omega_c^2 \{ \hat{x}, \hat{p} \} - 2 g\times 2 \omega_c \sqrt{2 \omega_R } \hat{Q} \{\hat{x}, \hat{p}\} - g_4 \times 4 \omega^2_c \{ \hat{x}^3, \hat{p}\}
\end{align}
where $\{ \hat{A} , \hat{B} \} = \hat{A}\hat{B} + \hat{B}\hat{A}$ is the anti-commutator. Within a Gaussian approximation theory, 
\begin{align}
    \langle \hat{x}^4 \rangle \approx& 3 \langle \hat{x}^2 \rangle^2 ,\\
    \langle \hat{x}^2 \hat{Q} \rangle \approx& \langle \hat{x}^2 \rangle \langle \hat{Q} \rangle ,\\
    \langle \{ \hat{x}^3, \hat{p} \} \rangle \approx& 3 \langle \hat{x}^2 \rangle \langle \{ \hat{x} , \hat{p} \} \rangle , \\
    \langle \{ \hat{x} , \hat{p} \} \hat{Q} \rangle \approx& \langle \{ \hat{x} , \hat{p} \} \rangle \langle \hat{Q} \rangle
\end{align}
 and the Raman coordinate and cavity fluctuations form a complete system of equations in terms of the variables, $\left\{ \langle \hat{Q} \rangle, \langle \hat{P} \rangle, \langle \hat{x}^2 \rangle, \langle \{\hat{x}, \hat{p}\} \rangle, \langle \hat{p}^2 \rangle \right\}$:
\begin{align}
    \frac{d \langle \hat{Q} \rangle }{dt} =& \langle \hat{P} \rangle, \\
    \frac{d \langle \hat{P} \rangle }{dt} =&   -\omega_R^2 \langle \hat{Q} \rangle - g\times 2 \omega_c \sqrt{2 \omega_R } \langle \hat{x}^2 \rangle  , \\
    \frac{d \langle \hat{x}^2 \rangle }{dt} =& \langle \{\hat{x}, \hat{p}\} \rangle ,\\
    \frac{d \langle \{\hat{x}, \hat{p}\} \rangle}{dt} =& 2 \langle \hat{p}^2 \rangle  - 2 \omega_c^2 \langle \hat{x}^2 \rangle - 4 g\times 2 \omega_c \sqrt{2 \omega_R } \langle  \hat{x}^2\rangle \langle \hat{Q} \rangle - 6 g_4 \times 4 \omega^2_c \langle \hat{x}^2 \rangle^2, \\
    \frac{d \langle \hat{p}^2 \rangle }{d t} =& - \omega_c^2 \langle \{ \hat{x}, \hat{p} \} \rangle  - 2 g\times 2 \omega_c \sqrt{2 \omega_R } \langle \hat{Q} \rangle \langle \{\hat{x}, \hat{p}\} \rangle  - 3 g_4 \times 4 \omega^2_c\langle \hat{x}^2\rangle \langle \{ \hat{x}, \hat{p}\} \rangle.
\end{align}
To make progress we first compute the equilibrium correlations, $\langle \hat{Q} \rangle = Q_0$, $\langle \hat{x}^2\rangle = x_0^2$ and $\langle \hat{p}^2\rangle = p_0^2$ within the Gaussian self-consistent approximation theory by taking the derivative of all quantities equal to zero in the above expressions which produces: 
\begin{align}
    Q_0 =& - \frac{g\times 2 \omega_c \sqrt{2 \omega_R }}{\omega_R^2} x_0^2, \\
    p_0^2 =& \omega_c^2 x_0^2 - 2 \frac{(g\times 2 \omega_c \sqrt{2 \omega_R })^2}{\omega_R^2} x_0^4 + 3 g_4 \times 4 \omega^2_c x^4_0, \\
    \langle \{ x , p \} \rangle_0 =& 0, 
\end{align}
The above expressions allows us to express the equilibrium expectations values only in terms of $x^2_0$. The value of $x_0^2$ is determined by the equilibrium state of the system, which we compute either through Truncated Wigner or through the self-consistent gaussian theory described in the previous section and depends on temperature. For the purpose of the Raman dynamics however, $x_0^2$, can be considered as an external input, and we can extract the coherent dynamics of the Raman variable as a function of $x_0^2$. 

We linearize around the equilibrium to find the collective modes. We consider dynamics of a coherent shift of the Raman mode $\langle \hat{Q} \rangle = Q_0 + Q_1$ and $\langle \hat{P} \rangle = P_1 $, together with fluctuations of cavity mode:  $\langle x^2 \rangle = x_0^2 + x_1^2 $, $\langle \{ x , p\} \rangle = \{ x , p \}_1 $ and $\langle p^2 \rangle = p_0^2 + p_1^2 $. Note that odd expectation values of photon operators are zero because Raman excitation is even under inversion and cannot excite electric and magnetic fields coherently. The linearized equations are given by: 
\begin{align}
    \partial_t  Q_1=& P_1, \\ 
     \partial_t P_1 =&  - \omega_R^2 Q_1 - \left( 2 g \omega_c \sqrt{2 \omega_R } \right) x^2_1, \\
    \partial_t x^2_1 =& \{ x , p \}_1, \\ 
   \partial_t \{ x , p \}_1 =&  - 8 g \omega_c \sqrt{2 \omega_R } x_0^2 Q_1  - 2 \omega_c^2\left(  1 +  \left(- 16\frac{g^2}{\omega_R} + 24 g_4 \right) x_0^2 \right) x_1^2 + 2 p^2_1, \\
   \partial_t p^2_1  =& \left(- \omega_c^2 - 4 g \omega_c \sqrt{2 \omega_R } Q_0 - 12 g_4 \omega^2_c x_0^2 \right)  \{x, p \}_1. \label{eq:HomogGaus}
\end{align}
Considering an oscillating ansatz of the type $Q \sim e^{ i \omega t} $, we find two distinct solutions corresponding to the hybridized Raman mode with photon fluctuations (Raman polariton branches):
\begin{equation}
\begin{split}    
\omega_{\pm}^2 =& \frac{1}{2 \omega_R} \Bigg(4 \omega_c^2 \omega_R +\omega_R^3+ 72 g_4 \omega^2_c \omega_R x_0^2   - 64 g^2\omega^2_c  x_0^2   \\
&\pm \sqrt{
  -16 \omega_c^2 \omega_R^3 \left( -24 g^2 x_0^2 + 18 g_4 x_0^2 \omega_R + \omega_R \right) + \left( - 64 g^2 x_0^2 \omega_c^2 + \left(4 + 72 g_4 x_0^2 \right)\omega_c^2 \omega_R + \omega_R^3 \right)^2
}\Bigg)
\end{split}
\end{equation}


\section{Nonlinearities, dissipation and thermal effects on the parametric resonances}

For the sake of completeness here we show the effects of increasing the nonlinear interaction and decay rates on the parametric resonance discussed in the main text. We also show that the amplification and localization of cavity fluctuations and Raman mode also appear by considering the whole system in a thermal state at relative low temperatures.

\begin{figure}[!tbh]
\hspace*{-0.7cm} 
\includegraphics[width=0.75\textwidth]{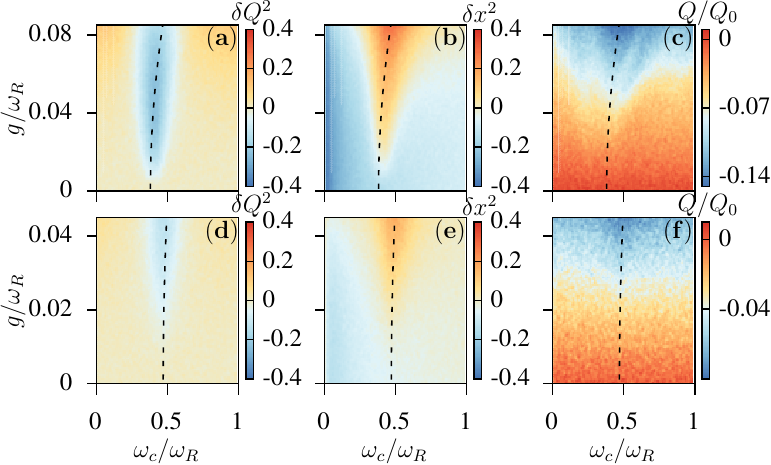}
\caption{Nonlinearities and dissipation effects on the parametric resonances. (a), (b), (c) The same as Fig. 3 of the main text but increasing the quartic nonlinearity to $g_4=0.04\omega_R$ keeping the same decay rates. (d), (e), (f) The same as Fig. 3 of the main text but in this case the decay rates have been increased to $\kappa=\gamma=0.06\omega_R$ keeping the same $g_4$. Dashed lines are $\omega_R=2\bar{\omega}_c$.} 
\label{fig:nonliearanddiss}
\end{figure}

Fig.~\ref{fig:nonliearanddiss} (a-c) shows $\delta Q^2$, $\delta x^2$ and $Q$ for a larger value of $g_4$. The main effect is a shift on the parametric resonance to the left while how much localize the Raman mode is, how much it is shifted and how much the cavity field is amplified remain practically the same. This shift to the left results from the analytical resonant condition $\omega_R/2=\bar{\omega}_c=\omega_c-12 g^2/\omega_R+3g_4$ which is the dashed line that matches nicely with the numerical TWA simulations (in color). For larger values of couplings $g$ and $g_4$, not only the linear dependence of the parametric resonance on $g_4$ is well described by this analytical expression, but also the slow quadratic dependence on $g$.

In Fig.~\ref{fig:nonliearanddiss}(d-f) we show how by increasing $\kappa$ and $\gamma$ four times while keeping $g_4$ constant, the resonance weakens with the Raman mode being less localized and cavity field less amplified in the steady state. Here the Raman and cavity fluctuations are modified by $\sim 20\%$ in resonance. Also the onset of the resonance is pushed to higher coupling strengths $g$ which is reminiscent of the physics of a periodically driven parametric oscillator in the presence of damping, where a critical amplitude of the external drive is needed to overcome dissipation and get into the zone of amplification~\cite{Landau}. For these larger decay rates the Raman shift decreases and becomes more independent on the cavity frequency (see Fig.~\ref{fig:nonliearanddiss} (f)). 

 We have checked that even for stronger nonlinearities (higher value of $g_4$) and/or strong dissipation the parametric resonance can still be seen so there is a resonant regime in which Raman mode fluctuations decrease in favor of cavity field amplification.

\begin{figure}[!tbh]
\hspace*{-0.7cm} 
\includegraphics[width=0.75\textwidth]{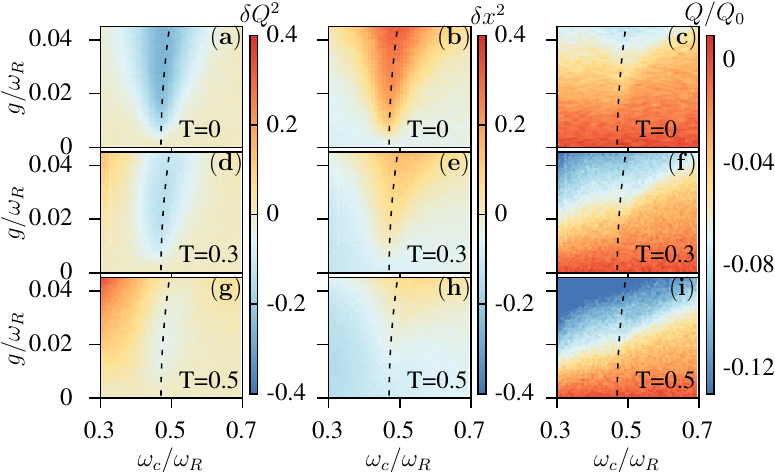}
\caption{Thermal effects on the parametric resonances. In panels (a), (b), (c) we repeat Fig. 3 of the main text (quantum noise limit) to be contrasted with (d), (e), (f) and (g), (h), (i)  where the different observables $\delta Q^2$, $\delta x^2$ and $Q/Q_0$ are computed considering a thermal state with $T=0.3\hbar\omega_R/k_B$ and  $T=0.5\hbar\omega_R/k_B$ respectively.}
\label{fig:thermalphasediag}
\end{figure}

Finally in Fig.~\ref{fig:thermalphasediag} we plot the variation of the Raman fluctuations $\delta Q^2$ and cavity fluctuations $\delta x^2$ as well as Raman coordinate $Q$ in the quantum noise limit ((a),(b),(c)) and for a thermal state at $T=0.3 \hbar\omega_R/k_B$ ((d),(e),(f)) and $T=0.5 \hbar\omega_R/k_B$ ((g),(h),(i)). It is clear that for thermal states at relative low temperature, the localization of the Raman mode, amplification of cavity fluctuations and the shift in the Raman coordinate also occur. Interestingly, when comparing the thermal fluctuating cases with the quantum one ((a),(b),(c)), it turns out that thermal fluctuations are less efficient in localizing the Raman mode and amplifying the cavity fluctuations. However they work better than the quantum fluctuating case for shifting the Raman coordinate. This indicates that our main findings are based on the fluctuating character of the system and not on its classical or quantum nature. However, we would like to emphasize that the parametric resonant phenomenon is more pronounced in the quantum limit which is clear by contrasting the results in panels ((g),(h),(i)) with those of panels ((a),(b),(c)).

An interesting observation is that a reduction of the Raman fluctuations can be interpreted as a cooling of the Raman mode in the context of thermal states. The Raman fluctuations in thermal equilibrium and in the absence of a coupling to the cavity mode is given by $\left< Q^2 \right>_0=Q_0\coth (\omega_R/2k_B T)/2$.  So essentially a reduction in the Raman fluctuations can be interpreted as a decrease on the phonon temperature. 
Considering for instance $T=0.5\hbar\omega_R/k_B$ and a coupling $g=0.01\omega_R$, $\delta Q^2\sim -0.05$ (see Fig.~\ref{fig:thermalphasediag} (g)) which could be interpreted as a relative temperature reduction of $ \frac{\Delta T}{T} = -0.09 $.

\section{Raman phonon-cavity coupling strength}

Following the references~\cite{Alex2016,Liang_14}, Raman phonons are coupled the electric field of light $\vec{E}(x)$ through the Raman tensor \underline{\underline{R}}. The Hamiltonian reads
\begin{equation}
    H = \int d^3 x  \frac{\epsilon_0 \vec{E}(x) \cdot \underline{\underline{\epsilon}}(Q) \cdot \vec{E}(x)}{2}  \approx H_0 +  \epsilon_0 \sum_i \frac{\vec{E}_i \cdot \underline{\underline{R}} \cdot\vec{E}_i}{2} Q_i,
    \label{eq:harl}
\end{equation}
where $\epsilon_0$ is the vacuum permittivity and $\underline{\underline{\epsilon}}(Q)$ is the phonon-coordinate dependent polarizability tensor (dielectric tensor). Expanding linearly in the phonon coordinate $Q$ one obtains a first term $H_0 = \int d^3 x  \frac{\epsilon_0 \vec{E}(x) \cdot \underline{\underline{\epsilon}}(Q = 0) \cdot \vec{E}(x)}{2}$ and the Raman-light coupling which is given by the second term in the right hand side of Eq.~\eqref{eq:harl}. We employ the dipole approximation for the cavity mode where we assume that the electric field is constant over one unit cell and given by $\vec{E}_i =  \frac{1}{V_{cell}}\int_{V^i} d^3 x \vec{E}(x)$ where $V_{cell}$ is the volume of the unit cell and $V^i$ the volume of the i-th unit cell. The Raman tensor is defined as 
\begin{equation}
R_{\alpha \beta} = V_{cell} \sum_{\mu = 1}^{N}\sum_{l = 1}^{3} \frac{\partial \epsilon_{\alpha \beta}}{\partial r_l(\mu)} \frac{e_l(\mu)}{\sqrt{M_\mu}}
\end{equation}
where $r_l(\mu)$ is the position of the $\mu$th atom along the direction $l$,  $\frac{\partial \epsilon_{\alpha \beta}}{\partial r_l(\mu)}$ is the first derivative of the dielectric tensor over the atomic displacement, $e_l(\mu)$ is the displacement of the $\mu$th atom along the direction $l$ of the Raman phonon and $M_\mu$ is the mass of the $\mu$th atom. Considering that the cavity mode has a constant electric field over the entire sample, $\hat{E}_{cav ,i} = E_{0} \vec{e}_{c} (\hat{a} + \hat{a}^\dag)$, the Raman-light coupling only involves the $q = 0$ phonon. Thus $\sum_i \hat{Q}_i = \sqrt{N} \hat{Q}_{q = 0} = \sqrt{N} \frac{\hat{b} + \hat{b}^\dag}{\sqrt{2 \omega_R}}  $, where $N = \frac{V_{samp}}{V_{cell}}$ is the total number of unit cells ($V_{samp}$ is the volume of the sample), giving rise to the coupling:
\begin{align}
    \frac{g}{\omega_R} =& \frac{\epsilon_0 E^2_{0} \sqrt{V_{samp} V_{cell} }}{2 \omega_R} \tilde{R}, \\
    \tilde{R} =& \vec{e}_{c} \cdot \sum_{\mu = 1}^{N}\sum_{l = 1}^{3} \frac{\partial \underline{\underline{\epsilon}}}{\partial r_l(\mu)} \frac{e_l(\mu)}{\sqrt{M_\mu} \sqrt{2 \omega_R}} \cdot \vec{e}_{c},
\end{align}
where $\tilde{R}$ is the dimensionless Raman coupling and $\vec{e}_{c}$ is the polarization vector of the cavity field. Following references~\cite{Alex2016,Liang_14} $\tilde{R} \sim 1 - 10$ for TMDs. The electric field of cavities is given by the relationship:
\begin{equation}
    E_{0}^2 = \frac{\omega_{c}}{\epsilon_0 V_{eff}},
\end{equation}
where on parametric resonance $\omega_c = \omega_R/2$, leading to the final result:
\begin{equation}
    \frac{g}{\omega_R} = \frac{1}{4} \tilde{R} \frac{\sqrt{V_{samp} V_{cell}}}{V_{eff}}.
\end{equation}
These arguments are rather crude and detailed research needs to be carried out for different cavity designs and Raman-active materials on a case by case basis.

\end{document}